\documentclass{article}

\usepackage[final]{neurips_2021}

\usepackage[utf8]{inputenc} % allow utf-8 input
\usepackage[T1]{fontenc}    % use 8-bit T1 fonts
\usepackage{hyperref}       % hyperlinks
\usepackage{url}            % simple URL typesetting
\usepackage{booktabs}       % professional-quality tables
\usepackage{amsfonts}       % blackboard math symbols
\usepackage{amsthm}
\usepackage{comment}
\usepackage{nicefrac}       % compact symbols for 1/2, etc.
\usepackage{microtype}      % microtypography
\usepackage{xcolor}         % colors
\usepackage[compact]{titlesec}
\titlespacing{\section}{0pt}{2ex}{1ex}

\usepackage[english]{babel}
\usepackage[utf8]{inputenc}
\usepackage[compact]{titlesec}

\usepackage{cmap}
\usepackage[T1]{fontenc}
\usepackage{bm}
\usepackage{amsmath}
\usepackage{amsfonts}
\usepackage{amssymb}
\usepackage{amsbsy}
\usepackage{amsthm}
\usepackage{dsfont}

\usepackage{mathtools}
\usepackage{xspace}
\usepackage{mleftright,xparse}

\usepackage{rotating}
\usepackage{float}
\usepackage{tikz}

\usepackage{listings}

\usepackage{enumitem}
\usepackage{multirow}
\usepackage{array}

\usepackage{accents}

\usepackage[outline]{contour}% http://ctan.org/pkg/contour
\usepackage{xcolor}

\usepackage{chngcntr}
\usepackage{soul}
\usepackage{arydshln}

\usepackage{thm-restate}
\newcommand{\calD}{\ensuremath{\mathcal{D}}}

\newcommand{\calN}{\ensuremath{\mathcal{N}}}

\newcommand{\calP}{\ensuremath{\mathcal{P}}}

\NewDocumentCommand\xDeclarePairedDelimiter{mmm}
 {%
  \NewDocumentCommand#1{som}{%
   \IfNoValueTF{##2}
    {\IfBooleanTF{##1}{#2##3#3}{\mleft#2##3\mright#3}}
    {\mathopen{##2#2}##3\mathclose{##2#3}}%
  }%
 }

\xDeclarePairedDelimiter\inner{\langle}{\rangle}
\xDeclarePairedDelimiter\ang{\langle}{\rangle}
\xDeclarePairedDelimiter\abs{\lvert}{\rvert}
\xDeclarePairedDelimiter\set{\{}{\}}
\xDeclarePairedDelimiter\p{(}{)}
\let\b=\relax
\xDeclarePairedDelimiter\b{[}{]}
\xDeclarePairedDelimiter\round{\lfloor}{\rceil}
\xDeclarePairedDelimiter\floor{\lfloor}{\rfloor}
\xDeclarePairedDelimiter\ceil{\lceil}{\rceil}
\xDeclarePairedDelimiter\norm{\lVert}{\rVert}
\theoremstyle{plain}            % following are "theorem" style
\newtheorem{theorem}{Theorem}[section]
\newtheorem{lemma}[theorem]{Lemma}
\newtheorem{inftheorem}[theorem]{Informal Theorem}
\newtheorem{corollary}[theorem]{Corollary}

\theoremstyle{definition}       % following are def style
\newtheorem{definition}[theorem]{Definition}

\newtheorem{assumption}[theorem]{Assumption}

\theoremstyle{remark}           % following are remark style
\newtheorem{remark}[theorem]{Remark}

\numberwithin{equation}{section}

\contourlength{0.1pt}
\contournumber{10}

\def\poly{\mathrm{poly}}

\newif\ifnotes\notestrue
\ifnotes
\usepackage{color}
\definecolor{mygrey}{gray}{0.50}
\newcommand{\notename}[2]{{\textcolor{red}{\footnotesize{\bf (#1:} {#2}{\bf
) }}}}

\else

\newcommand{\notename}[2]{{}}

\fi

\DeclareMathOperator*{\Exp}{{\mathbb{E}}}

\DeclareMathOperator*{\Cov}{{\mathrm{Cov}}}
\DeclareMathOperator*{\Var}{\mathrm{Var}}

\mathchardef\mdash="2D % Define a "math hyphen"

\newcommand{\eps}{\varepsilon}

\renewcommand{\epsilon}{\varepsilon}

\newcommand{\paragr}[1]{\noindent \textbf{#1}}

\DeclareMathOperator*{\argmin}{argmin}

\def\compactify{\itemsep=0pt \topsep=0pt \partopsep=0pt \parsep=0pt}
\let\latexusecounter=\usecounter

{\begin{itemize}%
\setlength{\itemsep}{0pt}%
\setlength{\topsep}{0pt}%
\setlength{\partopsep}{0 in}%
\setlength{\parskip}{0 pt}}%
{\end{itemize}}

\definecolor{niceRed}{RGB}{190,38,38}
\definecolor{blueGrotto}{HTML}{059DC0}
\definecolor{royalBlue}{HTML}{057DCD}
\definecolor{navyBlueP}{HTML}{0B579C}
\definecolor{limeGreen}{HTML}{81B622}

\title{Efficient Truncated Linear Regression with Unknown Noise Variance}

\author{%
  Constantinos Daskalakis \\
  \thanks{CD and PS were supported by NSF Awards IIS-1741137, CCF-1617730, and CCF-1901292, by a Simons Investigator Award, by the Simons Collaboration on the Theory of Algorithmic Fairness, by a DSTA grant, 11 and by the DOE PhILMs project (No. DE-AC05-76RL01830)}
  EECS and CSAIL, MIT\\
  \texttt{costis@csail.mit.edu} \\ 
  \And 
  Patroklos Stefanou \\
  EECS and CSAIL, MIT \\
  \texttt{stefanou@mit.edu} \\
  \And 
  Rui Yao \\
  EECS and CSAIL, MIT\\
  \texttt{rayyao@mit.edu} \\
  \And 

  Manolis Zampetakis \thanks{MZ was supported by NSF DMS-2023505 (FODSI)} \\
  EECS, UC Berkeley \\
  \texttt{mzampet@berkeley.edu} \\
}

\usepackage{pdfpages}
\usepackage{wrapfig}
\usepackage[font=small]{caption, subcaption}
\usepackage{listings}
\definecolor{codegreen}{rgb}{0,0.6,0}
\definecolor{codegray}{rgb}{0.5,0.5,0.5}
\definecolor{codepurple}{rgb}{0.58,0,0.82}
\definecolor{backcolour}{rgb}{0.95,0.95,0.92}
\lstdefinestyle{mystyle}{
    backgroundcolor=\color{backcolour},   
    commentstyle=\color{codegreen},
    keywordstyle=\color{magenta},
    numberstyle=\tiny\color{codegray},
    stringstyle=\color{codepurple},
    basicstyle=\ttfamily\footnotesize,
    breakatwhitespace=false,         
    captionpos=b,                    
    keepspaces=true,                 
    numbers=left,                    
    numbersep=5pt,                  
    showspaces=false,                
    showstringspaces=false,
    showtabs=false,                  
    tabsize=2
}
\hypersetup{
  colorlinks = true,
  urlcolor = {blueGrotto},
  linkcolor = {royalBlue},
  citecolor = {navyBlueP}
}

\definecolor{niceRed}{RGB}{190,38,38}
\definecolor{blueGrotto}{HTML}{059DC0}
\definecolor{royalBlue}{HTML}{057DCD}
\definecolor{navyBlueP}{HTML}{0B579C}
\definecolor{limeGreen}{HTML}{81B622}

\usepackage{color-edits}
\addauthor{mz}{niceRed}
\usepackage{algorithm}
\usepackage{algpseudocode}
\renewcommand{\algorithmicrequire}{\textbf{Input:}} % Use Input in the format of Algorithm
\renewcommand{\algorithmicensure}{\textbf{Output:}} % Use Output in the format of Algorithm
\usepackage{subcaption}
\begin{document}
\newcommand{\supp}{\mathrm{supp}}

\maketitle

\begin{abstract}
Truncated linear regression is a classical challenge in statistics, wherein 
a label, $y = w^T x + \varepsilon$, and its corresponding feature vector, $x \in \mathbb{R}^k$, are only observed if the label 
falls in some subset $S \subseteq \mathbb{R}$; otherwise the existence of the
pair $(x, y)$ is hidden from observation. Linear regression with truncated observations has remained a challenge, in its general form, since the early
works of~\cite{tobin1958estimation,amemiya1973regression}. When the distribution of the
error is normal with known variance, recent work of~\cite{daskalakis2019computationally} provides computationally and statistically efficient estimators of the linear model, $w$. In this paper, we provide the first computationally and statistically efficient
estimators for truncated linear regression when the noise variance is unknown, estimating both the linear model and the variance of the noise. Our estimator is based on an efficient implementation of Projected Stochastic Gradient Descent on the negative log-likelihood of the truncated sample. Importantly, we show that the error of our estimates is asymptotically normal, and we use this to provide explicit confidence regions for our estimates.

% We provide two estimators, using two different instantiations Projected Stochastic Gradient Descent (PSGD), with or without replacement, on the negative log-likelihood of the truncated sample. Each estimator attains better convergence rates in different ranges of parameters. % and therefore we provide the first estimation for the truncated linear 
% regression problem that also has explicit confidence region. The method that 
% we present in this paper and hence be used not only for estimation but also for 
% inference.
\end{abstract}

\section{Introduction} \label{sec:intro}

A common challenge facing statistical estimation is the systematic omission of relevant data from the data used to train models. Data omission is a prominent form of dataset bias, which may occur for a variety of reasons, including incorrect experimental design or data collection campaigns, which prevent certain sub-populations from being observed, instrument errors or saturation, which make certain measurements unreliable, societal biases, which may suppress the realization of certain samples, as well as legal or privacy constraints, which might prevent the use of some data. In turn, it is well-understood that failure to account for the systematic omission of data in statistical inference may lead to incorrect models, and this has motivated a long line of research in statistics, econometrics, and a range of other theoretical and applied fields, targeting statistical inference  that is robust to missing data; see e.g.~\cite{Galton1897,Pearson1902,PearsonLee1908,fisher31,tukey49,tobin1958estimation,amemiya1973regression,hausman1977social,heckman79,breen1996regression,hajivassiliou1998method,Pearl_data_missingness,BalakrishnanCramer}.

In this paper, we revisit the classical challenge of {\em truncated linear regression}. As in the standard linear regression setting, we assume that the world produces pairs $(x^{(i)},y^{(i)})_i$, where each $x^{(i)} \in \mathbb{R}^k$ is a feature vector drawn from some distribution $\calP$ and its corresponding label $y^{(i)}$ is linearly related to $x^{(i)}$ according to  $y^{(i)}={w^*}^{ T}x^{(i)}+\varepsilon^{(i)}$, for both some unknown 
% but fixed across different~$i$,  
coefficient vector $w^* \in \mathbb{R}^k$ and random noise $\varepsilon^{(i)} \sim {\cal N}(0,{\sigma^*}^2)$, where ${\sigma^*}^2$ is unknown and $\varepsilon^{(i)}$ is $i.i.d$ over all $i$. The difference to the standard setting is that we do not get to observe all pairs $(x^{(i)},y^{(i)})$ that the world produces. Rather, a pair $(x^{(i)},y^{(i)})$ is only observed if $y^{(i)} \in S$, where $S\subset \mathbb{R}$ is a fixed ``observation set,'' that is known or given to us via oracle access, i.e.~we can query an oracle about whether or not some point belongs to $S$. If $y^{(i)} \notin S$, then $(x^{(i)},y^{(i)})$ is removed from observation. Our goal is to estimate $w^*$ and ${\sigma^*}^2$ from the pairs that survived removal, called ``truncated samples.'' 
%\footnote{In the related, easier challenge of {\em censored linear regression}, the setting is almost  identical, except  we   get to see the features of  pairs $(x^{(i)},y^{(i)})$ whose $y^{(i)} \notin S$, i.e.~those pairs are not completely removed from observation, but only their label is removed from observation. While we focus on truncated linear regression,  our results directly apply to the easier, yet still challenging, censored setting.} 

\paragraph{Overview of our results.}  While both truncated and censored linear regression are fundamental problems, which commonly arise in practice and have been studied in the literature since at least the 1950s~\citep{tobin1958estimation,amemiya1973regression}, there are no known statistical inference methods that are computationally efficient and enjoy statistical rates that scale with the dimension $k$ of the problem. Recent work of~\cite{daskalakis2019computationally} has obtained such methods in the simpler setting where the noise variance ${\sigma^*}^2$ is known. Our goal here is to extend those results in two important ways. First, we study settings where ${\sigma^*}^2$ is unknown, providing computationally and statistically efficient estimators for the joint estimation of $w^*$ and ${\sigma^*}^2$. Moreover, we establish the asymptotic normality of our estimators, and use this to derive confidence regions for our estimators. To achieve these stronger results we make some additional assumptions compared to \cite{daskalakis2019computationally}. Namely, we assume that the covariates are sampled from some prior distributions, and we also make a stronger assumption about the survival probability of our model as we discuss below.
Other works that consider learning and regression problems with censored or truncated samples are \cite{plevrakis2021learning, fotakis2020efficient, moitra2021learning, ilyas2020theoretical}.

We next provide informal statements of our results, postponing formal statements to Sections~\ref{sec: LL func} and~\ref{sec: confidence region}. It is important to note that, as  shown in prior work \citep{daskalakis2018efficient,daskalakis2019computationally}, for the model parameters to be identifiable from truncated samples, we need to place some assumptions that limit how aggressively the data is truncated. The assumptions we make are listed below, and stated more precisely in Section~\ref{s2}. They combine normalization assumptions (Assumption~\ref{asp:normBounds}), assumptions that are needed even without truncation (Assumption~\ref{asp:covarianceOfCovariates}), and assumptions that are needed for identifiability purposes in the presence of truncation (Assumption~\ref{asp:survivalProbability}).

% We need to do a regression problem for the dataset $(x,y)$ based on its linear model $y={w}^Tx+\varepsilon$. where $\varepsilon$ distributed in an unknown-variance centered normal distribution. Also, the truncation acts on $y$, that we will discard if and only if $y\notin S$ for some certain, prior given oracle-access set $S\subset \mathbb{R}$. 

% Unfortunately, we need to use stronger assumption then the previous work in \cite{daskalakis2019computationally}.

\begin{assumption} \label{asp:normBounds}
We assume that $\norm{w^*}_2^2 \le \beta$ %and the feature vectors are normalized, i.e., 
and $\norm{x^{(i)}}_2^2 \le 1$ for all $i$.
\end{assumption}

\begin{assumption} \label{asp:survivalProbability}
For all $(x^{(i)},y^{(i)})$ in the truncated dataset, conditioning on $x^{(i)}$, the probability that ${w^*}^T x^{(i)}+\varepsilon \in S$, with respect to the randomness in $\varepsilon \sim {\cal N}(0,{\sigma^*}^2)$, is larger than some $a$. 

% In other words, the observations in the truncated dataset did not survive removal by extreme luck.
\end{assumption}

\begin{assumption} \label{asp:covarianceOfCovariates}
If $(x^{(i)},y^{(i)})_{i=1}^n$ is the truncated dataset, we assume that the feature covariance matrix, $X={\frac{1}{n}}\sum_{i=1}^n{x^{(i)}}{x^{(i)}}^T$, satisfies: $\Exp_{x\sim\calP}[X]=\Exp_{x\sim\calP}[xx^T] \succeq b \cdot I$ for some $b$.
%; and (ii) there is no component ${x^{(i)}}{x^{(i)}}^T$ that contributes more than a $n/\log k$ fraction of $X$ in any direction.
\end{assumption}

With these assumptions we prove our main results for the joint estimation of $w^*$ and ${\sigma^*}^2$, listed as Informal Theorems~\ref{theorem:informalWithReplacement}--\ref{theorem:informalConfidenceRegion} below, and respectively Theorems~\ref{thm: main estimation} and~\ref{thm: asym normal} in later sections.

\begin{inftheorem}[Estimation] \label{theorem:informalWithReplacement}
    Suppose we are given $n$  samples $(x^{(i)},y^{(i)})_{i=1}^n$ from the truncated linear regression model with parameters $w^*$, ${\sigma^*}^2$, and suppose Assumptions~\ref{asp:normBounds}--\ref{asp:covarianceOfCovariates} hold. Then running Projected Stochastic Gradient Descent (Algorithm~\ref{algo:costis}) on the conditional negative log-likelihood of the sample, using %a judiciously chosen
    the projection set described in Definition~\ref{projset}, we obtain with probability at least $2/3$ (which can be boosted to as large a constant as desirable) estimations 
  $\hat w,\hat{\sigma}^2$ such that 
  $\norm{\hat{w} -w^*}_2 + \abs{\hat{\sigma}^2 - {\sigma^*}^2}\le \varepsilon$, as long as
  $n \ge \frac{C_1\log^4 (1/\varepsilon)}{\varepsilon^4}$ or  $n\ge \frac{C_2}{\varepsilon^2}$ (depending on the learning rate), where $C_1$ depends \textit{polynomially} on $\beta$, $1/a$ and $1/b$, and $C_2$ depends linearly in $1/b$ and exponentially in $\beta/a$, where $\beta$, $a$ and $b$ are respectively the constants in Assumptions~\ref{asp:normBounds}, \ref{asp:survivalProbability} and~\ref{asp:covarianceOfCovariates}.
\end{inftheorem}

% I changed the result of 1/\eps^4 to 1/\eps^2 by our settings

%\begin{inftheorem}[SGD w/out Replacement] \label{theorem:informalWithoutReplacement}
%    Suppose we are given $N$ samples $(x^{(i)},y^{(i)})_{i=1}^N$ from the truncated linear regression model with parameters $w^*$, ${\sigma^*}^2$, and suppose Assumptions 1--3 hold. Then running Projected Stochastic Gradient Descent (PSGD) without replacement on the negative log-likelihood of the sample for $n=N$ steps, as described in Algorithm \hyperref[*]{ ($\star$)} we obtain with at least some absolute constant probability (which can be tuned to as large a constant as is desirable) estimations   $\hat w,\hat{\sigma}^2$ such that   $||\hat w-w^*||_2^2+(\hat{\sigma}-\sigma^*)^2\le \varepsilon$, as long as $N \ge \frac{C_2}{\varepsilon^2}$, where $C_2$ depends exponentially on $1/a$, where $a$ is the constant in Assumption~1, as well as the ratio of the maximum norm of $w^*$ over ${\sigma^*}^2$, and polynomially on all other parameters.
  
%   the mass
%   of the survival set, the maximum norm of the covariates as well as the norm of
%   $w^*$ and ${\sigma^*}^2$.
  
%   \textit{polynomially} on
%   $1/a$ and $1/b$, where $a$ and $b$ are respectively the constants in Assumptions~1 and 3(i), the maximum norm of $w^*$ and ${\sigma^*}^2$.
% \end{inftheorem}

\begin{inftheorem}[Asymptotic Normality] \label{theorem:informalConfidenceRegion}
    Let $\hat{w}$, $\hat{\sigma}^2$ be the estimates of Informal Theorem~\ref{theorem:informalWithReplacement}. There
  exists a matrix $A$ that depends on $\hat{w}$, $\hat{\sigma}^2$ (and is explicitly provided in Theorem~\ref{thm: asym normal}) such that
  the random vector $\sqrt{n}(\binom{\hat{w}}{\hat{\sigma}^2}-\binom{w^*}{{\sigma^*}^2})$ converges asymptotically to a 
  normal distribution with zero mean and covariance matrix $A$. 
  Hence the $1 - \alpha$ confidence region of $(\hat{w}, \hat{\sigma}^2)$ can 
  be computed using the matrix $A$ and the quantiles of the chi-squared 
  distribution.
\end{inftheorem}

\paragraph{Techniques.}  Our estimates are computed by running Projected Stochastic Gradient Descent (PSGD) on the population conditional negative log-likelihood of the truncated sample; a convex function of the parameters, after an appropriate reparameterization. We provide two analyses of PSGD resulting in two different bounds that are more effective in a different range of parameters, as we discuss next.

Our first analysis hinges on establishing a lower bound on the convexity of the conditional negative log-likelihood function. We can't however, expect that a strong bound holds over the whole parameter domain. For this reason, we add a projection step to our method, identifying a projection set wherein we show that the true parameters lie and wherein the conditional negative log-likelihood is strongly convex. This approach is similar to that of \cite{daskalakis2019computationally}, except that the unknown noise variance introduces many challenges in identifying an appropriate projection set. The main result  bounding the strong convexity of the conditional negative log likelihood function within the projection set is stated as Theorem~\ref{thm:convexity, bounded step var}. The main disadvantage of this analysis is that, when the variance is unknown, the bound on the strong convexity of the conditional negative log-likelihood becomes small and the resulting dependence of the sample size on the parameters $\beta$ and $1/a$ of the problem is exponential. On the other hand, the dependence of the sample size on $1/b$ is linear and its dependence on the required estimation accuracy is $\Theta(1/\eps^2)$, which is optimal. Our other analysis follows an approach that has not been pursued in  prior work, and in particular does not rely on a global lower bound on the strong convexity of the conditional negative log-likelihood. Instead, we only bound the strong convexity at the optimum and use an upper bound on the smoothness of the objective function to prove the convergence of PSGD in terms of function value. Again, it is impossible to prove a bound on the smoothness over the whole domain, because large noise variance can lead to very bad smoothness bounds. Thus we use the same projection set as above and show that, within this set, we can get an upper bound
on the smoothness that depends polynomially on all parameters, $\beta, 1/a, 1/b$, of the 
problem. The main drawback of this method is that the dependence on the accuracy is  $\Theta(\log^4(1/\eps)/\eps^4)$. It is worth noting that obtaining an algorithm whose sample complexity has an optimal $\Theta(1/\eps^2)$ dependence on the estimation accuracy, and also scales polynomially in $1/a$ is a fundamental bottleneck, which is already present  in the much simpler special case of our problem of estimating the variance of a 1-dimensional truncated Gaussian, studied in \cite{daskalakis2018efficient}.\footnote{While \cite{daskalakis2018efficient} does not explicitly analyze the dependence of the sample complexity on $\alpha$; by tracing through their bounds, one can see that the dependence of the sample complexity on $\alpha$ is exponential.}

% To analyze SGD without replacement, we need to obtain a lower bound on the strong convexity of the negative log-likelihood function. We can't however, assume that such bound holds over the whole domain. For this reason, we add a projection step to our method, identifying a projection set where we show that the true parameters lie. This approach is similar to that of \cite{daskalakis2019computationally}, except that the unknown noise variance introduces lots of challenges both in identifying an appropriate projection set and in proving the fast convergence of PSGD without replacement within this set. The main result  bounding the strong convexity of the negative log likelihood function within the projection set is stated as Theorem \hyperref[10]{ 10}. Based on this bound, we are able to use recent results on the convergence of PSGD without replacement~\cite{shamir2016without} to argue that our method converges fast to the true parameters. The main bottleneck of this method is that when the variance is unknown the bound on the strong convexity of the negative log-likelihood becomes small and the resulting dependence of the sample size on the parameters of the problem is weaker. On the other hand, the dependence of the sample size on the required estimation accuracy $\eps$ is $\Theta(1/\eps^2)$, which is optimal.

Finally, to help build intuition around the intricacies in estimating truncated linear regression models with unknown noise variance, let's consider why natural approaches of reducing this problem to estimating  truncated linear regression models with {\em known} noise variance -- which can be done using the algorithm  of~\cite{daskalakis2019computationally} -- fail. A simple such reduction might compute the empirical variance $\sigma_0^2$ of the  truncated sample, and plug that into the algorithm of~\cite{daskalakis2019computationally}  to estimate the linear model, pretending that $\sigma_0^2$ is the true noise variance. A more sophisticated approach might grid over candidate noise variances, plug each of these candidates into the algorithm of~\cite{daskalakis2019computationally} to estimate a linear model, and then select the (linear model, noise variance) pair that is most consistent, e.g.~the pair whose noise variance is closest to the empirical variance of the residuals between the linear model's predictions and the observed labels. Both of these approaches fail, as suggested by the example below. Suppose that the ground truth  is $y=w^* x + w_0^* + \varepsilon$, where $x$ is one-dimensional, $(w^*,w_0^*)=(0,0)$, and the noise variance is ${\sigma^*}^2=1$. Suppose also that we create a truncated dataset by, repeatedly, sampling $x_i$ from ${\cal N}(0,1)$, generating $y_i$ according to the ground truth, and truncating the resulting pair $(x_i,y_i)$ unless $y_i \ge \tau$, where $\tau$ is some threshold. It is clear that, if we guess the true noise variance to be the empirical one, then as $\tau$ gets bigger, the empirical variance gets smaller, thus the linear model that would be estimated by treating the empirical variance as the true noise variance would have intercept that is close to $\tau$ rather than $0$. So this naive approach would fail. Let us consider the more sophisticated one that grids over candidate noise variances. In particular, let us compare what happens when (1) we guess the noise variance to be $\sigma_1^2={\sigma^*}^2$, and run the algorithm of~\cite{daskalakis2019computationally} with this guess to get an estimate $(\hat{w},\hat{w}_0)$; and (2) when we guess the noise variance to be $\sigma_2^2\equiv \sigma_0^2$ (the empirical variance) and run the algorithm of~\cite{daskalakis2019computationally} with   this guess to get an estimate $(\hat{w}',\hat{w}_0')$. It is easy to see that, as $\tau$ gets large, $\Big|{\sigma^*}^2 - {1 \over N}\sum_i (y_i - \hat{w}x_i-\hat{w}_0)^2 \Big| \gg \Big|\sigma_0^2 - {1 \over N}\sum_i (y_i - \hat{w}'x_i-\hat{w}_0')^2\Big|$ (here, $N$ is the number of samples. This holds even if $N$ goes to infinity). Thus the selection criterion proposed above will fail to prefer (1) over (2), even though the guess of the noise variance under (1) is the correct one. Our approach in this paper amounts to using a different criterion, namely selecting the pair with the largest conditional log-likelihood. But analyzing when and how well this works, as well as obtaining an efficient algorithm to find the best pair is the main contribution of this work.

{\bf Roadmap.} We present the truncated linear regression setting and our assumptions in Section \ref{s2}. We state and prove our main estimation results Section \ref{sec: LL func} and \ref{sec: overview}, and prove the asymptotic normality of our estimators in Section \ref{sec: confidence region}. Finally, in Section~\ref{sec:experiments} we assess the performance of our methods.

\section{Models and assumptions} \label{s2}

We study the truncated linear regression setting studied in~\cite{daskalakis2019computationally}, with the additional complexity that the variance of the noise model is unknown. In particular, let $S \subset \mathbb{R}$ be a measurable
subset of the real line, to which oracle access is provided. Assume we have $N=3n$ truncated samples $(x^{(i)}, y^{(i)})_{i=1}^N$ 
(we will split the samples into three, see \ref{sec:proofEnd}), each
generated according to the following procedure, for some unknown $w^* \in \mathbb{R}^k$ and $\sigma^* \in \mathbb{R}$ that are fixed across different $i$'s:
\begin{enumerate}
  \item sample $x^{(i)}$ from some distribution $\calP_0$ with support $\supp(\calP_0) \subseteq \mathbb{R}^k$;
  \item generate $y^{(i)}$ according to the following model, where $\varepsilon \sim \mathcal{N}(0, {\sigma^*}^2)$;
  \begin{equation}\label{model}
      y^{(i)} = {w^*}^T x^{(i)} + \varepsilon
  \end{equation} 
  \item if $y^{(i)}\in S$ then return $(x^{(i)}, y^{(i)})$ as the $i$th sample, otherwise repeat from step 1. 
\end{enumerate}
\vskip -0.5em
After truncation, denote the distribution of $x$ to be $\calP$ with the $\supp(\calP)=\supp(\calP_0)$. The sampled $x$ is still $i.i.d$, so we could only consider the distribution $\calP$. As already established in prior work~\citep{daskalakis2018efficient, daskalakis2019computationally, kontonis2019efficient}, some assumptions must be placed on $S$ and its interaction with the unknown parameters $(w^*, {\sigma^*}^2)$ and the $x^{(i)}$'s in order for the parameters to be identifiable. We state our assumptions after a useful  definition. 
% survival probability of a sample in our case and then we state formally the assumptions that we make for the rest of the paper. 
% We will make the assumptions listed below, which are a mixture of normalization assumptions (Assumption~\ref{asm: norm}), assumptions that are needed even without truncation (Assumption~\ref{asm: thickness}), and assumptions that we need for identifiability purposes in the presence of truncation (Assumption~\ref{asm: const svvl prob}).
% Before stating our assumptions, we define the survival
% probability of a given feature vector $x$, under parameters $(w,\sigma^2)$, and a given truncation set $S$.

% and th orderrecent works on truncated statistics
% \cite{daskalakis2019truncatedregressiprior workn} we may need some assumptions on $S$. From
% the previous work 
 
% we already know that additional assumptions on $S$ are necessary for any 
% reasonable estimation to be possible. 

\begin{definition}[Survival Probability] \label{def:survival}
Let $S\subseteq \mathbb{R} $ be  measurable,  $x, w \in \mathbb{R}^k$, and $\sigma \in \mathbb{R}$. We define the survival probability 
$\alpha(w, \sigma, x, S)$ 
%of the sample with feature vector $x$ 
%with respect to parameters $w,\sigma$ and trunctation set $S$ 
as
$\alpha(w, \sigma, x, S)=\mathbb{P}\left\{Y\in S\right\};~\text{where}~Y\sim\mathcal{N}(w^Tx,\sigma^2).$
When $S$ is clear from context we may omit $S$ from the arguments of $\alpha$ and simply write $\alpha(w, \sigma, x)$.
\end{definition}

We are now ready to state our assumptions. Our first assumption gives an upper bound on the norm of the true parameters and the feature vectors.
%Since the bound on the former appears in the sample complexity, this normalization assumption is standard.
% on These bounds are necessary in order to bound 
% the variance of the stochastic gradients as well as the strong convexity of
% negative log-likelihood.

\begin{assumption}[Normalization/Data Generation] \label{asm: norm}
    Assume that $\norm{w^*}_2^2 \le \beta$, and $\norm{x}_2^2 \le 1$ for all 
  $x \in \supp(\calP)$.
  %and that we have sample access to the, otherwise unknown, 
  %distribution $\calP$. 
  We also assume oracle access for set $S$, i.e.~we have an oracle which determines, for any $y$, whether $y \in S$. 
\end{assumption}

Our next assumption is the survival probability lower bound following \cite{daskalakis2018efficient, daskalakis2019computationally}. 
% of the observed features $x^{(1)},\ldots,x^{(N)}$ is large enough, i.e.~that the observations in the dataset did not avoid truncation by extreme luck.

\begin{assumption}[Constant Survival Probability] \label{asm: const svvl prob} For all 
$x \in \supp(\calP)$, we have $\alpha(w^*,\sigma^*,x)>a$.
\end{assumption}

% \medskip

% \costis{Here, the $n$ refers to the number of data that taken into account for stochastic gradient descent, we will see later that in the SGD with replacement, we will only take a part of the data points. Also, we may need to have a prior bound on samples and true parameter. Or, the difference for means in the truncated Gaussian will cause great difference of log-likelihoods, and thus will not give a proper bound for the final result.} 

% \medskip

% \costis{Apart from those assumptions, another assumption for the matrix have been made, in order to assume covariance matrix of $x^{(i)}$’s has high enough variance in every direction. This is used for finally estimating the strong convexities of log-likelihood functions.}

Our final assumption is about the spectrum of the covariance matrix
of the features and is classical in 
%many
some other linear regression settings, such as in \cite{daskalakis2019computationally}.

\begin{assumption}[Thickness of Feature Covariance] \label{asm: thickness} 
Let $X=\frac{1}{n}\sum_{i=1}^n {x^{(i)}}{x^{(i)}}^T$ be the $k\times k$ feature vector covariance matrix. We assume that, for some positive real number $b$, we have
$\Exp_{x\sim\calP}[X]=\Exp_{x\sim\calP}[xx^T]\succeq b\cdot I$.
%~~ \text{and} ~~ X\succeq\frac{\log k}{n}{x^{(i)}}{x^{(i)}}^T, \forall i\in[n]$.
Here we abuse the notation that the $(x^{(i)}, y^{(i)})_{i=1}^n$ are samples used for PSGD. 
\end{assumption}
\vskip -0.5em
\section{Main Result}\label{sec: LL func}
 
  When the variance parameter $\sigma^{*2}$ is unknown, the conditional log-likelihood function
of the plain linear regression becomes non-concave, and is therefore not 
straightforward to optimize. For this reason, we follow the common practice of reparameterizing the problem with respect to the parameters 
$v = w/\sigma^2$ and $\lambda = 1/\sigma^2$. The algorithm that we use is Projected SGD on
the population negative log-likelihood function $\bar{\ell}(v,\lambda)$ as presented in
Algorithm \ref{alg:**}. We formally define the loss 
function $\bar{\ell}$ in Section \ref{sec:loglikelihood}.

% \begin{algorithm}[tb]
%   \caption{Projected SGD Without Replacement for $\delta$-Strongly Convex Functions ($\star$)}
%   \label{alg:*}
% \begin{algorithmic}
%   \INPUT data $x^{(i)},y^{(i)}$, $x^{i}\in \mathbb{R}^k$, $y^{(i)}\in \mathbb{R}$, $1\le i\le n$
%   \STATE Do ordinary least square, have estimation $w,\sigma^2$
%   \STATE Reparametrize to the initial $v^{(0)}=w/\sigma^2,\lambda^{(0)}=1/\sigma^2$
%   \FOR {$i=1$ {\bfseries to} $n$}
%   \STATE Sample gradient $u^{(i)}$ such that: 
%   \STATE $\mathbb{E} [u^{(i)}| (v,\lambda)^{(i-1)}]=\nabla l((v~\lambda)^{(i-1)}); x^{(i)},y^{(i)})$
   
%   \COMMENT {Estimate the gradient}
%   \STATE $r^{(i)}\gets (v,\lambda)^{(i-1)} - \frac{1}{\delta\cdot i}u^{(i)}$
%   \STATE $(v,\lambda)^{(i)} \gets \arg\min_{(v,\lambda)\in D} |(v,\lambda) - r^{(i)}|$
   
%   \COMMENT {Projection Step}
%   \ENDFOR
%   {\bfseries return} $ (\bar v,\bar\lambda)\gets \frac{1}{M} \sum_i^M  (v,\lambda)^{(i)}$
% \end{algorithmic}
% \end{algorithm}

% Note that the algorithm with replacement \hyperref[**]{($\star\star$)}
% only uses $n = O(\sqrt{N})$ data points out of the $N$ available points.
% The reason for this is that we could not select the same $x^{(i)}$
% twice because we do not have two independent $y^{(i)}$ for the same 
% $x^{(i)}$. 

\begin{algorithm}[htb]
   \caption{Projected SGD on $\bar{\ell}$} \label{alg:**}
\begin{algorithmic}[1]
   \Require $(x^{(i)}, y^{(i)})$ for $i \in [n]$, learning rates $\eta_t$
   %,  convex set $D$ over pairs  (from Def~\ref{projset})%, $\texttt{average} \in \set{\textbf{true}, \textbf{false}}$
   
   \State $(w_0, \sigma_0^2) \leftarrow $ OLS estimates using other $n$ samples; % Do ordinary least square, have estimation $w,\sigma^2$
    $(q, \lambda^{(0)}) \gets (w_0/\sigma_0^2, 1/\sigma_0^2)$
    
    \Comment{\textit{(a) Initialization part 1}}
    
    \State In terms of $\sigma_0^2$ define convex set $D$ over pairs $(v,\lambda)$ as described in Definition~\ref{projset}
    
   \State $v^{(0)} \gets \argmin_{v : (v, \lambda^{(0)}) \in D} \norm{v - q}_2$ \Comment{\textit{(a) Initialization part 2}}
%   \FOR {$i=1$ {\bfseries to} $10$}
%   \STATE Sample $j^{(i)}$ from $[N]$ for $i \in [n]$  where $n = \lfloor\sqrt{N}\rfloor$
%   \IF {all sampled $j^{(i)}$ are distinct}
%   \STATE relabel $x^{(i)},y^{(i)}\gets x^{j^{(i)}},y^{j^{(i)}}$
%   {\bfseries break for}
%   \ENDIF
%   \ENDFOR
   
%   \COMMENT{Sample $\sqrt{N}$ indices for 10 times, until we get distinct indices. If we fail to do this, we just use the sampled data}
   \For{$t = 1, \dots, n$}
   
   \State Sample gradient $u^{(t)}$ such that $\Exp \b{u^{(t)}| (v, \lambda)^{(t - 1)}} = \nabla \bar{\ell} (v^{(t - 1)}, \lambda^{(t - 1)})$   
   
   \Comment{\textit{(b) Unbiased Gradient Estimate, sampling method see Section \ref{sec: sampling}}}
   
   \State $(v^{(t)},\lambda^{(t)})\gets (v^{(t - 1)}, \lambda^{(t - 1)}) - \eta_t \cdot u^{(t)}$
   \State $(v^{(t)}, \lambda^{(t)}) \gets \arg\min_{(v,\lambda)\in D} \norm{(v,\lambda) - (v^{(t)},\lambda^{(t)})}_2$
      \Comment{\textit{(c) Efficient Projection}}

   \EndFor
%   \IF{\texttt{average}}
%   \STATE $(\hat{v}, \hat{\lambda}) \gets \p{\frac{1}{n} \sum_{t = 1}^n v^{(t)}, \frac{1}{n} \sum_{t = 1}^n \lambda^{(t)}}$
%   \ELSE
   \State $(\hat{v}, \hat{\lambda}) \gets \p{v^{(n)}, \lambda^{(n)}}$;
%   \ENDIF
   \Return $(\hat{w}, \hat{\sigma}^2) \gets \p{\hat{v}/\hat{\lambda}, 1/\hat{\lambda}}$
\end{algorithmic}\label{algo:costis}
\end{algorithm}

  Our goal is to apply the above algorithm to the conditional negative log-likelihood function
of the truncated linear regression model. To execute the above algorithm
in polynomial time, we need to solve the following algorithmic problems.
\begin{enumerate}
  \item[(a)] \textbf{initial feasible point:} compute an initial feasible point in some projection set $D$,
  \item[(b)] \textbf{unbiased gradient estimation:} sample an unbiased estimate of $\nabla \bar{\ell}(v^{(t - 1)}, \lambda^{(t - 1)})$,
  \item[(c)] \textbf{efficient projection:} design an algorithm to project to the set $D$.
\end{enumerate}

\noindent Solving the computational problems (a)-(c) is the first step in the
proof of our main result. First we define the appropriate projection set to use in our
algorithm.

\begin{definition}[Projection Set]\label{projset} 
%Let $(\bar{x}^{(1)}, \bar{y}^{(1)})$, $\dots$, $(\bar{x}^{(n)}, \bar{y}^{(n)})$
%be $n$ samples from the truncated regression problem. We use the notation 
%$(\bar{x}^{(i)}, \bar{y}^{(i)})$ to distinguish these samples from the samples 
%that we use for the rest of the SGD Algorithm \ref{alg:**}. 
We define the
projection set to be the $(v,\lambda)$ satisfying
%%%%%%%%%%%%%%%%%%%%%%%%
%% old projection set %%
%%%%%%%%%%%%%%%%%%%%%%%%
%\begin{align}
%    D_r = \left\{(v,\lambda)=\left({w \over \sigma^2},{1 \over \sigma^2}\right)\in \mathbb{R}^k \times \mathbb{R}~\Big|~ \frac{1}{n}\sum_{i=1}^n\frac{(\bar{y}^{(i)}-w^T \bar{x}^{(i)})^2}{\sigma^2}\bar{x}^{(i)}{\bar{x}^{(i)}}^T\preceq r\bar{X}, 
%    \frac{1}{8(5-2\log a)}\le \frac{{\sigma}^2}{\sigma_0^2}\le \frac{96}{a^2}, \norm{w}_2^2\le\beta\right\} \notag
%\end{align}
%%%%%%%%%%%%%%%%%%%%%%%%
%% new projection set %%
%%%%%%%%%%%%%%%%%%%%%%%%
\begin{align}
    D = \left\{(v,\lambda)=\left({w \over \sigma^2},{1 \over \sigma^2}\right)\in \mathbb{R}^k \times \mathbb{R}~\Big|~ 
    \frac{1}{8(5+2\log(1/a))}\le \frac{{\sigma}^2}{\sigma_0^2}\le \frac{96}{a^2}, \norm{w}_2^2\le\beta\right\} \notag
\end{align}
where 
%$\bar{X} = \frac{1}{n}\sum_{i=1}^n \bar{x}^{(i)}{\bar{x}^{(i)}}^T$ and 
$\sigma_0^2$ is the estimated variance from solving the classical linear
regression problem (using $n$ samples) on the data ignoring truncation. 
%We typically take $r^{*}=5-2\log a$, and we may use
%$D$ or $D_r$ to refer to $D_{r^*}$.
\end{definition}

This projection set contains two parts: the first part that restricts the weights
and the second part that restricts the variance. For the second part, we use the
variance predicted using ordinary least
squares.

Now we are ready to formally state our main theorem about the estimation of the 
parameters $w$ and $\sigma$ from truncated samples under the Assumptions
\ref{asm: norm}- \ref{asm: thickness}.

\begin{theorem} \label{thm: main estimation} Let $(x^{(1)}, y^{(1)}), \cdots, (x^{(n)}, y^{(n)})$ be 
$n$ samples from the linear regression model \eqref{model} with ground truth
$w^*, {\sigma^*}^2$. If Assumptions \ref{asm: norm}- \ref{asm: thickness} hold, then we can instantiate the learning rates used in Algorithm~\ref{alg:**} such that, with success probability $\ge 2/3$, the output estimates $\hat{w}, \hat{\sigma}^2$ satisfy either one of the following bounds:
\begin{equation}\label{eq: 3.1}
    \norm{\hat{w} -w^*}_2 + \abs{\hat{\sigma}^2 - {\sigma^*}^2}\le \frac{\poly(\frac{\sigma^* \cdot \beta}{a \cdot b}, \frac{1}{a\cdot \sigma^*}) \cdot \log(n)}{n^{1/4}}
\end{equation}  
\begin{equation}\label{eq: 3.2}
    \norm{\hat{w} -w^*}_2 + \abs{\hat{\sigma}^2 - {\sigma^*}^2}\le \frac{\poly(\sigma^*,\frac{1}{b}) \cdot \exp\p{\poly(\frac{\beta}{a \cdot \sigma^*})}}{ \sqrt{n}}
\end{equation}
for some polynomials that we make explicit in the detailed proof of the theorem (Section \ref{sec: detailed proof}).
%here we can choose $\eta=\frac{b\varepsilon}{\mathrm{poly}(1/a,\beta,\sigma_0+1/\sigma_0)}$ and 
%$n \ge \mathrm{poly}(1/a,\beta,\sigma_0+1/\sigma_0)/\varepsilon^4 b^2$, 
%where $\sigma_0$ is the estimated variance using ordinary least squares. 

%Moreover, the running time of the algorithm is $\poly(n, \sigma^*) \cdot \exp\p{\poly(\frac{\beta}{a \cdot \sigma^*})}$ and if the survival set $S$ is a union of $k$ intervals then the running time of the algorithm becomes $\poly(k, n, \sigma^*, 1/\sigma^*, 1/a, \beta, 1/b)$.
\end{theorem}

\begin{remark}
    It is easy to see that the probability of success in Theorem 
  \ref{thm: main estimation} can be boosted to $1 - \delta$ with an additional cost of 
  order $\log(1/\delta)$ in the rates. We can achieve this using a 
  folklore boosting technique for parameter estimation problems. In 
  particular, we can run the algorithm of Theorem \ref{thm: main estimation} $\log(1/\delta)$
  times independently and then we can pick the estimate that is $\eps$
  close to at least the half of the rest of the estimates, where $\eps$ is
  the target estimation error. It is easy to see that with this boosting 
  technique we can get error at most $2 \eps$ with probability of success 
  at least $1 - \delta$.
\end{remark}

To prove Theorem \ref{thm: main estimation} we provide two different analyses of the Algorithm
\ref{alg:**}. The first analysis uses only the fact that the stochastic gradients 
of the population conditional negative log-likelihood function have bounded second moment and
is based on the following theorem.

\begin{theorem}[Theorem 2 of \cite{shamir2013stochastic}] \label{thm: lr -1/2}
    Let $\bar{\ell}$ be a convex function with a minimizer $(v^*, \lambda^*)$ and
  suppose that there exists a constant $\rho$ such that
  \begin{enumerate}
    \item[\emph{(i)}] \emph{\textbf{bounded step variance :}} $\Exp\b{\norm{u^{(t)}}_2^2} \le \rho^2$ where $u^{(t)}$ is the sampled step variance in algorithm \ref{alg:**},
    \item[\emph{(ii)}] \emph{\textbf{bounded domain:}}  $\max_{(v, \lambda) \in D} \norm{(v - v^*, \lambda - \lambda^*)}_2^2 \le \rho^2$, 
    \item[\emph{(iii)}] \emph{\textbf{feasibility of optimal:}} the minimizer $(v^*, \lambda^*) \in D$
  \end{enumerate}
  then if we apply Algorithm \ref{alg:**} on $\bar{\ell}$ with learning rates 
  $\eta_t = c/\sqrt{t}$ where $c$ is an absolute constant then for every 
  $t \in [n]$ it holds that
  $\Exp\b{\bar{\ell}(v^{(t)}, \lambda^{(t)}) - \bar{\ell}(v^*, \lambda^*)} \le \frac{\poly(\rho) \cdot \log(t)}{\sqrt{t}}.$
\end{theorem}
\vskip -0.5em
For the second analysis of Algorithm \ref{alg:**} we also need to bound the strong
convexity of $\bar{\ell}$. This is the reason that we get the exponential 
dependence on some of the parameters of the problem, but we can also get a better 
consistency rate. For this second analysis we use the following theorem.

\begin{theorem}[Lemma 1 of \cite{rakhlin2011making}] \label{thm: lr -1}
   Let $\bar{\ell}$ be a convex function with a minimizer $(v^*, \lambda^*)$ and
  suppose that there exist constants $\rho$, $\zeta$ such that
  \begin{enumerate}
    \item[\emph{(i)}] \emph{\textbf{bounded  step variance:}} $\Exp\b{\norm{u^{(t)}}_2^2} \le \rho^2$ where $u^{(t)}$ is the sampled step variance in algorithm \ref{alg:**}
    \item[\emph{(iii)}] \emph{\textbf{feasibility of optimal:}} the minimizer $(v^*, \lambda^*) \in D$
    \item[\emph{(iv)}] \emph{\textbf{strong convexity:}} $\bar{\ell}$ is $\zeta$-strongly convex
  \end{enumerate}
  then if we apply Algorithm \ref{alg:**} on $\bar{\ell}$ with learning rates 
  $\eta_t = 1/(\zeta \cdot t)$, then for every $t \in [n]$ it holds that
  \[ \Exp\b{\norm{(v^{(t)}, \lambda^{(t)}) - (v^*, \lambda^*)}_2^2} \le \frac{4 \rho^2}{\zeta^2 \cdot t}. \]
\end{theorem}

%\medskip
In sum, the main challenges for proving Theorem \ref{thm: main estimation} are solving  
computational problems (a)--(c) and proving mathematical properties (i)--(iv). 
\vskip -0.5em
\section{Overview of the proof of Theorem \ref{thm: main estimation}.} \label{sec: overview}
% Now we present the outline of proof of Theorem \hyperref[thm: main estimation]{ 3}.
We provide an overview of the proof of Theorem~\ref{thm: main estimation}, postponing most technical details to the supplementary material. The main steps of the proof are the following.
\begin{enumerate} 
    \item In Section \ref{sec:loglikelihood} we derive the negative population conditional
    log-likelihood, its gradient, and its Hessian
    matrix.
    \item In Section \ref{sec:computational} we discuss the computational problems (a) - (c).
    % \item In Section \ref{} we present the statements that prove the smoothness for
    % the loss function.
    \item In Section \ref{sec:properties} we formally state our results
    that prove the properties (i), (ii), and (iv).
    \item In Section \ref{sec:feasibility} we state our results  
    for property (iii), the feasibility of
    the optimal solution. 
    % i.e.
    % we have a high probability that the projection set contains the true 
    % parameters, and also the number of samples that our algorithm may use 
    % without the need to use two independent $y^{(i)}$ for the same $x^{(i)}$, as
    % we discussed earlier in the paper.
    \item Finally in Section \ref{sec:proofEnd} we use all the
    above results to complete the proof of \ref{thm: main estimation}.
\end{enumerate}

%\paragraph{Overview proof of Theorem \hyperref[4]{ 4}.} 
% Towards proving Theorem \hyperref[4]{ 4}, we follow the following steps.

%\begin{enumerate}
%  \item We use the results from Sections 3.1, 3.2 and 3.5 that we discussed
%  above.
%  \item In Section 3.4 we present the statements that prove the bounded step 
%  variance and strong convexity of the log-likelihood
%  function.
%  \item Finally in Section 3.7 we use all the mentioned results to prove Theorem
%  \hyperref[4]{ 4}.
%\end{enumerate}

\subsection{The Negative Population Conditional Log-Likelihood Function of Truncated Linear Regression} \label{sec:loglikelihood}

\noindent In this section we define the objective function that we use to apply our PSGD algorithm. This objective function is derived by taking the negative expected value of the log-likelihood of $y$ conditional on the value of $x$. Given a sample $(x, y)$, the negative conditional log-likelihood that $y$ is a sampled from the truncated linear regression with parameters $w$, $\sigma^2$ given the value of $x$ is equal to
\begin{align}
    \ell(w,\sigma^2; x, y)
    %=& -\frac{1}{2\sigma^2}(y-w^Tx)^2\notag\\
    %-&\log\left(\int_S \exp(-\frac{1}{2\sigma^2}(z-w^Tx)^2)\mathrm{d}z\right)\notag \\
    = \frac{1}{2\sigma^2}(y^2-2yw^Tx)+ \log\left(\int_S \exp(-\frac{1}{2\sigma^2}(z^2-2zw^Tx))\mathrm{d}z\right)\notag 
\end{align}
If we reparameterize $\ell$ with $\lambda = 1/\sigma^2$
and $v = w/\sigma^2$, we have
\begin{equation}\label{eq:reparam'ed NLL}
\begin{split}
    \ell(v, \lambda; & x,y)= \frac{1}{2}(\lambda y^2-2yv^Tx)  + \log\left(\int_S \exp(-\frac{1}{2}(\lambda z^2-2zv^Tx))\mathrm{d}z\right)
\end{split}
\end{equation}

We define the distributions $F_x=\mathcal{N}({w^*}^T x,{\sigma^*}^2,S)$ and
$Q_x=\mathcal{N}({w}^T x, {\sigma}^2,S)$, and the \textit{negative population conditional log-likelihood function} $\bar{\ell}$ as follows
\begin{align}\label{eq:negative population LL}
  \bar{\ell}(v, \lambda) = & \Exp_{x \sim \calP}\b{\Exp_{y \sim F_x}\b{\ell(v, \lambda;x,y )}}.
    % =&-\frac{1}{n}\sum_{i}\mathbb{E}_{y\sim F_i}(-\frac{1}{2}\left(\lambda y^2-2yv^Tx^{(i)})\right)\notag\\
    % -&\log\left(\int_S \exp(-\frac{1}{2}(\lambda z^2-2zv^Tx^{(i)}))\mathrm{d}z\right)
\end{align}
We can then easily compute 
\[ \frac{\partial \bar{\ell}}{\partial v}= \Exp_{x \sim \calP}\b{\Exp_{z \sim Q_x}\b{z \cdot x} - \Exp_{y \sim F_x}\b{y \cdot x}}~~~~ \frac{\partial \bar{\ell}}{\partial \lambda}= \frac{1}{2} \Exp_{x \sim \calP} \b{\Exp_{y \sim F_x}\b{y^2} - \Exp_{z \sim Q_x}\b{z^2}} ,  \]

% \begin{align}
%     \frac{\partial^2 \bar{\ell}}{\partial v^2}&=\Exp_{x \sim \calP}\b{\Cov_{z \sim D_x}\b{z \cdot x, z \cdot x}} \notag\\
%     \frac{\partial^2 \bar{\ell}}{\partial v\partial\lambda}&= - \frac{1}{2} \Exp_{x \sim \calP}\b{\Cov_{z \sim D_x}\b{z^2, zx^{(i)}}} \notag\\
%     \frac{\partial^2\bar{\ell}}{\partial\lambda^2}&=\frac{1}{4} \Exp_{x \sim \calP}\b{\Var_{z \sim D_x}\b{z^2}}\notag
% \end{align}
From these we also compute the Hessian matrix which is
\begin{equation}\label{eq:Hessian exp}
\mathbf{H}(v,\lambda)=\Exp_{x \sim \calP}\left[\begin{pmatrix}
     \Var_{z\sim Q_x}[z] x x^T & -\Cov_{z \sim Q_x}\b{\frac{1}{2}z^2,z} x\\
     -\Cov_{z\sim Q_x}\b{\frac{1}{2} z^2, z} x^T & \Var_{z \sim Q_x}[\frac{1}{2}z^2]
\end{pmatrix}\right]
\end{equation}

Here, $\mathbf{H}$ is the expectation of covariance matrices, which is positive definite. Thus, $\bar{\ell}$ is convex. 

%For the proof of Theorem \hyperref[4]{4} we need to define
%\begin{equation}\label{05}
%    f_i(v,\lambda)=q(v,\lambda)+c_iv^Tx^{(i)}+d_i\lambda
%\end{equation}
%where
%$$q(v,\lambda)=\frac{1}{n}\sum_{i=1}^n\log\left(\int_S \exp(-\frac{1}{2}(\lambda z^2-2zv^Tx^{(i)}))\mathrm{d}z\right),$$
%$$c_i=\mathbb{E}_{F_i}(-y),\qquad d_i=\mathbb{E}_{F_i}(\frac{1}{2}y^2).$$
%It is easy to see that we can rewrite $L_{\mathcal{D}}$ as 
%$\frac{1}{n} \sum_{i = 1}^n f_i$.

\subsection{Computational Problems} \label{sec:computational}

The computational problems (a) - (c) can be tackled in the following way
\begin{enumerate}
  \item[(a)] We start with computing $(w_0, \sigma_0)$ by applying OLS to the 
  truncated data.
  \item[(b)] As we  see from the expression of the gradient of $\bar{\ell}$, we
  can compute an unbiased estimate of the gradient using one of our samples 
  $(x^{(i)}, y^{(i)})$, our sample access to the distribution $\calP$ of $x$'s, and
  by doing rejection sampling until we find a point with $(x, z)$ with $z \in S$. To 
  bound the time needed for this step, we need to understand the mass of the
  survival set $S$ for some $x \sim \calP$ and some vector of parameters 
  $(v, \lambda) \in D$. See Supplementary Material \ref{sec:survival}.
  \item[(c)] We show in the supplementary material that $D$ is a convex set and
  define an efficient algorithm to project our estimates to $D$. It is elaborated in Supplementary Material \ref{sec:proj algo}
\end{enumerate}

For more details on the way we solve the computational problems (a) - (c) we refer
to the supplementary material.

%\subsubsection{Survival Probability of Feasible Points} \label{sec:survival}

%For many of the following proofs we need to have an estimation of
%$\alpha(w,\sigma,x)$ for $w$ feasible.

%\begin{lemma} \label{7}Let $x,w,w'\in \mathbb{R}^k$ and $\sigma,\sigma'>0$. Then,
%\begin{align}
%    &\log\left(\frac{1}{\alpha(w,\sigma,x)}\right)\le \max\left(1,\frac{2{\sigma'}^2}{\sigma^2}\right)\log\frac{1}{ \alpha(w',\sigma',x)}
%    \max\left(0,\frac{4{\sigma'}^2}{\sigma^2}-2\right)+\frac{2{\sigma'}^2}{\sigma^2}\left(\frac{(w^Tx-{w'}^Tx)^2}{2{\sigma'}^2}\right)\notag
%\end{align}
%\end{lemma}

%\begin{lemma}\label{8}\emph{(Lemma 3 in \cite{daskalakis2019computationally})} Let
%$x, w\in \mathbb{R}^k$, $\sigma>0$, then 
%$\mathbb{E}_{y \sim \mathcal{N}(w^Tx,\sigma^2,S)}(y-w^Tx)^2 \le (4-2\log \alpha(w,\sigma,x))\sigma^2$.
%\end{lemma}

% \subsection{Bounded Smoothness}

% \noindent Now we prove the assumption of Theorem \hyperref[5]{3.4}.

% \begin{theorem}\label{smoothness} The Hessian of ${L_{\mathcal{D}}(v,\lambda)}$ satisfies
% $\mathbf{H}^2 \preceq \gamma^2 I$ where 
% $\gamma=\mathrm{poly}(1/a,\frac{1}{\sigma_0},{\sigma_0},\beta)$. Hence
% $L_\mathcal{D}$ is $\gamma$ smooth.
% \end{theorem}

% This part I will move it to the proof, or supplementary material

\subsection{Bounded Step Variance, Bounded Domain, and Strong Convexity} \label{sec:properties}

\noindent In this section we present the statements that establish the properties 
(i), (ii), (iv) that we need in order to apply Theorem \ref{thm: lr -1/2} and Theorem \ref{thm: lr -1}.

\begin{theorem} \label{thm:convexity, bounded step var} For every $(v, \lambda) \in D$, we have 
\begin{equation}
    \Exp\b{\norm{(y - z) x}_2^2} \le \mathrm{poly}(1/a)(1+\beta^2+\sigma_0^4), ~~\mathbf{H}(v, \lambda) \succeq b \cdot \exp\left(-\mathrm{poly}(1/a)(1+\frac{\beta}{\sigma_0^2})\right)\begin{pmatrix} \sigma_0^2 I & 0\\
0 & \sigma_0^4
\end{pmatrix}\notag
\end{equation}
where, $x \sim \calP$, $y \sim F_x$ and $z \sim Q_x$ and $\mathbf{H}$ from
\eqref{eq:Hessian exp}.
\end{theorem}

The proof of Theorem \ref{thm:convexity, bounded step var} can be found in Supplementary Material \ref{sec:convexity, bounded step var}. Also, it is easy to see that the diameter of $D$ is bounded: $ \norm{v}_2 < {\norm{w}_2}/{\sigma^2}\le 8(5-2\log a){\sqrt{\beta}}/{{\sigma_0}^2}$ and $\abs{\lambda}<1/{\sigma^2}\le 8(5-2\log a)/{\sigma_0^2}$
% Finally for the coefficients $c_i$, $d_i$ we have that
% \begin{align}
%     &|c_i|=|\mathbb{E}_{F_i}(y)|\le |{x^{(i)}}^T{w^*}|+\sqrt{\mathbb{E}_{F_i}(y-{x^{(i)}}^T{w^*})^2}\notag \\
%     &\le\beta+{\sigma^*}\sqrt{4-2\log a}\le, \beta+\frac{24}{a}{\sigma_0}\sqrt{4-2\log a}\notag \\
%     &|d_i|=|\frac{1}{2}\mathbb{E}_{F_i}(y^2)|\le |{x^{(i)}}^T{w^*}|^2+\mathbb{E}_{F_i}(y-{x^{(i)}}^T{w^*})^2 \notag\\
%     &\le \beta^2+{\sigma^*}^2(4-2\log a)\le \beta^2+\frac{96}{a^2}\sigma_0^2(4-2\log a).\notag
% \end{align}

\subsection{Feasibility of Optimal Solution} \label{sec:feasibility}

  We need to calculate the probability for the optimal point to be in the projection
set, i.e.,  $w^*,\sigma^*\in D$. 

\begin{lemma}\label{lem:prob in proj set} Under Assumptions~\ref{asm: const svvl prob} and~\ref{asm: norm} we have that, $\mathbb{P}((w^*,\sigma^*) \in D)\ge 13/16$.
\end{lemma}

The proof of this Lemma can be found in Supplementary 
Material \ref{sec:prob in proj set}.

\subsection{Proof of Theorem \ref{thm: main estimation}} \label{sec:proofEnd}

 The algorithm that proves Theorem \ref{thm: main estimation} starts with splitting the samples $n$
into three parts of equal size in order to avoid dependence. The first part of $n/3$ samples is used to compute the OLS estimates $w_0$, 
where $w_0$ is used in the initialization step.
The second part of $n/3$ samples is used to compute the OLS estimate of 
$\sigma_0$, where $\sigma_0$
is used both to initialize and  define the projection set $D$,
and is used at every step where a projection to $D$ is needed. 
The number of sample to define $D$ is
$\poly(\frac{\sigma^*,\beta}{a,b},\frac{1}{\sigma^*})$, and our $n$ itself will reach this bound (see Section \ref{sec:prob in proj set}.)
Finally, the 
third part of the $n/3$ samples is used in our main PSGD algorithm. For the rest of
the proof we abuse the notation and we use $n \gets n/3$ for simplicity. The next thing that we need to decide is the learning rates that we are going to
use for the PSGD. We will analyze PSGD under two possible choices of its learning rate schedule, namely $\eta_t = c/\sqrt{t}$ and $\eta_t = 1/(\zeta \cdot t)$, and  will show that it produces estimates  for the ground-truth parameters that scale as $A_1 = \frac{\poly(\frac{\sigma_0 \cdot \beta}{a \cdot b}, \frac{1}{a\cdot \sigma_0}) \cdot \log(n)}{n^{1/4}}$ and  $A_2 = \frac{\poly(\sigma_0) \cdot \exp\p{\poly(\frac{\beta}{a \cdot \sigma_0})}}{b \sqrt{n}}$ respectively, for some polynomials that we will make explicit in Section \ref{sec: detailed proof}.

\paragr{Analysis for learning schedule $\eta_t = c/\sqrt{t}$.} First, by combining Theorem \ref{thm: lr -1/2} and Theorem \ref{thm:convexity, bounded step var}, and noticing the relationship between $\sigma_0$ and $\sigma^*$ in the projection set, we get the following corollary whose proof can be found in Section \ref{sec: detailed proof}.

\begin{corollary}\label{cor: diff of func lr -1/2}
Suppose that $(v^*,\lambda^*)=(w^*/{\sigma^*}^2,1/{\sigma^*}^2) \in D$ and $\bar{\ell}$ is defined as in \eqref{eq:negative population LL}. Then, the output $(\hat{v}, \hat{\lambda})$ of Algorithm \ref{alg:**} satisfies
$\Exp\left[{\bar\ell(\hat{v},\hat{\lambda})-\bar\ell(v^*,\lambda^*)}\right]\le \frac{\poly(\frac{{\sigma^*} \cdot \beta}{a \cdot b}, \frac{1}{{\sigma^*}}) \cdot \log(n)}{n^{1/2}}.$
\end{corollary}
%This can be proven by observing that, due to the projection set, ${\sigma^*}$ is $\mathrm{poly}(1/a,\beta,\sigma^*)$, and $\rho=\poly(1/{\sigma^*}^2,1/a,\beta)$.

%Our next step is to transform the optimal function values to values closest to the true parameters. To do so we use the strong convexity of $\bar\ell$ at the optimum.

%I changed the sentence

Our next step is utilize the above bound to so that our estimates achieve small errors in the parameter space. For this we use the strong convexity of $\bar\ell$ at the optimum.

\begin{lemma}
\label{lem: Hessian at min} The Hessian at the true $(v^*, \lambda^*)$
satisfies 
$\mathbf{H}(v^*, \lambda^*)\succeq \frac{b\times\mathrm{poly}(a)}{\mathrm{poly}(1/a,\beta/{\sigma^*}^2)}\begin{pmatrix} {\sigma^*}^2 I & 0\\
0 & {\sigma^*}^4
\end{pmatrix}$.
\end{lemma}

The proof of Lemma \ref{lem: Hessian at min} can be found in Supplementary Material \ref{sec: proof of strong convexity}.
%\noindent \textbf{Main Proof of Theorem \hyperref[thm: main estimation]{3}: }
We also utilize the following lemma to calculate an upper bound on $\bar\ell({v}^*,{\lambda}^*)$:
\begin{lemma}\label{lem: bound on square}\emph{(Lemma 3 in \cite{daskalakis2019computationally})} Let
$x, w\in \mathbb{R}^k$, $\sigma>0$, then 
$\mathbb{E}_{y \sim \mathcal{N}(w^Tx,\sigma^2,S)}(y-w^Tx)^2 \le (4-2\log \alpha(w,\sigma,x))\sigma^2$.
\end{lemma}
By the above lemma, we have: 
\begin{align}
    \bar\ell({v}^*,{\lambda}^*) = \Exp_{x\in \calP} \Exp_y \left[\frac{(y-{w^*}^Tx)^2}{2{\sigma^*}^2}+\log \alpha(w^*,\sigma^*,x)\right]
    \le 4-2\log \min_{x\in\calP}\alpha(w^*,\sigma^*,x) \le 4-2\log a\notag 
\end{align}

Applying Markov's inequality we have
\begin{equation}\label{eq: 4.4}
    \mathbb{P}\left(\bar\ell(\hat{v},\hat{\lambda})-\bar\ell(v^*,\lambda^*)  > 7\mathbb{E}\b{\bar\ell(\hat{v},\hat{\lambda})-\bar\ell(v^*,\lambda^*)}\right) < \frac{1}{7}.
\end{equation}

The inequality implies that, with at least $6/7$ probability,  the actual difference is at most $7$ times its expectation. Also, the probability that $(v^*,\lambda^*)\in D$ is at least $13/16$ by Lemma \ref{lem:prob in proj set}. Combining these two, we have at least $13/16-1/7>2/3$ probability to hold: 
\begin{equation}\label{eq: 4.5}
    {\bar\ell(\hat{v},\hat{\lambda})-\bar\ell(v^*,\lambda^*)} < \frac{\poly(\frac{\sigma^* \cdot \beta}{ab}, \frac{1}{\sigma^*}) \cdot \log(n)}{n^{1/2}}
\end{equation}
% Note that if $(v,\lambda)$ and $(v^*,\lambda^*)$ are close, the Hessian matrix can estimate the convexity. Also, we present in the proof of the confidence interval, that $(\hat v,\hat\lambda)$ converge to $(\lambda^*,\sigma^*)$ almost surely. So after a constant of time at a high probability, we have $||(v,\lambda)-(v^*,\lambda^*)||$ is small enough. Strictly speaking, there is a region $\varepsilon$ such that for all $v,\lambda,$ $||(v,\lambda)-(v^*,\lambda^*)||<\varepsilon$, 
Now applying Lemma \ref{lem: Hessian at min}, since $\hat{v},\hat{\lambda}$ converge to $v^*,\lambda^*$, when they have small distance, we have
\begin{equation}\label{eq: 4.6}
    \norm{\hat{v}-v^*}_2^2+(\hat{\lambda}-\lambda^*)^2\le  \frac{2}{\delta}\left({\bar\ell(\hat{v},\hat{\lambda})-\bar\ell(v^*,\lambda^*)}\right) \le  \frac{\poly(\frac{\sigma^* \cdot \beta}{ab}, \frac{1}{\sigma^*}) \cdot \log(n)}{\delta \cdot n^{1/2}}
\end{equation} 
%\[ \bar\ell({v},\hat{\lambda})-\bar\ell(v^*,\lambda^*)\ge \frac{\delta}{4}( ||v-v^*||^2+(\lambda-\lambda^*)^2). \]
where $\mathbf{H}(v^*, \lambda^*)\succeq \delta I$, and $ \delta^{-1} = \poly\left(\frac{\sigma^* \cdot \beta}{a \cdot b}, \frac{1}{\sigma^*}\right)$.

Finally, doing some simple calculations for the reparameterization ---see  Section~\ref{sec: detailed proof} in supplementary material--- we have that
\begin{equation}\label{eq: 4.7}
    \norm{\hat{w} - w^*}_2+\abs{\hat{\sigma}^2 - {\sigma^*}^2}\le \sqrt{2\left(\norm{\hat{w} - w^*}_2^2+\abs{\hat{\sigma}^2 - {\sigma^*}^2}^2\right)}
    %\norm{\frac{\hat{v}}{\hat{\lambda}}-\frac{v^*}{ \lambda^*}}^2\notag \\
    %\le& 2\left( \norm{\frac{\hat{v}}{\hat{\lambda}}-\frac{\hat{v}}{\lambda^*}}^2+ \norm{\frac{\hat{v}}{{\lambda}^*}-\frac{v^*}{\lambda^*}}^2\right)\notag\\
    %&\le\frac{\norm{\hat{v}}^2}{{\lambda^*}^2\hat{\lambda}^2}(\hat\lambda-\lambda^*)^2+\frac{||\hat{v}-v^*||^2}{{\lambda^*}^2}\notag\\ 
    %&\le \mathrm{poly} (1/a,{\sigma^*},\beta)\ \norm{\hat{v},\hat{\lambda}-(v^*,\lambda^*)}_2^2 \notag \\
    %&|\hat{\sigma}^2-{\sigma^*}^2|^2=\norm{\frac{1}{\hat{\lambda}}-\frac{1}{
    %\lambda^*}}^2\le\frac{1}{{\lambda^*}^2\hat{\lambda}^2}(\bar\lambda-\lambda^*)^2\notag\\
    \le \poly(1/a,{\sigma^*})\ \norm{(\hat{v},\hat{\lambda}) - (v^*,\lambda^*)}_2.
\end{equation}
Combining inequalities \eqref{eq: 4.4}, \eqref{eq: 4.5}, \eqref{eq: 4.6}, \eqref{eq: 4.7},  inequality \eqref{eq: 3.1} of Theorem \ref{thm: main estimation} follows.

% Since the coefficients can be absorbed into $\poly(\frac{\sigma^* \cdot \beta}{a}, \frac{1}{\sigma^*})$, we can say 
% $$||\hat{w}-w^*||_2^2+||\hat{\sigma}^2-{\sigma^*}^2||_2^2\le\frac{\poly(\frac{\sigma^* \cdot \beta}{a}, \frac{1}{\sigma^*}) \cdot \log(n)}{n^{1/2}}$$
% Take the square root of both sides and using Cauchy Schwartz by 
% $$\sqrt{||\hat{w}-w^*||_2^2+||\hat{\sigma}^2-{\sigma^*}^2||_2^2}\ge\frac{\norm{\hat{w} -w^*}_2 + \abs{\hat{\sigma}^2 - {\sigma^*}^2}}{\sqrt{2}}$$
% we can have the first inequality of Theorem \hyperref[thm: main estimation]{3.2} proved

%%%%%%Split 2 inequalities%%%%%%%

\noindent \paragr{Analysis for learning schedule $\eta_t = 1/(\zeta t)$.} We combine Theorem \ref{thm: lr -1} and Theorem \ref{thm:convexity, bounded step var} to get the
following corollary whose proof can be found in Section \ref{sec: proof of 4.6}.

\begin{corollary}\label{cor: diff of func lr -1}
 Suppose the true parameters $(v^*,\lambda^*)=(w^*/{\sigma^*}^2,1/{\sigma^*}^2) \in D$ and $\bar{\ell}$ is defined as equation \eqref{eq:negative population LL}.  Then, for $\hat{v},\hat{\lambda}$ being the output of Algorithm \ref{alg:**}, it holds that
\begin{align}
  \Exp&\left[\norm{(\hat{v},\hat{\lambda}) - (v^*,\lambda^*)}_2^2\right]\le \mathrm{poly}\left(\beta,\sigma^*+\frac{1}{\sigma^*},\frac{1}{a}\right)\times \exp\left(\mathrm{poly}(1/a)\left(1+\frac{\beta}{{\sigma^*}^2}\right)\right) \frac{1}{b^2n}.\notag
\end{align}

\end{corollary}

%\noindent Applying  Markov's inequality we have
%\[\mathbb{P}\left(\norm{(\hat{v},\hat{\lambda})-(v^*,\lambda^*)}_2^2<120\mathbb{E}\b{\norm{\hat{v},\hat{\lambda}-(v^*,\lambda^*)}_2^2}\right)<1/120\]
%Similarly to the proof of the first inequality of Theorem \ref{thm: main estimation}, we have that with probability $2/3$ it holds that 
Similarly ---details can be found in Section~\ref{sec: detailed proof} of the supplementary material--- to those used to prove inequality~\eqref{eq: 3.1} above, we get that with probability $\ge 2/3$ it holds that 
% \begin{align}
%   \norm{(\hat{v},\hat{\lambda})-(v^*,\lambda^*)}_2^2\le& \mathrm{poly}(\beta,{\sigma^*}+\frac{1}{\sigma_0},\frac{1}{a})\notag\\
%   \times &\exp\left(\mathrm{poly}(1/a)(1+\frac{\beta}{\sigma_0^2})\right) \frac{1}{b^2n}.\notag
% \end{align}
% The coefficient $120$ is absorbed in the $\mathrm{poly}$ term. 
% %Also using the strong convexity bound from Theorem \hyperref[10]{4.3}, we can prove
%$f(\hat{v},\hat{\lambda})-f(v^*,\lambda^*)\ge \frac{\delta}{2}( ||v-v^*||^2+(\lambda-\lambda^*)^2)$, and hence
%\begin{align}
%    ||v-v^*||_2^2+(\lambda-\lambda^*)^2\le& \mathrm{poly}(\beta,\sigma_0+\frac{1}{\sigma_0},\frac{1}{a})\notag \\\times& \exp\left(\mathrm{poly}(1/a)(1+\frac{\beta}{\sigma_0^2})\right) \frac{1}{b^2n}\notag
%\end{align}
%The $2/\delta$ is absorbed in the $\exp$ term. 
% Thus, we can rewrite as 
$\norm{(\hat{v},\hat{\lambda})-(v^*,\lambda^*)}_2^2 \le % &\mathrm{poly}(\beta,\sigma_0+\frac{1}{\sigma_0},\frac{1}{a})\notag\\
    % \times& \exp\left(\mathrm{poly}(1/a)(1+\frac{\beta}{\sigma_0^2})\right) \frac{1}{b^2n}\notag\\
    \frac{\poly({\sigma^*}) \cdot \exp\p{\poly(\frac{\beta}{a \cdot ({\sigma^*})})}}{b^2 \cdot n}.$
Then, we can prove \eqref{eq: 3.2} using the definition of
the projection set $D$. 

\section{Inference and Confidence Regions}\label{sec: confidence region}

We now prove the asymptotic normality of our estimates (Theorem \ref{thm: asym normal}), which we use to obtain their confidence regions (Corollary \ref{cor: asym normal}). For the proofs we refer to the Section \ref{sec: proof of conf reg}.

\begin{theorem} \label{15}\label{thm: asym normal} 
    Let $(\hat{w},\hat{\sigma}^2)$ be the output of Algorithm 
  \ref{alg:**} with $\eta_t = 1/(\zeta \cdot t)$. Then, the vector 
  $\sqrt{n}S^{-1/2}(\binom{\hat{w}} {\hat{\sigma}^2}) - \binom{{w}^*}{{{\sigma}^*}^2})$, 
  asymptotically converges to $\mathcal{N}(0,I_{k+1})$, where
  \begin{enumerate}
    \item[-] $S = \frac{1}{\zeta} R^T \Sigma R$, $R = \begin{pmatrix}
                         {1}/{\hat\sigma^2}I & -{\hat w}/{\hat\sigma^4} \\
                         0 & -{1}/{\hat\sigma^4}
                       \end{pmatrix}$,
    \item[-] $\Sigma$ is the matrix that satisfies
             $\p{\hat{\mathbf{H}}-\frac{\zeta}{2}I}\Sigma+\Sigma\p{\hat{\mathbf{H}}-\frac{\zeta}{2}I} = \Gamma(\hat{v}, \hat{\lambda}) = \Gamma\left(\frac{\hat{w}}{\hat{\sigma}^2}, \frac{1}{\hat{\sigma}^2}\right)$,
    \item[-] $\hat{\mathbf{H}}$ is the empirical estimation of the Hessian
             matrix defined by \eqref{eq:Hessian exp} at the point $(\hat{v}, \hat{\lambda})$,
    \item[-] $\Gamma(v,\lambda)=\frac{1}{n}\sum_{i=1}^n \frac{\partial l(v,\lambda;x^{(i)},y^{(i)})}{{\partial (v,\lambda)}}\frac{\partial l(v,\lambda;x^{(i)},y^{(i)})}{{\partial (v,\lambda)}}^T$.
  \end{enumerate}
\end{theorem}

\begin{corollary} \label{cor: asym normal} 
Let $q_{\alpha}$ be the $1-\alpha$ quantile of  distribution $\chi^2_{k+1}$. Then the $1-\alpha$ asymptotic confidence region of $(w,\sigma^2)$ is
 $\left(\begin{pmatrix}w\\\sigma^2\end{pmatrix}-\begin{pmatrix}\hat w\\\hat\sigma^2\end{pmatrix}\right)^TR^T\Sigma R\left(\begin{pmatrix}w\\\sigma^2\end{pmatrix}-\begin{pmatrix}\hat w\\\hat\sigma^2\end{pmatrix}\right)\le \frac{q_{\alpha}}{\zeta \cdot n}.$
\end{corollary}
\section{Synthetic Experiments} \label{sec:experiments}
%\vskip -1em
  \begin{wrapfigure}[21]{l}{.3\textwidth}
 \begin{minipage}{\linewidth}
     \centering
     \captionsetup[subfigure]{justification=centering}
     \includegraphics[width=\linewidth]{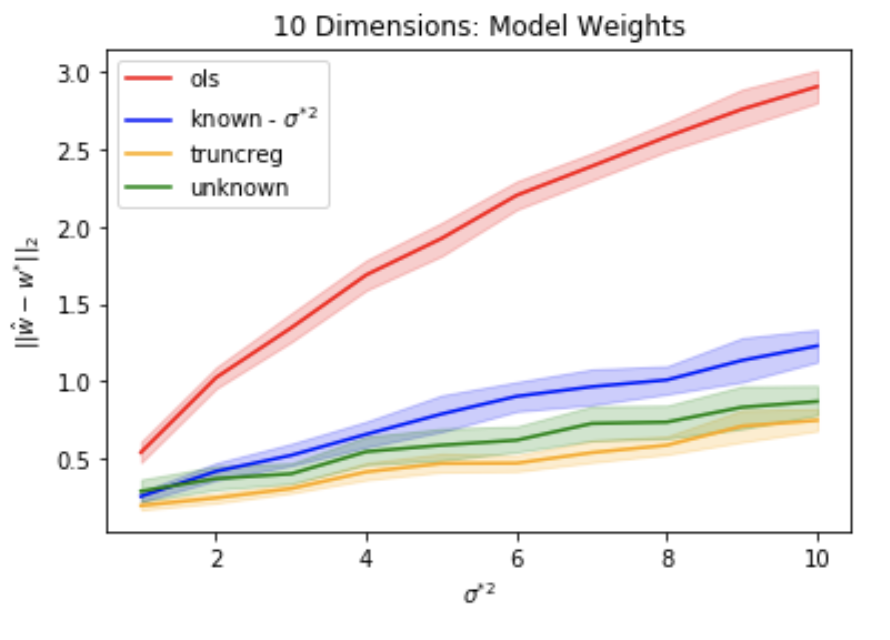}
     \subcaption{}
     \includegraphics[width=\linewidth]{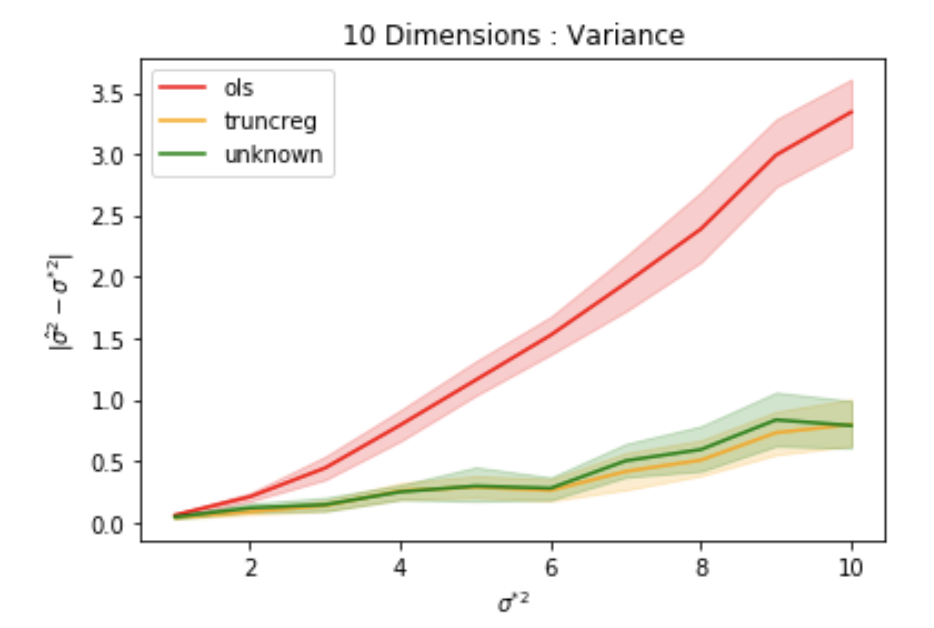}
     \subcaption{}
 \end{minipage}
     \caption{Figure (a) shows $\hat w$ estimation errors and Figure (b) shows the $\hat \sigma^{*2}$ estimation errors for each $\sigma^{*2}$.} 
 \end{wrapfigure}
 
 In Figure 1, we show how our algorithm performs against models with large noise variances. For this experiment, we sample a 10 dimensional ground-truth weight vector $w^{*} \overset{i.i.d.}\sim \mathcal{U}(-1, 1)$ and generate $10000$ $(x^{(i)}, y^{(i)})$ pairs according to $x^{(i)} \overset{i.i.d.}\sim \mathcal{U}(-5, 5)$ and $y^{(i)} = {w^{*}}^{T} x^{(i)} + \epsilon^{(i)}$; where $\epsilon^{(i)} \sim \mathcal{N}(0, \sigma^{*^2})$. We fix $w^{*}$ and $x^{(0)}, ..., x^{(n)}$ across all trials, but vary ${\sigma^*}^2$. After adding Gaussian noise to our ground-truth predictions, we left truncate at zero, removing all of the pairs who's $y^{(i)}$ is negative; retaining approximately $50\%$ of the original samples. We vary $\sigma^{*2}$ over the interval $[1, 10]$ and evaluate how well our procedure recovers $w^{*}$ and $\sigma^{*2}$ in comparison to OLS,  \cite{daskalakis2019computationally}, given $\sigma^{*2}$, and \textit{truncreg}, an R package for truncated regression. All three methods  
%  noise variance is known, \cite{daskalakis2019computationally} performs the best, but both our algorithm and \textit{truncreg} perform
%

 remove a significant amount of bias in the presence of truncation. We emphasize though that \textit{truncreg} cannot be applied in settings where the truncation is more complicated that a simple interval whereas our method can still be applied. 

%  In Figure 2 (a), we show that our method removes significant amounts of bias even when given a complex k-interval filtering mechanism. For this experiment, $w^{*}$ is a two dimensional vector of ones, and $1500$ $(x^{(i)}, y^{(i)})$ pairs are sampled according to $x^{(i)} \overset{i.i.d}\sim \mathcal{N}(0, I/2)$ and $\epsilon^{(i)} \overset{i.i.d.}\sim \mathcal{N}(0, 5)$. We then truncate according to $S = ([-2, -1] \cup [0, .5] \cup [1, 2])$, and retain $500$ samples post truncation.

%  In Figure 2 (a), we show that our procedure also works even when the linear model's noise distribution is fat-tailed. In this experiment, we sample a 5 dimensional $w^{*} \overset{i.i.d}\sim \mathcal{U}(-1, 1)$, and generate our data according to $x^{(i)} \overset{i.i.d.}\sim \mathcal{U}(-5, 5)$ and $\epsilon^{(i)} \overset{i.i.d.}\sim \text{Laplace}(0, 5)$ ($\sigma^{*2}=5$). We then left truncate at varying points over the interval [-2.5, 1.0].

 In Figure 2, we show that our theoretical error bounds hold in practice. For this experiment, we set $w^{*}$ to a ten dimensional weight vector of ones, sample $n$ $(x^{(i)}, y^{(i)})$ pairs, where $x^{(i)} \overset{i.i.d.}\sim \mathcal{N}(0, I)$ and $\epsilon^{(i)} \overset{i.i.d}\sim \mathcal{N}(0, 1)$, and right truncate at zero. We generate our data once. However, for each experiment, we randomly select a $k$ samples that exceed zero, and run our experiment. We vary $k$ over the interval $[10, 5000]$. Our results show that the error bound asymptotically approaches zero at a rate 
 %of $O(n^{\frac{1}{4}})$, as we increase the number of surviving samples; 
 that reaffirms our theoretical analysis.
\vskip -1em
\begin{figure}[h]
    %  \begin{subfigure}[b]{0.3\linewidth}
    %  \centering
    %      \captionsetup[subfigure]{justification=centering}
    %      \includegraphics[width=\linewidth]{Union_2D.png}
    %      \subcaption{}
    %  \end{subfigure}
    %  \hfill
    %  \begin{subfigure}[a]{0.3\linewidth}
    %  \centering
    %     \includegraphics[width=\linewidth]{RealLaplace.png}
    %     \subcaption{}
    % \end{subfigure}
    % \hfill
    %  \begin{subfigure}[]{0.5\linewidth}
        \centering
        \includegraphics[width=.6\linewidth]{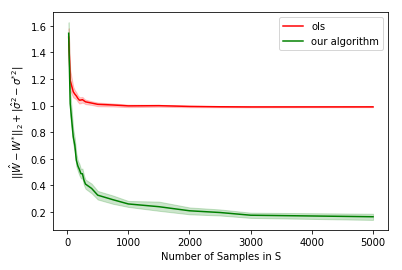}
    % \end{subfigure}
     \caption{Comparison of the proposed method with ordinary least squares.}
 \end{figure}
\vskip -1em

\section{Semi-Synthetic Experiment}
 Here we show results for an additional experiment that we conducted on semi-synthetic data. For this experiment, we used the \cite{pm10} dataset. The \cite{pm10} dataset was originally collected by the Norwegian Public Roads Administration for a study of air pollution at a road in Oslo, Norway. The dataset consists of 500 observations. Interestingly enough, it is common for environmental data to be truncated because of problems in reliably measuring low concentrations. 

For this experiment, we apply left truncation to the dataset, varying a truncation parameter $C$ over the interval [1.0, 4.0].  We run our procedure a total of 10 times and retain the trial that has the smallest gradient as our algorithm's prediction. We conduct the experiment with the same hyperparameters as for the synthetic data experiments. Below, we report our results.

In Figure 3 (a), we show the $L^{2}$ distance between the model parameters of each method and the ground-truth, which we take to be the OLS estimation without truncation. In this figure, we show that our method performs very comparably to \cite{truncreg}. We also note that, at high values of the truncation, both methods exhibit non-monotonicity in their estimation error, which is presumably due to model misspecification and the truncation being too aggressive.

In Figure 3 (b), we report the $L^{1}$ distance between the predicted noise variance and the ground-truth noise variance. Again, our method performs very comparably to \cite{truncreg}, and we observe non-monotonicity at high values of the truncation.

% In Figure 5, we show $R^{2}$ that each of the model's has on the entire dataset after running the respective procedure. Again, we show that our method removes a significant amount of the bias caused by the truncation. We do acknowledge that once we truncate beyond 3, we start to yield $R^{2}$ values that are negative. However, this is possible in the presence truncation, since truncating a dataset can greatly skew our model's empirical estimates. Although out estimates have negative $R^{2}$ , we emphasize that both our and \cite{truncreg}'s predictions  reduce a significant amount of the bias. 

\begin{figure}[htp]   
    % \hfill
    \begin{subfigure}[b]{0.41\linewidth}
    \includegraphics[width=\linewidth]{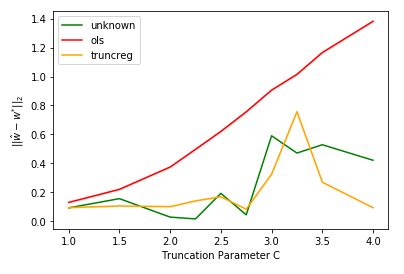}
    % \caption{Comparison of $L^{2}$ distance between predicted model parameters and ground-truth parameters.}
    \subcaption[]{}
    \end{subfigure}
    \hfill%
    \begin{subfigure}[b]{0.41\linewidth}   
    \includegraphics[width=\linewidth]{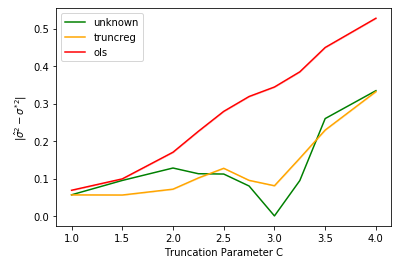}
    % \caption{Comparison of $L^{1}$ distance between noise variance estimates and ground-truth noise variance.}
    \subcaption[]{}
    \end{subfigure}
    \caption{Our method's performance on the semi-synthetic PM10 dataset.}
\end{figure}

\section*{Acknowledgements}

CD and PS were supported by NSF Awards CCF-1901292,  DMS-2022448 (FODSI) and  DMS-2134108, by a Simons Investigator Award, by the Simons Collaboration on the Theory of Algorithmic Fairness, by a DSTA grant, and by the DOE PhILMs project (No. DE-AC05-76RL01830). MZ was supported by NSF DMS-2023505 (FODSI).

% \bibliography{neurips_2021.bbl}
\bibliographystyle{alpha}

\newpage
\setcounter{page}{1}
\appendix

\section{Projection Set} \label{sec:projection}

If it is not specified, all the norms for vectors are $L^2$ norms, by default. 

\subsection{Proof for the convexity of the projection set} \label{sec:conex projset}
%%%%%%%%%%%%%%%%%%%%%%%%
%% The projection set %%
%%    is updated      %%
%%%%%%%%%%%%%%%%%%%%%%%%
Recall that the projection set is 
%\begin{align}
%    D_r = \left\{(v,\lambda)=\left({w \over \sigma^2},{1 \over \sigma^2}\right)\in \mathbb{R}^k \times \mathbb{R}~\Big|~ \frac{1}{n}\sum_{i=1}^n\frac{(\bar{y}^{(i)}-w^T \bar{x}^{(i)})^2}{\sigma^2}\bar{x}^{(i)}{\bar{x}^{(i)}}^T\preceq r\bar{X}, 
%    \frac{1}{8(5-2\log a)}\le \frac{{\sigma}^2}{\sigma_0^2}\le \frac{96}{a^2}, \norm{w}_2^2\le\beta\right\} \notag
%\end{align}
\begin{align}
    D_r = \left\{(v,\lambda)=\left({w \over \sigma^2},{1 \over \sigma^2}\right)\in \mathbb{R}^k \times \mathbb{R}~\Big|~    \frac{1}{8(5-2\log a)}\le \frac{{\sigma}^2}{\sigma_0^2}\le \frac{96}{a^2}, \norm{w}_2^2\le\beta\right\} \notag
\end{align}
So we have
%$$\frac{1}{n}\sum_{i=1}^n\frac{(\bar{y}^{(i)}-w^T \bar{x}^{(i)})^2}{\sigma^2}\bar{x}^{(i)}{\bar{x}^{(i)}}^T=\frac{1}{n}\sum_{i=1}^n\left(\frac{\bar{y}^{(i)}-w^T \bar{x}^{(i)}}{\sigma^2}\right)^2\frac{1}{1/\sigma^2}\bar{x}^{(i)}{\bar{x}^{(i)}}^T=\frac{1}{n}\sum_{i=1}^n\left(\bar{y}^{(i)}\lambda-v^T \bar{x}^{(i)}\right)^2\frac{1}{\lambda}\bar{x}^{(i)}{\bar{x}^{(i)}}^T$$
%Also we have
$$\norm{w}^2=w^Tw=(v/\lambda)^T(v/\lambda)=v^Tv/\lambda^2=\norm{v}^2/\lambda^2$$
Therefore, if we transform it into a direct representation with $v,\lambda$, we get 
$$
    %D_r = \left\{(v,\lambda)\in \mathbb{R}^k \times \mathbb{R}_{>0}~\Big|~ \frac{1}{n}\sum_{i=1}^n(\bar{y}^{(i)}\lambda-v^T \bar{x}^{(i)})^2\bar{x}^{(i)}{\bar{x}^{(i)}}^T\preceq \lambda r\bar{X}, \frac{a^2}{96\sigma_0^2}\le\lambda\le\frac{8(5-2\log a)}{\sigma_0^2}, \norm{v}^2\le\beta\lambda^2\right\} 
    D_r = \left\{(v,\lambda)\in \mathbb{R}^k \times \mathbb{R}_{>0}~\Big|~  \frac{a^2}{96\sigma_0^2}\le\lambda\le\frac{8(5-2\log a)}{\sigma_0^2}, \norm{v}^2\le\beta\lambda^2\right\} 
$$
Where $\mathbb{R}_{>0}$ denotes the set of positive real numbers. We prove that this is a convex set by showing that it is an (infinite) intersection of convex sets.

%First, notice that $\frac{1}{n}\sum_{i=1}^n(\bar{y}^{(i)}\lambda-v^T \bar{x}^{(i)})^2\bar{x}^{(i)}{\bar{x}^{(i)}}^T\preceq \lambda r\bar{X}$ is equivalent to that for any unit vector $z\in \mathbb{R}^k$, we have $z^T\frac{1}{n}\sum_{i=1}^n(\bar{y}^{(i)}\lambda-v^T \bar{x}^{(i)})^2\bar{x}^{(i)}{\bar{x}^{(i)}}^Tx\le z^T\lambda r\bar{X}z$, or, $\frac{1}{n}\sum_{i=1}^n(\bar{y}^{(i)}\lambda-v^T \bar{x}^{(i)})^2(z^T\bar{x}^{(i)})^2\le \lambda rz^T\bar{X}z$. Notice that this can be written as form $(v^T ~\lambda)A\binom{v}{\lambda}+b^T\binom{v}{\lambda}\le 0$ where $A=\frac{1}{n}\sum_{i=1}^n(z^T\bar{x}^{(i)})^2\binom{\bar{x}^{(i)}}{\bar{y}^{(i)}}({\bar{x}^{(i)}}^T~\bar{y}^{(i)})$ is a sum of positive semi-definite symmetric matrix, thus $A$ is a positive semi-definite symmetric matrix. Also, $b=(0 ~ \cdots ~ 0 ~ rz^T\bar{X}z)^T$ (only the coordinate corresponding to $\lambda$ is nonzero.) Therefore, the constraint $(v^T ~\lambda)A\binom{v}{\lambda}+b^T\binom{v}{\lambda}\le 0$ is a (may be degenerated) ellipsoid, which is convex. So, the constraint $\frac{1}{n}\sum_{i=1}^n(\bar{y}^{(i)}\lambda-v^T \bar{x}^{(i)})^2\bar{x}^{(i)}{\bar{x}^{(i)}}^T\preceq \lambda r\bar{X}$ is a infinite intersection of convex constraints $\frac{1}{n}\sum_{i=1}^n(\bar{y}^{(i)}\lambda-v^T \bar{x}^{(i)})^2(z^T\bar{x}^{(i)})^2\le \lambda rz^T\bar{X}z$, for all unit vector $z$. So, this matrix constraint is convex.

The first constraint $\frac{a^2}{96\sigma_0^2}\le\lambda\le\frac{8(5-2\log a)}{\sigma_0^2}$ is a region between two hyper planes, which is also a convex constraint.

We prove that the second constraint $\norm{v}^2\le\beta\lambda^2$ is convex. Suppose we have two vectors such that $\norm{v_1}^2\le\beta\lambda_1^2$ and $\norm{v_2}^2\le\beta\lambda_2^2$. We need to prove that $\norm{\mu v_1+(1-\mu )v_2}^2\le\beta(\mu\lambda_1+(1-\mu)\lambda_2)^2$ for all $\mu\in(0,1)$. With this condition and knowing that $\lambda_1,\lambda_2>0$, we know that $\norm{v_1}\le\sqrt{\beta}\lambda_1$ and $\norm{v_2}\le\sqrt{\beta}\lambda_2$.  By the triangle inequality, we have
$\norm{\mu v_1+(1-\mu )v_2}^2\le(\mu\norm{v_1}+(1-\mu)\norm{v_2})^2\le(\mu\sqrt{\beta}\lambda_1+(1-\mu)\sqrt{\beta}\lambda_2)^2=\beta(\mu\lambda_1+(1-\mu)\lambda_2)^2$. So, we proved the convexity of the third constraint.

Since the three constraints are all convex and the projection set is the intersection of the three constraints, the projection set is convex.

\subsection{Algorithm for projecting to the projection set}{\label{sec:proj algo}}

We have an explicit formula for projecting to the projection set. For convenience, we denote $\lambda_{\min}=\frac{a^2}{96\sigma_0^2}$ and $\lambda_{\max}=\frac{8(5-2\log a)}{\sigma_0^2}$. Thus, the projection formula projects $(v,\lambda)\in D_r$ by minimizing $\norm{(v_0,\lambda_0)-(v,\lambda)}$) as below:
$$\left\{\begin{array}{cc}
     (v_0,\lambda_0)&  \lambda_{\min}\le\lambda_0\le\lambda_{\max}, ~ \norm{v_0}^2\le\beta\lambda_0^2\\
     (v_0,\lambda_{\max})& \lambda_0\ge\lambda_{\max}, ~ \norm{v_0}^2\le\beta\lambda_{\max}^2\\
     \left(\frac{\sqrt{\beta}\lambda_{\max}}{\norm{v_0}}v_0,\lambda_{\max}\right)& \lambda_0\ge\lambda_{\max}, ~ \norm{v_0}^2\ge\beta\lambda_{\max}^2\\
     (v_0,\lambda_{\min})& \lambda_0\le\lambda_{\min}, ~ \norm{v_0}^2\le\beta\lambda_{\min}^2\\
     \left(\frac{\sqrt{\beta}\lambda_{\min}}{\norm{v_0}}v_0,\lambda_{\min}\right)& \lambda_0\le\lambda_{\min}, ~ \sqrt{\beta}\lambda_{\min}\le \norm{v_0}\le\sqrt{\beta}\lambda_{\min}+\frac{\lambda_{\min}-\lambda_0}{\sqrt{\beta}}\\
     \left(\frac{\sqrt{\beta}\lambda_{\max}}{\norm{v_0}}v_0,\lambda_{\max}\right)& \lambda_0\le\lambda_{\max}, ~  \norm{v_0}\ge\sqrt{\beta}\lambda_{\max}+\frac{\lambda_{\max}-\lambda_0}{\sqrt{\beta}}\\
     \left(\frac{\beta\norm{v_0}+\sqrt{\beta}\lambda_0}{(\beta+1)\norm{v_0}}v_0,\frac{\sqrt{\beta}\norm{v_0}+\lambda_0}{\beta+1}\right)& \text{Otherwise}\\
     
\end{array}\right.$$
\section{Missing Proofs}

If it is not specified, all the norms for vectors are $L^2$ norms, by default. 

\subsection{Sampling the Gradient of the Objective Function} \label{sec: sampling}
Let $\mathcal{D}^*$ be the joint distribution of the observed pairs
$(x, y)$, where $(w^*, \sigma^*)$ are the ground truth parameters. Notice that we have a sampler that generates samples $(x,y)$ from $D^*$.  Also let's define
$\mathcal{D}_{(w, \sigma)}$ as the joint distribution of the pairs $(x, y)$, when
the vector of parameters is $(w,\sigma)$. We recall that $\calP$ is the marginal
distribution of $x$ of the joint distribution $\mathcal{D}^*$. Also, $\calP_0$ is
the original distribution of $x$ before truncation. 

So, after truncation, we have
\[ \mathcal{D}^*(\bar{x}, \bar{y}) = \frac{\mathbf{1}\{\bar{y} \in S\} \cdot \mathcal{N}(w^{* T} \bar{x}, \sigma^{* 2}; \bar{y}) \cdot \calP_0(\bar{x})}{\Exp_{x \sim \calP_0} \left[ \alpha (w^*,\sigma^{*}, \bar{x}; S)) \right]}\]
We note that $\mathcal{N}(w^{* T} \bar{x}, \sigma^{* 2}; \bar{y})$ is the probability density of $\bar{y}$; that is, $\exp\left({-\frac{(\bar{y}-w^{*T}x)^2}{2{\sigma^*}^2}}\right)$. To make it explicit, let's say that the relationship between $\calP$ and $\calP_0$ is,
\[ \calP(\bar{x}) = \int_\mathbb{R} \frac{\mathbf{1}\{\bar{y} \in S\} \cdot \mathcal{N}(w^{* T} \bar{x}, \sigma^{* 2}; \bar{y}) \cdot \calP_0(\bar{x})}{\Exp_{x \sim \calP_0} \left[\alpha (w^*,\sigma^{*}, \bar{x}; S)\right]} d \bar{y} = \frac{\alpha (w^*,\sigma^{*}, \bar{x}; S)}{\Exp_{x \sim \calP_0} \left[ \alpha (w^*,\sigma^{*}, \bar{x}; S)) \right]} \cdot \calP_0(\bar{x}) \]

%Therefore, after truncation, the marginal distribution of $x$ is $\bar{x}\sim \calP$. 

Now we can sample the gradient. The gradient is
\[ \frac{\partial \bar{\ell}}{\partial v}= \Exp_{x \sim \calP}\b{\Exp_{z \sim Q_x}\b{z \cdot x} - \Exp_{y \sim F_x}\b{y \cdot x}}~~~~ \frac{\partial \bar{\ell}}{\partial \lambda}= \frac{1}{2} \Exp_{x \sim \calP} \b{\Exp_{y \sim F_x}\b{y^2} - \Exp_{z \sim Q_x}\b{z^2}} ,  \]
Where $F_x=\mathcal{N}({w^*}^T x,{\sigma^*}^2,S)$ and
$Q_x=\mathcal{N}({w}^T x, {\sigma}^2,S)$. Further, note that
\begin{align*}
  &~~~~\Exp_{x \sim \calP}\b{\Exp_{y \sim \mathcal{N}(w^{*T}x, \sigma^{* 2}, S)}\b{y \cdot x}} \\& = \int_x \int_S (y \cdot x) \frac{\mathcal{N}(w^{*T}x, \sigma^{* 2}; \bar{y})}{\mathcal{N}(w^{*T}x, \sigma^{* 2}; S)} \cdot dy \cdot \calP(x) \cdot dx \\
    & = \int_x \int_S (y \cdot x) \frac{\mathcal{N}(w^{*T}x, \sigma^{* 2}; \bar{y})}{\mathcal{N}(w^{*T}x, \sigma^{* 2}; S)} \cdot dy \cdot \frac{\mathcal{N}(w^{* T} \bar{x}, \sigma^{* 2}; S)}{\Exp_{x \sim \calP_0} \left[ \mathcal{N}(w^{* T} x, \sigma^{* 2}; S) \right]} \cdot \calP_0(\bar{x}) \cdot dx \\
    & = \int_x \int_S (y \cdot x) \frac{\mathcal{N}(w^{*T}x, \sigma^{* 2}; \bar{y})}{\Exp_{x \sim \calP_0} \left[ \mathcal{N}(w^{* T} x, \sigma^{* 2}; S) \right]} \cdot \calP_0(\bar{x}) \cdot dy \cdot dx  = \Exp_{(x, y) \sim \mathcal{D}^*}\b{y \cdot x}
\end{align*}
At each gradient step, we sample a pair of $x,y$ and multiply them together. Similarly we sample $\frac{1}{2}y^2$, where $x \sim \calP$ and $y \sim F_x$, and ignore $x$ after sampling. For the second term, we use rejection sampling to generate the data $z\cdot x$ for $z\sim Q_x$ and $x\sim \calP$. Here, since we know the $w$ and $\sigma$, the generation process is simple. First, we generate $x$ using $\calD^*$ (we discard the $y$ value) and then sample $z=w^Tx+\varepsilon$ where $\varepsilon=\calN(0,\sigma^2)$. If $z\notin S$, we continue to sample $\varepsilon$ until we get $z\in S$. Similarly, we can sample $\frac{1}{2}z^2$ where $x \sim \calP,z \sim Q_x$ in this way, and we can ignore $x$ after sampling.

Notice that if $w,\sigma=w^*,\sigma^*$, we have $Q_x=F_x$, so
\[ \frac{\partial \bar{\ell}}{\partial v}(w^{*T}, \sigma^{*2}) = \Exp_{x \sim \calP}\b{\Exp_{z \sim Q_x}\b{z \cdot x} - \Exp_{y \sim F_x}\b{y \cdot x}} =0\]
and 
\[ \frac{\partial \bar{\ell}}{\partial \lambda}= \frac{1}{2} \Exp_{x \sim \calP} \b{\Exp_{y \sim F_x}\b{y^2} - \Exp_{z \sim Q_x}\b{z^2}}=0 .\]

\subsection{Auxilliary Lemmas for Survival Probability of Feasible Points} \label{sec:survival}

For many of the following proofs, we require an estimation of
$\alpha(w,\sigma,x)$ for a feasible $w$.

\begin{lemma} \label{lem: surv prob}Let $x,w,w'\in \mathbb{R}^k$ and $\sigma,\sigma'>0$. Then,
\begin{align}
    &\log\left(\frac{1}{\alpha(w,\sigma,x)}\right)\le \max\left(1,\frac{2{\sigma'}^2}{\sigma^2}\right)\log\frac{1}{ \alpha(w',\sigma',x)}
     + \max\left(0,\frac{4{\sigma'}^2}{\sigma^2}-2\right)+\frac{2{\sigma'}^2}{\sigma^2}\left(\frac{(w^Tx-{w'}^Tx)^2}{2{\sigma'}^2}\right)\notag
\end{align}
\end{lemma}
\begin{proof}

\noindent Now, let's define $D$, $D'$, and $D'_S$ as  $D=\mathcal{N}(w^Tx,\sigma)$,  $D'=\mathcal{N}(w'^Tx,\sigma')$, and $D'_S=\mathcal{N}(w'^Tx,\sigma',S)$. Then, we have: 
\begin{align}
    \alpha(w,\sigma,x)= \Exp_{y \sim D}\Big[{\mathbf{1}_{y\in S}}\Big] =& \Exp_{y \sim D'}\Bigg[{\mathbf{1}_{y\in S}}\exp\left(\frac{(y-w'^Tx)^2}{2\sigma'^2}-\frac{(y-w^Tx)^2}{2\sigma^2}\right)\Bigg]
    \notag \\
    =& \alpha(w',\sigma',x) \cdot \Exp_{y \sim D'_S}\Bigg[\exp\left(\frac{(y-w'^Tx)^2}{2\sigma'^2}-\frac{(y-w^Tx)^2}{2\sigma^2}\right) \Bigg] \notag \\
    \ge& \alpha(w',\sigma',x) \cdot \exp\left(\Exp_{y \sim D'_S} \Bigg [\frac{(y-w'^Tx)^2}{2\sigma'^2}-\frac{(y-w^Tx)^2}{2\sigma^2} \Bigg ]\right). \notag
\end{align}
Now, from Cauchy-Schwarz we have that
\begin{align}
    \frac{(y-w^Tx)^2}{2\sigma^2}\le\frac{2{\sigma'}^2}{\sigma^2}\left(\frac{(y-{w'}^Tx)^2}{2{\sigma'}^2}+\frac{(w^Tx-{w'}^Tx)^2}{2{\sigma'}^2}\right)\notag
\end{align}
and if we apply this to the above expression and take the logarithm of both sides of the inequality we get
\begin{align}
    & \log \alpha(w,\sigma,x)\notag \\
    \ge &\log \alpha(w',\sigma',x)+\Exp_{y \sim D'_S}\Bigg[\left(1-\frac{2{\sigma'}^2}{\sigma^2}\right)\frac{(y-w'^Tx)^2}{2\sigma'^2}
    \Bigg]
    -\Exp_{y \sim D'_S} \Bigg[\frac{2{\sigma'}^2}{\sigma^2}\left(\frac{(w^Tx-{w'}^Tx)^2}{2{\sigma'}^2}\right)\Bigg]
    \notag \\
    \ge &\log \alpha(w',\sigma',x)
    +\Exp_{y \sim D'_S} \Bigg[\min\left(0,1-\frac{2{\sigma'}^2}{\sigma^2}\right)(2-\log \alpha(w',\sigma',x))
    -\frac{2{\sigma'}^2}{\sigma^2}\left(\frac{(w^Tx-{w'}^Tx)^2}{2{\sigma'}^2}\right)\Bigg]
    \notag \\
    = &\min\left(-1,-\frac{2{\sigma'}^2}{\sigma^2}\right)(-\log \alpha(w',\sigma',x))
    -\min\left(0,2-\frac{4{\sigma'}^2}{\sigma^2}\right)-\frac{2{\sigma'}^2}{\sigma^2}\left(\frac{(w^Tx-{w'}^Tx)^2}{2{\sigma'}^2}\right)
    \notag 
\end{align}
The second to last line is followed from Lemma \ref{lem: bound on square} and the Lemma follows.
\end{proof}

\subsection{Proof of Theorem \ref{thm:convexity, bounded step var} and Lemma \ref{lem: Hessian at min}} \label{sec:convexity, bounded step var}

Denote $s=\frac{\max_{\sigma\in D}\sigma}{\min_{\sigma\in D}\sigma}=\sqrt{8(5-2\log a)\frac{96}{a^2}}$ be the largest possible difference between variances, where $D$ is the projection set defined in \ref{projset}. 

Next, to prove our strong convexity result, we use the following anti-concentration bound.

\begin{theorem}[Theorem 8 in \cite{carbery2001distributional})]
    \label{19}There is an absolute constant $\mathcal{C}$ such that if $p:\mathbb{R}^n\to \mathbb{R}$ is a polynomial of degree at most $d$, $0<q<\infty$, and $\mu$ is a \emph{\textit{log-concave probability measure}} on $\mathbb{R}^n$, then
\begin{align}
    &\left(\int |p(x)|^{q/d}\mathrm{d}\mu(x)\right)^{1/q}\alpha^{-1/d} \cdot \mu\left\{x\in\mathbb{R}^n:|p(x)|\le \alpha\right\} \le \mathcal{C} q\notag
\end{align}
\end{theorem}

We also need the following lemma for bounding the survival probability for any parameters in $D$.
\begin{lemma}
\label{18} For $w,\sigma\in D$, we can find a lower bound for the survival probability: $\frac{1}{\alpha(w,\sigma,x^{(i)})}\le \exp(\mathrm{poly}(1/a)(1+\frac{\beta}{\sigma_0^2}))$
\end{lemma}
\smallskip

\begin{proof}
From Lemma \ref{lem: surv prob}, when we plug $\sigma'=\sigma^*,w'=w^*$, we can derive that
\begin{align}
    \log\left(\frac{1}{\alpha(w,\sigma,x)}\right)\le &\max\left(1,\frac{2{\sigma^*}^2}{\sigma^2}\right)\log\left(\frac{1}{ \alpha(w^*,\sigma^*,x)}\right
    )\notag\\
    +&\max\left(0,\frac{4{\sigma^*}^2}{\sigma^2}-2\right)+\frac{2{\sigma^*}^2}{\sigma^2}\left(\frac{(w^Tx-{w^*}^Tx)^2}{2{\sigma^*}^2}\right)\notag
\end{align}
Since $\sigma,\sigma^*\in D$, we have $\frac{{\sigma^*}^2}{\sigma^2}\le s^2$. Thus, we have
\begin{align}
    &\log\left(\frac{1}{\alpha(w,\sigma,x)}\right)\notag\\
    \le &\max\left(1,\frac{2{\sigma^*}^2}{\sigma^2}\right)\log\left(\frac{1}{ \alpha(w^*,\sigma^*,x)}\right)
    +\max\left(0,\frac{4{\sigma^*}^2}{\sigma^2}-2\right)+\frac{2{\sigma^*}^2}{\sigma^2}\left(\frac{(w^Tx-{w^*}^Tx)^2}{2{\sigma^*}^2}\right)\notag \\
    \le &2s^2 \left(\log\frac{1}{ \alpha(w^*,\sigma^*,x)}+2+\left(\frac{(w^Tx-{w^*}^Tx)^2}{2{\sigma^*}^2}\right)\right)\notag \\
    \le &2s^2 \left(-\log a+2+\left(\frac{(\norm{w}^2+\norm{w^*}^2)\norm{x}^2}{{\sigma^*}^2}\right)\right)\notag\\
    \le &2s^2 \left(-\log a+2+\left(\frac{2\beta}{{\sigma^*}^2}\right)\right)\notag 
\end{align}
Since $\sigma_0$ and $\sigma^*$ only have a polynomial difference, we have $\log\Big(\frac{1}{\alpha(w,\sigma,x)}\Big)\le \mathrm{poly}(1/a)(1+\frac{\beta}{\sigma_0^2})$ which finishes the proof.
\end{proof}
\begin{lemma}
\label{20} Let $z\sim \mathcal{N}(\mu,\sigma^2,S)$ be a truncated normal variable. Assume $\alpha(\mu,\sigma^2,S)=a$. Then, we have that
$$\mathrm{Var}(z^2)= 2\mu^2\sigma^2(O(\log(a)))+\sigma^4 O(\log(a)^2)$$
\end{lemma}

\begin{proof}
Denote an affine transformation $S'=\{\frac{x-\mu}{\sigma}|x\in S\}$. So $D=\mathcal{N}(\mu,\sigma,S)$ is transformed into $D'=\mathcal{N}(0,1,S')$, while the survival probability is maintained. We then have
\begin{align}
    \mathrm{Var}_{z \sim D} (z^2)&=\mathrm{Var}_{y \sim D'}((\mu+\sigma y)^2)=\mathrm{Var}_{y \sim D'}(2\sigma\mu y+\sigma^2 y^2)
    \notag\\
    & \le 8 (\sigma\mu)^2 \mathrm{Var}_{y \sim D'}(y)+ 2(\sigma)^4 \mathrm{Var}_{y \sim D'}(y^2)\notag
\end{align}

By Lemma \ref{lem: bound on square} we have $\mathrm{Var}_{y \sim D'}[y]\le \mathbb{E}[y^2]\le 4-2\log a$, and
\begin{align}
    \mathrm{Var}_{D'}(y^2)\le \mathbb{E}_{D'}(y^4)\le \frac{\int_q^\infty x^4 e^{-\frac{x^2}{2}}\mathrm{d}x}{\int_q^\infty  e^{-\frac{x^2}{2}}\mathrm{d}x}= \frac{\int_q^\infty x^4 e^{-\frac{x^2}{2}}\mathrm{d}x}{a/2}\notag
\end{align}

where $q$ satisfies $\int_q^\infty e^{-\frac{x^2}{2}}\mathrm{d}x=a/2$.

Now, all we need to prove is that $q=O(\sqrt{-\log a})$. Notice that if $q>2\sqrt{-\log a}$ (in the case $a<\sqrt{2}/2$ and $q>1$), then we have 
\begin{align}
    a/2=&\int_q^{\infty}e^{-z^2/2}\mathrm{d}z=\int_{q^2/2}^{\infty}\frac{1}{\sqrt{z}}e^{-z}\mathrm{d}z
    <\frac{\sqrt{2}}{q}e^{-q^2/2}<\sqrt{2}a^2/q<a\notag
\end{align}

This is a contradiction.

By integration by parts, we obtain
\begin{align}
    &{\int _q^\infty z^4e^{-z^2/2}\mathrm{d}z}=q^3e^{-q^2/2}+3{\int _q^\infty z^2e^{-z^2/2}\mathrm{d}z}
    =q^3e^{-q^2/2}+3qe^{-q^2/2}+3\int _q^\infty e^{-z^2/2}\mathrm{d}z\notag
\end{align}

and
\begin{align}
    &\frac{\int _q^\infty z^4e^{-z^2/2}\mathrm{d}z}{\int_q^{\infty} e^{-z^2/2}\mathrm{d}z}=3+\frac{q^3e^{-q^2/2}+3qe^{-q^2/2}}{\int_q^{\infty} e^{-z^2/2}\mathrm{d}z}
    \le 3+\frac{q^3e^{-q^2/2}+3qe^{-q^2/2}}{\int_q^{q+1/q} e^{-(q^2+3)/2}\mathrm{d}z}\le 3+15q^2+5q^4\notag
\end{align}

\noindent Since when $a<\sqrt{2}/2$, we have both $q<2\sqrt{-\log a}$ and
$\mathrm{Var}_{Ds'}(y^2) = O(\log a^2)$. The lemma follows.
\end{proof} 

\subsubsection{Strong Convexity}\label{sec: proof of strong convexity}

Now, let's deal with the Hessian Matrix.

First, we consider the case without truncation, that is, 
$$\mathbb{E}_\mathcal{P}\left[\begin{pmatrix}\mathrm{Var}_{x}[z]xx^T & -\mathrm{Cov}_{x}\Big[\frac{1}{2}z^2,z\Big]x\\
-\mathrm{Cov}_{x}\Big[\frac{1}{2}z^2,z\Big]{x}^T & \mathrm{Var}_{x}\Big[\frac{1}{2}z^2\Big]
\end{pmatrix}\right]$$
The variance ($\text{Var}_x$) and covariance ($\text{Cov}_x$) are calculated from the untruncated normal $z\sim \mathcal{N}(w^Tx,\sigma^2)$.
For all $x$, we have 
\begin{align}
    &\binom{v}{\lambda}^T \begin{pmatrix}\mathrm{Var}_{x}[z]xx^T & -\mathrm{Cov}_{x}\Big[\frac{1}{2}z^2,z\Big]x\\
    -\mathrm{Cov}_{x}\Big[\frac{1}{2}z^2,z\Big]{x}^T & \mathrm{Var}_{x}\Big[\frac{1}{2}z^2\Big]
\end{pmatrix} \binom{v}{\lambda}\notag \\
    =& \sigma^2 ({x}^Tv)^2 - 2\sigma^2w^Txx^Tv\lambda
    + \frac{1}{2}\sigma^2 ( 2(w^Tx)^2+\sigma^2)\lambda^2:=I(x)\notag
\end{align}

%HERE! I used LLN, tried to mend the error of wrong assumption
Denote $\bar X=\Exp_{x\sim\mathcal{P}}\Big[\sum xx^T\Big]$. By assumption \ref{asm: thickness}, we have $\bar X\succeq bI$, and thus we have 
\begin{align}
    &\mathbb{E}_\mathcal{P}[I(x)]=\mathbb{E}_\mathcal{P} [\sigma^2 ({x}^Tv)^2 - 2\sigma^2w^Txx^Tv \lambda\frac{1}{2}\sigma^2 ( 2(w^Tx)^2+\sigma^2)\lambda^2]\notag\\ 
    = &\sigma^2v^TXv-2\sigma^2w^TXv\lambda+\frac{1}{2}\sigma^2 ( 2w^TXw+\sigma^2)\lambda^2\notag\\
    = & \sigma^2(x-w\lambda)^TX(x-w\lambda)+\frac{1}{2}\sigma^4\lambda^2\notag\\
    \ge& b\sigma^2\norm{x-w\lambda}^2+\frac{1}{2}\sigma^4\lambda^2\notag
\end{align}

So, it is larger than 
$$\frac{b}{2b(1+\norm{w}^2/\sigma^2)+1}(\sigma^2\norm{v}^2+\sigma^4\lambda^2)$$

Now we consider each summand again for the untruncated sum: we can write it as an integral in 2 variables with degree at most 4, as:
\begin{align}
    I(x)= &\binom{v}{\lambda}^T \begin{pmatrix}\mathrm{Var}_{x}[z]xx^T & -\mathrm{Cov}_{x}\Big[\frac{1}{2}z^2,z\Big]x\\
    -\mathrm{Cov}_{x}\Big[\frac{1}{2}z^2,z\Big]{x}^T & \mathrm{Var}_{x}\Big[\frac{1}{2}z^2\Big]
\end{pmatrix} \binom{v}{\lambda}
    \notag\\
    =&\iint_{\mathbb{R}\times \mathbb{R}}\frac{1}{2}(v^Tx)^2(y-z)^2-\frac{1}{2}\lambda v^Tx(y-z)^2(y+z)+\frac{1}{8}\lambda^2(y-z)^2(y+z)^2\mathrm{d}\mathbb{P}_x\notag
\end{align}

Where $\mathbb{P}_x$ is the untruncated joint normal distribution $\mathcal{N}(w^Tx,\sigma^2)^{\otimes 2}$. This can easily be verified as a log-concave measure. Next, we define polynomial $p(y,z)$ as \begin{align}
    p(y,z) =&\frac{1}{2}(v^Tx)^2(y-z)^2-\frac{1}{2}\lambda v^Tx(y-z)^2(y+z)+\frac{1}{8}\lambda^2(y-z)^2(y+z)^2\notag
\end{align}Let $\alpha={I(x)}\alpha(w,\sigma,x)^8/{2^{12}\mathcal{C}^4}$ where $\mathcal{C}$ is the constant defined in Theorem \ref{19}. Since joint normal is a log-concave distribution, using Theorem \ref{19}, plug in $d=q=4$ and $\alpha$ we get 

$${I(x)}^{1/4}\alpha^{-1/4}\mathbb{P}\{|p(y,z)|\le \alpha\}\le 4\mathcal{C}$$

Hence we have $\mathbb{P}\{|p(y,z)|\le \alpha\}\le\frac{\alpha(w,\sigma,x)^2}{2}$. Notice that the polynomial $p$ is non-negative, therefore, we have a probability of $\mathbb{P}\{p(y,z)>\alpha\}>1-\frac{\alpha(w,\sigma,x)^2}{2}$. Since $\mathbb{P}\{p(y,z)\in S\times S\}=\frac{\alpha(w,\sigma,x)^2}{2}$, we have $\mathbb{P}\left\{p(y,z)>\alpha|(y,z)\in S\times S \right\}>\frac{1}{2}$. So we can estimate
\begin{align}
    & \binom{v}{\lambda}^T \begin{pmatrix}\mathrm{Var}'_{x}[z]xx^T & -\mathrm{Cov}'_{x}\Big[\frac{1}{2}z^2,z\Big]x\\
    -\mathrm{Cov}'_{x}\Big[\frac{1}{2}z^2,z\Big]{x}^T & \mathrm{Var}'_{x}\Big[\frac{1}{2}z^2\Big]
\end{pmatrix} \binom{v}{\lambda} \notag \\
=& \frac{1}{\alpha(w,\sigma,x)^2}\mathbb{E}\Big[p(y,z)\Big]\ge \frac{1}{\alpha(w,\sigma,x)^2}\frac{\alpha(w,\sigma,x)^2}{2}\alpha=\frac{\alpha(w,\sigma,x)^8}{2^{13}\mathcal{C}^4}I^{(i)}\notag
\end{align}
Here, the variance ($\text{Var}'_x$) and covariance ($\text{Cov}'_x$) are calculated from the truncated normal distribution $z\sim \mathcal{N}(w^Tx,\sigma^2,S)$.

Now, by taking both the minimum survival probability, $\alpha(w,\sigma,x)$ and the sum of the inequality above, we show that the lower bound of the Hessian Matrix is:
$$r\begin{pmatrix} \sigma^2 I & 0\\
0 & \sigma^4
\end{pmatrix}:=\frac{\min_i \alpha(w,\sigma,x)^8}{2^{13}\mathcal{C}^4}\frac{b}{2b(1+\norm{w}^2/\sigma^2)+1}\begin{pmatrix} \sigma^2 I & 0\\
0 & \sigma^4
\end{pmatrix}$$
From the lower bound in Lemma \ref{18}, we have 
$$r\ge \exp\left(-\mathrm{poly}(1/a)(1+\frac{\beta}{\sigma_0^2})\right) \frac{b}{2b(1+\norm{w}^2/\sigma^2)+1}$$
Since $\sigma>\mathrm{poly}(1/a)\sigma_0$ and $\norm{w}^2<\beta$, we have 
$$\mathbf{H}\succeq b\exp\left(-\mathrm{poly}(1/a)(1+\frac{\beta}{\sigma_0^2})\right)\begin{pmatrix} \sigma_0^2 I & 0\\
0 & \sigma_0^4
\end{pmatrix} $$
Using the same argument, we can derive Lemma \ref{lem: Hessian at min}, calculating the lower bound for the strong convexity of the minimum point. We can put $\sigma=\sigma^*$, $w=w^*$ and $\min_i \alpha(w,\sigma,x)\ge a$ due to the assumptions.
$$\mathbf{H}(v^*,\lambda^*)\succeq b\frac{a^8}{2^{13}\mathcal{C}^4}\frac{1}{3+2b\norm{w^*}^2/{\sigma^*}^2}\begin{pmatrix} {\sigma^*}^2 I & 0\\
0 & {\sigma^*}^4
\end{pmatrix}$$

\textit{Remark}: By Lemma \ref{18}, we have $\alpha(w,\sigma,x)\ge \exp\left(-2s^2 \left(-\log a+2+\frac{2\beta}{{\sigma^*}^2}\right)\right)$. Notice that $\sigma^2\ge \frac{1}{18(5-2\log a)}\sigma_0^2$, we have
$$\frac{b}{2b(1+\norm{w}^2/\sigma^2)+1}\ge\frac{b}{2b(1+\beta/\sigma^2)+1}=\Omega(\frac{b}{b(1+\beta/\sigma^2)+1})$$
Therefore, we can yield a bound of
$$\mathbf{H}\succeq\Omega\left( \exp\left(-16s^2 \left(-\log a+2+\frac{2\beta}{{\sigma^*}^2}\right)\right)\frac{b}{b(1+\beta/\sigma^2)+1}\right)\begin{pmatrix} \sigma^2 I & 0\\
0 & \sigma^4
\end{pmatrix} $$
Here, we can ignore the constant factor, since it is unknown what the $\mathcal{C}$ is in \cite{CarberyW01}.
Also, we did not transform a general $\sigma$ in the projection set to $\sigma_0$ on purpose.
\subsubsection{Bounded Step Variance}
\label{sec: proof of step var}
Now, let's focus on the upper bound of the squared variance. To eliminate confusion let $y$ be the value of the dependent variable for which we are computing the gradient.

Let the $y\sim\mathcal{N}({w^*}^Tx,{\sigma^2}^*)$ and $z\sim\mathcal{N}(w^Tx,\sigma^2)$. So, we have
\begin{align}
     \mathbb{E}_{\mathcal{P}}\Big[\mathbb{E}\Big[||(y-z)x||^2\Big]\Big]\le& 4 \cdot \mathbb{E}_\mathcal{P}\Big[ \mathbb{E}[||(y-{w^*}^Tx)x||^2 + ||(z-{w}^Tx)x||^2]\Big] + 4 \cdot  ||x^T(w - w^*) x||^2 \notag
\end{align}

For the first term, we have that by Lemma \ref{lem: surv prob}, it holds 
that
\begin{align}
    &4\mathbb{E}_\mathcal{P} \Big[ \mathbb{E}\Big[||(y-{w^*}^Tx) x||^2\Big]\Big]
    \le4\mathbb{E}_\mathcal{P} \left[ \mathbb{E}\Big[(y-{w^*}^Tx)^2\Big]\norm{x}^2\right]\notag \\
    \le & 4(\sigma^*)^2( 2-4\log(\alpha(w^*,x,\sigma^*)))\norm{x}^2
    \le4(2-4\log a)(\sigma^*)^2\le 4(2-4\log a)(s\sigma_0)^2\notag
\end{align}

Similarly, we have 
\begin{align}
    &4 \mathbb{E}_{\mathcal{P}} \Big[||(z-{w}^Tx)x||^2\Big]
    \le 4 \mathbb{E}_{\mathcal{P}} \Big[(z-{w}^Tx)^2\Big]\norm{x}^2\notag \\
    \le & 4{\sigma}^2\Exp_{\mathcal{P}}\left[ 2-4\log(\alpha(w,x,\sigma))\norm{x}^2\right]\notag\\
    =&8\sigma^2+ 16\mathbb{E}_\mathcal{P}\left[ -{\sigma}^2\log(\alpha(w,x,\sigma))\norm{x}^2\right]
    \le 8\sigma^2+ 16\mathbb{E}_\mathcal{P}\left[ -{\sigma}^2\log(\alpha(w,x,\sigma))\right]\notag
\end{align}

By Lemma \ref{lem: bound on square}, we have
\begin{align}
    &-{\sigma}^2\log(\alpha(w,x,\sigma)) \notag\\
    \le &\max\left(\sigma^2,{2{\sigma^*}^2}\right)\log\frac{1}{ \alpha(w^*,\sigma^*,x)}
    +\max\left(0,{4{\sigma^*}^2}-2\sigma^2\right)+2{\sigma^*}^2\left((w^Tx-{w^*}^Tx)^2\right)\notag \\
    \le &{\sigma^*}^2(-s^2\log a+4+2\beta)\notag
\end{align}

Hence, we have 
\begin{align}
    &4\mathbb{E}_\mathcal{P} \mathbb{E}\left[\norm{(z-{w}^Tx)x}^2\right]
    \le 8\sigma^2+ 16\mathbb{E}_\mathcal{P}\left[ -{\sigma}^2\log(\alpha(w,x,\sigma))\norm{x}^2\right]\notag \\
    \le & 8s^2{\sigma_0}^2+16s^2{\sigma_0}^2(4+2\beta-s^2\log a) \notag
\end{align}

And for the last term, we have 
\begin{align*}
    4 \cdot  ||x^T(w - w^*) x||^2 \le 16 \beta^2 
\end{align*}
because $||w||^2, ||w^*||^2 \le \beta^2$ and $||x|| \le 1$.

Now, let's deal with the squared gradient of $\lambda$. Notice that for $y\sim\mathcal{N}({w^*}^Tx,{\sigma^2}{}^*,S)$, we have:
\begin{align}
    \mathbb{E}[y^4] = & (\mathbb{E}[y^2])^2+\mathrm{Var}(y^2)\notag\\ 
    = & 2(\mathbb{E}[(y-{w^*}^Tx)^2]+({w^*}^Tx)^2)^2+\mathrm{Var}(y^2)\notag\\ 
    \le& 2((2-4\log \alpha(w^*, \sigma^*, x)){\sigma^*}^2+({w^*}^Tx)^2)^2 
    + 2({w^*}^Tx)^2{\sigma^*}^2O(\log \alpha(w^*, \sigma^*, x))\notag \\
    + & {\sigma^*}^4O(\log \alpha(w^*, \sigma^*, x)^2)\notag\\ 
    \le & 2(O(\log a)\mathrm{poly}(1/a)\sigma_0^2+\beta)^2
    +  O(\log a)\mathrm{poly}(1/a){\beta}\sigma_0^2
    +  \sigma_0^4O(\log a)\mathrm{poly}(1/a)\notag\\ 
    = &  \mathrm{poly}(1/a)({\sigma_0}^4+\sigma_0^2\beta+\beta^2)
    \le \mathrm{poly}(1/a)({\sigma_0}^4+\beta^2)\notag
\end{align}
The first inequality is due to \ref{lem: bound on square} and \ref{20}. The second inequality comes from the following three facts: because of the projection set, we have $\sigma^*=\mathrm{poly}(1/a)\sigma_0$, $({w^*}^Tx)^2\le \norm{w}^2_2\norm{x}^2_2\le \beta$ by \ref{asm: norm}, and $\alpha(w^*, \sigma^*, x)\ge a$ by our assumption \ref{asm: const svvl prob}. The third inequality combines the term and by AM-GM inequality: $\beta\sigma_0\le (\beta^2+\sigma_0^4)/2$.

Now we are dealing with $z\sim\mathcal{N}({w}^Tx,{\sigma^2},S)$, which is more complicated, but still has the same order bound.
\begin{align}
    \mathbb{E}[z^4]= & (\mathbb{E}[z^2])^2+\mathrm{Var}(z^2)\notag\\ 
    = & 2(\mathbb{E}[(z-{w}^Tx)^2]+({w}^Tx)^2)^2+\mathrm{Var}(z^2)\notag\\ 
    \le& 2((2-4\log \alpha(w, \sigma, x)){\sigma}^2+({w}^Tx)^2)^2
    +  2({w}^Tx)^2{\sigma}^2O(\log \alpha(w, \sigma, x))\notag \\
    + & {\sigma}^4O(\log \alpha(w, \sigma, x)^2)\notag
\end{align}
The inequalities hold for the same reason as bounding $\mathbb{E}[y^4]$. By \ref{18} We know that $-\log {\alpha(w,\sigma,x^{(i)})}\le \mathrm{poly}(1/a)(1+\frac{\beta}{\sigma_0^2})$. Since we are estimating the parameters inside the projection set, we can derive $\sigma=\mathrm{poly}(1/a)\sigma_0$ and by \ref{asm: const svvl prob}, we have $(w^Tx)^2\le \norm{w}_2^2\norm{x}_2^2\le \beta$, giving us
\begin{align}
    & 2((2-4\log \alpha(w, \sigma, x)){\sigma}^2+({w}^Tx)^2)^2\notag \\
    \le & 2((2+\mathrm{poly}(1/a)(1+\frac{\beta}{\sigma_0^2}))^2\sigma^2)+(w^Tx)^2)\notag \\
    = & 2((2+\mathrm{poly}(1/a)(1+\frac{\beta}{\sigma_0^2}))^2\sigma_0^2)\mathrm{poly}(1/a)+\beta)\notag \\
    = & \mathrm{poly}(1/a) (\sigma_0^2+\beta)^2 = \mathrm{poly}(1/a)(\sigma_0^4+\beta^2)\notag
\end{align}
The last inequality is due to the AM-GM inequality, since $\beta\sigma_0^2\le 1/2(\beta^2+\sigma_0^4)$ and
\begin{align}
    &  2({w}^Tx)^2{\sigma}^2O(\log \alpha(w, \sigma, x)) + {\sigma}^4O(\log \alpha(w, \sigma, x)^2)\notag\\
    \le & 2\beta\sigma_0^2O(\mathrm{poly}(1/a)(1+\frac{\beta}{\sigma_0^2}))+\mathrm{poly}(1/a)\sigma_0^4(\mathrm{poly}(1/a)(1+\frac{\beta}{\sigma_0^2}))^2\notag \\
    = & 2\beta O(\mathrm{poly}(1/a)({\beta}+{\sigma_0^2})) + \mathrm{poly}(1/a)((\sigma_0^2+{\beta}))^2\notag \\
    = & \mathrm{poly}(1/a)(\beta^2+\beta\sigma_0^2+\sigma^4)
    =  \mathrm{poly}(1/a)(\sigma_0^4+\beta^2)\notag
\end{align}
Combining these, we get the result:
\begin{align}
    & \mathbb{E}_{\mathcal{P}} \mathbb{E}\Big[{y}^2/2-{z}^2/2\Big]^2 \le \frac{1}{4}\mathbb{E}_\mathcal{P}\left[\mathbb{E}({y}^4)+\mathbb{E}({z}^4)\right] \notag \\
    \le& \mathbb{E}_\mathcal{P}\left[\mathrm{poly}(1/a)(\sigma_0^4+\beta^2)\right] 
    =  \mathrm{poly}(1/a)(\sigma_0^4+\beta^2)=  \mathrm{poly}(1/a)(1+\sigma_0^4+\beta^2)\notag
\end{align}
The gradient of $v$ is also within this bound. Thus, the bounded step variance is proved.

\textit{Remark:} From the proof we can say 
$$\mathbb{E}[y^4]=O(1-\log a)({\sigma^*}^4+\beta^2)$$
And similarly, we have the bound for $\mathbb{E}(z^2)$ using Lemma \ref{18}
\begin{align}
    \mathbb{E}[z^4]\le& 2((2-4\log \alpha(w, \sigma, x)){\sigma}^2+({w}^Tx)^2)^2
    +  2({w}^Tx)^2{\sigma}^2O(\log \alpha(w, \sigma, x))\notag \\
    + & {\sigma}^4O(\log \alpha(w, \sigma, x)^2)\notag\\
    = & 2((2+8s^2 (-\log a+2+\frac{2\beta}{{\sigma^*}^2})){\sigma}^2+({w}^Tx)^2)^2\notag\\
    +&2({w}^Tx)^2{\sigma}^2O(2s^2 (-\log a+2+\frac{2\beta}{{\sigma^*}^2}))+{\sigma}^4O(4s^4 (-\log a+2+\frac{2\beta}{{\sigma^*}^2})^2)\notag
\end{align}
Notice that because $\sigma^2/{\sigma^*}^2\le s^2=O(\frac{1-\log a}{a^2})$, and also $(w^Tx)^2\le \norm{w}^2\norm{x}^2\le \beta$, the highest ``degree'' containing $a$ is $a^{-8}(1-\log a)^{10}$ ($a^{-4}(1-\log a)^{4}$ comes from $s^4$, additional $(1-\log a)^2$ from the coefficients, and additionally $a^{-4}(1-\log a)^{4}$ comes from a hidden $s^4$ produced by $\sigma^4/{\sigma^*}^4$). Also, we can write $(w^Tx)^2\sigma^2=O(\sigma^4+(w^Tx)^4)$. Finally, we can write 
$$\mathbb{E}[z^4]\le O\Bigg(\frac{(1-\log a)^{10}}{a^8}(\beta^2+{\sigma^*}^4)\Bigg)$$
The terms in $\mathbb{E}[\norm{(y-z)x}^2]$ have lower degrees of both in $1/a$, $\beta$ ,and $\sigma^*$. So, we can only add one to $\beta^2+{\sigma^*}^4$ as in the proof above. Therefore, we can write the total gradient as a bound of
$$O\Bigg(\frac{(1-\log a)^{10}}{a^8}(1+\beta^2+{\sigma^*}^4)\Bigg)$$
This bound is greater than the bounded domain.

\subsection{Proof of Lemma \ref{lem:prob in proj set}} \label{sec:prob in proj set}

We estimate the projection set with the following procedure:

(1) We use OLS on $m$ samples $({x}^{(i)},{y}^{(i)})$, and get a estimated weight $w_0$

(2) Then we take another $m$ samples $(\bar{x}^{(i)},\bar{y}^{(i)})$ and calculate the mean $\sigma_0^2=\frac{1}{m}\sum_{i=1}^m(\bar{y}^{(i)}-w_0^T\bar{x}^{(i)})^2$.

%We only need to deal with the constraint $\mathbb{P}(\sigma^*\in D)$, or, $\mathbb{P}\left(\frac{1}{16(4-2\log a)}\sigma_0^2 \le {\sigma^*}^2\le \frac{96}{a^2}\sigma_0^2\right)$. We claim that 

%$$\mathbb{P}\Bigg(\frac{1}{16(4-2\log a)}\sigma_0^2 \le {\sigma^*}^2\le \frac{96}{a^2}\sigma_0^2\Bigg)>7/8$$

%Notice that $\Exp[\sigma_0^2]\le \Exp[\frac{1}{n}\sum_i(y^{(i)}-{w^*}^Tx^{(i)})^2]=\mathbb{E}(y-w^{*T} x)^2\le (4-2\log a){\sigma^*}^2$ By Markov's inequality we show that $\mathbb{P}({\sigma^*}^2>\frac{1}{16(4-2\log a)}\sigma_0^2)<\frac{1}{16}$. We now show the upper bound of the projection set. In other words, we prove  that $\sigma_0$ is large enough with high probability.

First, we try to figure out the lower bound of $\sigma_0^2$. We prove a claim:  $m$ is large (at least $m \ge 3$), for any $w$, there are at least $n/6$ (with probability $>15/16$) samples such that  $|\bar y^{(i)}-{w}^T\bar x^{(i)}|>\frac{a}{4}{\sigma^*}$. 
Notice that $\sigma_0^2=\frac{1}{m}\sum{|\bar y^{(i)}-{w}^T\bar x^{(i)}|}^2$ for some empirical $w=w_0$. 
To finish the proof, we scale the whole distribution by multiplying by $\frac{1}{\sigma^*}$. Since we proved this, we have
$\sigma_0^2=\frac{1}{m}(\frac{n}{6}\times\frac{a^2}{16}{\sigma^*}^2)\ge\frac{a^2}{96}{\sigma^*}^2$. Now we can show that  $\sigma^*=1$
WLOG. In the proof of Lemma 9 in the paper \cite{daskalakis2019computationally}, we notice that in the worst case, $y_i-{w^*}^Tx^{(i)}$ are part of the distribution $D(a)$ (where $D(a)$ dominates $\mathrm{Unif}(a/2,a/2)=U(a)$).
for all $t>0$ and $D\sim D(a)$ and $U\sim U(a)$, $\mathbb{P}(|D|<t)\le\mathbb{P}(|U|<t)$ holds. 
Here, this means that even for the densest $[-\frac{a}{4},\frac{a}{4}]$ it only take $1/2$. So, any $a/2$ window of $y$ takes at most $1/2$ of the probability. So we have 
$$\mathbb{P}\Bigg(|y^{(i)}-{w^*}^Tx^{(i)}|>\frac{a}{4}\Bigg)>1/2$$

Now, all we have left to prove is that if $X,X_i\sim\mathrm{Ber}(1/2)$ $\mathbb{P}$,
then $\mathbb{P}(X_1+\cdots +X_n <n/6)<1/16$ if $n>4$. Thus, we need to prove that $\binom{n}{0}+\cdots+\binom{n}{r}<2^n/16$ where $r=\lfloor (n-1)/6\rfloor$.

For $n\le 18$ it is easy to see this result by checking one by one. Regardless, for $13\le n\le 18$, we have $(r+1)\binom{n}{r}<2^n/16$. For $n\ge 19$, we use induction to prove the stronger bound of $(r+1)\binom{n}{r}<2^n/16$. Now, suppose that this stronger statement holds for $n$, and we want to prove the same bound for $n+6$. We already know that $(r+1)\binom{n}{r}<2^n/16$. So, if we want to prove $(r+2)\binom{n}{r+1}<2^{n+6}/16$, we need to show that
$$\frac{r+2}{(r+1)^2}\frac{(n+6)\cdots(n+1)}{(n-r+5)\cdots(n-r+1)}<64$$

Further, we know that 
$$\frac{r+2}{r+1}\le \frac{4}{3}<2$$
$$\frac{n+1}{r+1}\le \frac{n+1}{n/6} \le 6\times \frac{14}{13}<7$$
\begin{align}
    & \frac{(n+6)\cdots(n+2)}{(n-r+5)\cdots(n-r+1)} \notag\\
    \le& \frac{(n+6)(n+5)(n+4)(n+3)(n+2)}{(5n/6+5)(5n/6+4)(5n/6+3)(5n/6+2)(5n/6+1)}\notag\\
    \le& \frac{19\times18\times 17\times 16\times 15}{15\times14\times13\times12\times11}<4\notag
\end{align}
So, we get $$\frac{r+2}{(r+1)^2}\frac{(n+6)\cdots(n+1)}{(n-r+5)\cdots(n-r+1)}<2\times7\times4<64$$
Therefore, we prove the claim. Now we turn to the upper bound on the $\sigma_0^2$. Denote the least square estimator of $m$ samples to be $w_1$. So we have
$$\Exp[\sigma_0^2]= \Exp\Bigg[\frac{1}{m}\sum_i(y^{(i)}-{w_0}^Tx^{(i)})^2\Bigg]\le \frac{1}{2}\Exp\Bigg[\sum_i(y^{(i)}-w_1^{T} x^{(i)})^2+(w_0^{T} x^{(i)}-w_1^{T} x^{(i)})^2\Bigg]$$
Since $w_1$ is the OLS estimator, we have 
$$\Exp\Bigg[\frac{1}{m}\sum_i(y^{(i)}-w_1^{T} x^{(i)})^2\Bigg]\le \Exp\Bigg[\frac{1}{m}\sum_i(y^{(i)}-{w^*}^{T} x^{(i)})^2\Bigg]\le \frac{1}{m}\sum_i (4-2\log a){\sigma^*}^2=(4-2\log a){\sigma^*}^2$$
The first inequality is because of OLS, and the second inequality is due to Lemma \ref{lem: bound on square}. Now we need to bound $\frac{1}{2}\Exp[\frac{1}{m}\sum_i(w_0^{T} x^{(i)}-w_1^{T} x^{(i)})^2]$. 
Notice that 
$$(w_0^{T} x^{(i)}-w_1^{T} x^{(i)})^2=((w_0^T-w_1^T)x^{(i)})^2\le \norm{w_0^T-w_1^T}^2\norm{x^{(i)}}^2\le \norm{w_0^T-w_1^T}^2$$
So, we need this value to have a small norm. Also since we are doing OLS on the same distribution of data $(x,y)$, we can prove that there is a concentration of samples around the empirical estimator. 

Moreover, notice that the weight formula for linear regression is
$\Big(\sum_{i}x^{(i)}{x^{(i)}}^T\Big)^{-1}\sum_i{x^{(i)}y^{(i)}}$ 
(here we abuse the notation, we are talking about general $m$ samples.)
Or, we do an average, yielding $(\frac{1}{m}\sum_{i}x^{(i)}{x^{(i)}}^T)^{-1}\frac{1}{m}\sum_i{x^{(i)}y^{(i)}}$.
Therefore, if we have infinite samples, we get a final estimator
$w_{\infty}=(\mathbb{E}(xx^T))^{-1}\mathbb{E}(yx)$ for $(x,y)\sim D^*$, where $D^*$ is defined in Section \ref{sec: sampling}. We also have $w_0,w_1\to w_{\infty}$ if $m\to \infty$. Now, we are going to show how will they converge.

First, we investigate $\frac{1}{m}\sum_{i}x^{(i)}{x^{(i)}}^T$. We need a theorem in \cite{tropp2015introduction}. 

\begin{theorem} (Combination of Theorem 1.6.2 and Section 1.6.3 in
\cite{tropp2015introduction})
    Let $x_1$,...,$x_n$ be $i.i.d.$ random vectors with dimension $p$. Let $Y=\frac{1}{n}\sum x_ix_i^T$ and $A=\Exp[x_ix_i^T]$. Assume that each one is uniformly bounded $\norm{x_k }\le B$ for each $k = 1,\cdots,n$. 
    Introduce the sum $Z =Y-A$ 
    %and let $v(Z)$ denote the matrix variance statistic of the sum: $v(Z) = \max(\norm{E(Z Z^T)}_2,\norm{E(Z^TZ)}_2)=\max(\norm{\sum_{i=1}^n \mathbb{E}(S_kS_k^T)}_2,\norm{\sum_{i=1}^n \mathbb{E}(S_k^TS_k)}_2)$ 
    Then $\mathbb{P}{(\norm{Z}_2\ge t)} \le 2p\cdot\exp\left(\frac{-nt^2/2}{B\norm{A}_2+Bt/3}\right)$ for all $t \ge 0$. Here $\norm{A}_2$ is the spectral norm. 
\end{theorem}

Let $X_0=\mathbb{E}(xx^T)$. We can choose a big enough $m$ such that with probability $1-\frac{1}{64}$,
the spectral norm of $\frac{1}{m}\sum_{i}x^{(i)}{x^{(i)}}^T-X_0$ does not exceed $\delta$; $\delta$ is another constant determined later. 
This $\delta$ is smaller than $b$ and $X_0\succeq bI$ by the assumptions. 
So, $\frac{1}{m}\sum_{i}x^{(i)}{x^{(i)}}^T$ is positive definite and we can take the inverse.

Now, let's calculate the inverse. Let $\bar{X}=\frac{1}{m}\sum_{i}x^{(i)}{x^{(i)}}^T$. 
Notice that 
$$\bar{X}^{-1}-X_0^{-1}=X_0^{-1}(X_0-\bar{X})\bar{X}^{-1}$$
has spectral norm at most $\frac{\delta}{b(b-\delta)}$.
Since $X_0$ has a spectral norm $\ge b$, $X_0-\bar{X}$ has spectral norm $\le \delta$ therefore $\bar{X}$ has spectral norm $\ge b-\delta$.

Then we investigate $\bar{x}=\frac{1}{m}\sum_{i}y^{(i)}{x^{(i)}}$. Let $x_0=\Exp[yx]$. We use Chebyshev's inequality in the vector form. Notice that
$$\mathbb{P}(\norm{\bar{x}-x_0}\ge \delta')\le\frac{\Exp[\norm{\bar{x}-x_0}^2]}{\delta'^2}$$

In our case, we know that all $y,x$ are independently chosen. So,  we have
\begin{align*}
    &\Exp[\norm{\bar{x}-x_0}^2\le \Exp[\norm{\bar{x}}^2]=\frac{1}{m}\Exp[\norm{y\cdot x}^2]\\
    =&\frac{1}{m}\Bigg(\Exp\Bigg[\frac{1}{2}\Big(\norm{(y-{w^*}^Tx)^2\cdot x}^2+\norm{({w^*}^Tx)x}^2\Big)\Bigg]\Bigg)\le \frac{1}{2m}((4-2\log a){\sigma^*}^2+\beta)
\end{align*}
Here, we use both Lemma \ref{lem: bound on square} and assumption \ref{asp:normBounds} that $\norm{x}\le 1,\norm{w}^2\le\beta$.
From this we can also derive that 
$$\mathbb{E}[\norm{x_0}]\le\sqrt{\mathbb{E}[\norm{x_0}^2]}=\sqrt{(4-2\log a){\sigma^*}^2+\beta}$$
We also make $m$ large enough to have the probability of at most $1/64$, so that 
$\norm{\bar{x}-x_0}\ge \delta'$.

Notice that we have the estimator $w=\bar{X}^{-1}\bar{x_0}$. This gives us 
\begin{align*}
    &\norm{\bar{X}^{-1}\bar{x}-{X_0}^{-1}x_0}=\norm{\bar{X}^{-1}(\bar{x}-x_0)+(\bar{X}^{-1}-{X_0}^{-1})x_0}\\
    \le&\norm{\bar{X}^{-1}(\bar{x}-x_0)}+\norm{(\bar{X}^{-1}-{X_0}^{-1})x_0}\le\frac{1}{b-\delta}\delta'+\frac{\delta}{b(b-\delta)}\sqrt{(4-2\log a){\sigma^*}^2+\beta}
\end{align*}
(Notice for any vector $v$ and square matrix $A$ if $\norm{v}\le a,\norm{A}_2\le b$, then $\norm{Av}\le ab$. This is because the $L_2$ norm for a vector equals to its spectral norm, and we have $\norm{AB}_2\le\norm{A}_2\norm{B}_2$) .

Since there is a $1/64$ probability that $\norm{\bar{x}-x_0}\le \delta'$ does not hold, 
and there is also a $1/64$ probability that $\norm{\bar{X}-X_0}_2\le \delta'$ does not hold, this $\norm{\bar{X}^{-1}\bar{x}-{X_0}^{-1}x_0}\le\frac{1}{b-\delta}\delta'+\frac{\delta}{b(b-\delta)}\sqrt{(4-2\log a){\sigma^*}^2+\beta}$ holds with a probability of at least $1-1/32$. That means, $w_0$ and $w_1$ each have a $1-1/32$ probability of being close to the convergence limit, and a probability of at most $1-1/16$ that both of them are close.

Take $\delta=\frac{{\sigma^*}b^2}{4\sqrt{(4-2\log a){\sigma^*}^2+\beta}}$ and $\delta'={b{\sigma^*}}/{4}$. 
So, we have at most $1-1/16$ probability that both $w_0$ and $w_1$ are
$$\frac{1}{b-\delta}\delta'+\frac{\delta}{b(b-\delta)}\sqrt{(4-2\log a){\sigma^*}^2+\beta}\le\frac{{\sigma^*}}{2}$$
close to $w_\infty$. Therefore, the norm between $w_0$ and $w_1$ is at most ${\sigma^*}$.

To achieve $\delta$ difference of $\bar{X}-X_0$, we may need $2p\cdot\exp\left(\frac{-nt^2/2}{B\norm{A}_2+Bt/3}\right)\le 1/64$. 
In our case, we have $p=k, t=\delta$, $B=1$ and $\norm{A}_2$ also has a upper bound $1$. So, we need at least 
$\log(128k)\frac{1+\delta/3}{\delta^2}\le\log(128k)\frac{32((4-2\log a){\sigma^*}^2+\beta)}{b^4{\sigma^*}^2}$ samples. 

Also, to achieve $\delta'$ difference of $\bar{x}-x_0$, we may need $\frac{1}{2m}((4-2\log a){\sigma^*}^2+\beta)\frac{1}{({b{\sigma^*}}/{4})^2}\le 1/64$. 
So, we need at least 
$\frac{512((4-2\log a){\sigma^*}^2+\beta)}{{b^2{\sigma^*}^2}}$ samples.

These two bounds of samples are covered with "poly" constraints of the informal theorem \ref{theorem:informalWithReplacement} and the first inequality of theorem \ref{thm: main estimation}.

Overall, we have 
\begin{align*}
    &\Exp[\sigma_0^2]= \Exp\Bigg[\frac{1}{m}\sum_i(y^{(i)}-{w_0}^Tx^{(i)})^2\Bigg]\le \frac{1}{2}\Exp\Bigg[\sum_i(y^{(i)}-w_1^{T} x^{(i)})^2+(w_0^{T} x^{(i)}-w_1^{T} x^{(i)})^2\Bigg]\\
    \le &\frac{1}{2}((4-2\log a){\sigma^*}^2+\norm{w_0^T-w_1^T}^2)\le \frac{1}{2}((4-2\log a){\sigma^*}^2+{\sigma^*}^2)=\frac{1}{2}((5-2\log a){\sigma^*}^2)
\end{align*}

By Markov's Inequality, we have a probability of $1/16$ that $\sigma_0^2\le 16(\mathbb{E}(\sigma_0^2))$. Now, we have $1/8$ probability ($1/16$ for Markov, $1/16$ for the concentration) to hold
$\sigma_0^2\le 16(\mathbb{E}(\sigma_0))\le 8(5-2\log a){\sigma^*}^2$. Therefore, for the whole projection set, it takes $\ge 1-1/8-1/16=13/16$ probability to hold.
%%%%%%%%%%%%%%%%%%%%%%%%%%
%By proving this, we now have 

%$$\mathbb{P}\Bigg(\frac{a^2}{96}{\sigma^*}^2<\sigma_0^2<{16(4-2\log a)}{\sigma^*}^2\Bigg)>7/8$$

%and thus $$\mathbb{P}\Bigg(\frac{1}{16(4-2\log a)}\sigma_0^2 \le {\sigma^*}^2\le \frac{96}{a^2}\sigma_0^2\Bigg)>7/8$$

%The probability for the $w^*$ and $\sigma^*$ satisfy the $\sum(\bar y^{(i)}-w^T{\bar x}^{(i)})^2 {\bar x}^{(i)}{{{\bar x}^{(i)}}}{}^T\preceq r\sigma^2\sum \bar x^{(i)}{\bar x^{(i)}}{}^T$ can be derived from the proof of Lemma 10 in \cite{daskalakis2019computationally}. We can select a larger $t=10$ in the original proof to get that $\mathbb{P}(\sum(\bar y^{(i)}-w^T\bar x^{(i)})^2\bar x^{(i)}{\bar x^{(i)}}{}^T\preceq r\sigma^2\sum \bar x^{(i)}{\bar x^{(i)}}{}^T)\ge 4/5$. 
%Since $\norm{w}^2\le \beta$ is in the assumptions, this constraint in the projection set can be taken as granted. Thus, we can conclude that $\mathbb{P}(w^*,\sigma^*\in D_r)\ge 7/8+4/5-1=27/40$

\subsection{Proof of the Smoothness}\label{sec: smoothness}

The smoothness can be given by this bound:

\begin{theorem}
    \label{thm: smoothness} The Hessian of ${\bar \ell(v,\lambda)}$ satisfies
$\mathbf{H}^2(v,\lambda) \preceq \gamma^2 I$ where 
$\gamma=\mathrm{poly}(1/a,\frac{1}{\sigma_0},{\sigma_0},\beta)$. Hence
$\bar\ell$ is $\gamma$ smooth.
\end{theorem}

Denote that $s=\frac{\max_{\sigma\in D_r}\sigma}{\min_{\sigma\in D_r}\sigma}=\sqrt{8(5-2\log a)\frac{96}{a^2}}$  is the largest possible difference between the variances. Since we already have $\mathbf{H}$ is symmetric and positive definite, $\mathbf{H}^2$ is symmetric and positive definite. So, by the Cauchy-Schwarz inequality, we have
\begin{align}
    &\mathbf{H}^2=\left(\Exp_{x\sim\mathcal{P}}\Big[\mathbf{H}_x\Big]\right)^2 \preceq \Exp_{x\sim\mathcal{P}}\Big[\mathbf{H}_x^2\Big]\notag
\end{align}
Since for each sample ($x^{(i)}$) we can calculate
\begin{align}
    &\mathbf{H}_x^2
    =  \begin{pmatrix}Ax_x^T & -Bx\\
    -Bx^T & C
    \end{pmatrix}^2\notag \\
    = &\begin{pmatrix}  (A^2+B^2\norm{x}^2)xx^T & -(A+C)B\norm{x}^2x\\
    -(A+C)B\norm{x}^2{x^{(i)}}^T & (B\norm{x}^2)^2+C^2
    \end{pmatrix}\notag \\
    \preceq  &2\begin{pmatrix}  (A^2+B^2\norm{x}^2)xx^T & 0\\
    0& (B\norm{x}^2)^2+C^2
    \end{pmatrix}\notag \\
    \preceq  &2\begin{pmatrix}  (A^2+B^2) I & 0\\
    0 & B^2+C^2
    \end{pmatrix}\notag\\
    \preceq &2\begin{pmatrix}  (A^2+AC) I & 0\\
    0 & AC+C^2
    \end{pmatrix}\notag 
\end{align}
where $$A=\mathrm{Var}_{D}[z], B=\mathrm{Cov}_{D}\Big[\frac{1}{2}z^2,z\Big],C=\mathrm{Var}_{D}\Big[\frac{1}{2}z^2\Big]$$
and the distribution $D=\mathcal{N}(w^Tx,\sigma^2,S)$,

we can deduce the first inequality this way. Since the matrix $\mathbf{H}_x^2$ is positive definite, it is also positive semi-definite. Further, if the matrix $\begin{pmatrix}A&B\\B^T&C\end{pmatrix}$ is positive semi-definite, the matrix  $\begin{pmatrix}A&-B\\-B^T&C\end{pmatrix}$ is also positive- semi-definite. Thus, we conclude
$$\begin{pmatrix}A&B\\B^T&C\end{pmatrix}\preceq 2\begin{pmatrix}A&0\\0&C\end{pmatrix}$$ The second inequality comes from $B^2>0$ and $d=\norm{x}^2_2<1$. Finally, the third inequality comes from $B^2<AC$.

Thus, we can write $$\gamma^2=2\mathbb{E}_{\mathcal{P}}\Big[A+C\Big]^2=2\mathbb{E}_{\mathcal{P}}\left[\mathrm{Var}_{D(x)}([z]+\mathrm{Var}_{D(x)}[\frac{1}{2}z^2)]\right]^2$$

where $D(x)=\mathcal{N}(w^Tx,\sigma^2,S)$. Also, by Lemma \ref{20}, we have 
\begin{align}
    &(A+C)^2
    =\mathrm{poly}\Big(\log\frac{1}{\alpha(w,\sigma,x)}\sigma^2+\log^2\frac{1}{\alpha(w,\sigma,x)}\sigma^4
    +\log\frac{1}{\alpha(w,\sigma,x)}(w^Tx)^2\sigma^2\Big)\notag
\end{align}
Since $\log\frac{1}{\alpha(w,\sigma,x)}$ can be written as $\mathrm{poly}(1/a)(1+\beta/\sigma_0^2)$ by Lemma \ref{18},  $\sigma$ can be written as $\mathrm{poly}(\sigma_0,1/a)$, and $|w^Tx|\le \beta$. Then finally we can write 
$$\gamma^2=\mathrm{poly}\Big(1/a,\beta,\sigma_0+\frac{1}{\sigma_0}\Big)$$

Now, we will prove the smoothness. For any two vectors $\theta_1=(v_1,\lambda_1)$ and $\theta_2=(v_2,\lambda_2)$, we define the unit vector with same direction of $(v_2,\lambda_2)-(v_1,\lambda_1)$ as $\mathbf{u}$. Since the projection set written in $v$, $\lambda$ is convex, we claim that all of the points along the line segment $(v_2,\lambda_2)(v_1,\lambda_1)$ are within the projection set. Hence,
\begin{align}
    &\norm{\nabla \bar\ell(v_2,\lambda_2)-\nabla \bar\ell(v_1,\lambda_1)}=\norm{\int_{0}^{\norm{\theta_2-\theta_1}} \left.\frac{\partial \nabla \bar\ell}{\partial \mathbf{u}}\right|_{(v_1,\lambda_1)+\mathbf{u}t} \mathbf{u}\mathrm{d}t}\notag \\
    \le& \int_{0}^{\norm{\theta_2-\theta_1}}\norm{ \left.\frac{\partial \nabla \bar\ell}{\partial \mathbf{u}}\right|_{(v_1,\lambda_1)+\mathbf{u}t} \mathbf{u}}\mathrm{d}t 
    = \int_{0}^{\norm{\theta_2-\theta_1}}\sqrt{\mathbf{u}^T\mathbf{H}((v_1,\lambda_1)+\mathbf{u}t)^T\mathbf{H}((v_1,\lambda_1)+\mathbf{u}t)\mathbf{u}}\mathrm{d}t \notag \\
    \le& \int_{0}^{\norm{\theta_2-\theta_1}}\sqrt{\mathbf{u}^T\gamma^2 I \mathbf{u}} \mathrm{d}t =\gamma\norm{\theta_2-\theta_1}\notag
\end{align}

Thus, we derive that $L_{\mathcal{D}}$ is $\gamma$-smooth. 

\subsection{Proof of Corollary \ref{cor: diff of func lr -1}}\label{sec: proof of 4.6}

\begin{proof}
By Theorem \ref{thm:convexity, bounded step var}, we can choose $\rho$ to be 
$\mathrm{poly}(\beta,\sigma_0+\frac{1}{\sigma_0},\frac{1}{a})$, and 
$\zeta=\exp\left(-\mathrm{poly}(1/a)(1+\frac{\beta}{\sigma_0^2})\right)\min\{\sigma_0^2,\sigma_0^4\}$.
Then using Section \ref{sec:feasibility} the assumptions of Theorem \ref{thm: lr -1}are satisfied
and applying Theorem \ref{thm: lr -1}the Lemma follows. 
\end{proof}

\subsection{Detailed proof of Theorem \ref{thm: main estimation}}\label{sec: detailed proof}

The notation is the same as in Section~\ref{sec:proofEnd}, and we present the details of the two cases considered in that section. 
Here, the estimation below does not contain the number of samples for $n$ to to define projection $D$. In Section \ref{lem:prob in proj set}, 
we need the number of samples as $O\Big(\frac{(1-\log a){\sigma^*}^2+\beta}{b^4{\sigma^*}^2}\Big)$. This is bounded above by the right hand side of \ref{thm: main estimation}: 
$n$ must be at the order of 
$\poly(\frac{\sigma^* \cdot \beta}{a \cdot b}, \frac{1}{a\cdot \sigma^*})$ and the order of 
$\poly(\sigma^*,\frac{1}{b}) \cdot \exp\p{\poly(\frac{\beta}{a \cdot \sigma^*})}$.

\paragr{Case $\eta_t = c/\sqrt{t}$.} First, combining Theorem \ref{thm: lr -1/2} and Theorem \ref{thm:convexity, bounded step var} and the projection set, we get the corollary \ref{cor: diff of func lr -1/2}. We can derive the corollary as follows: by Theorem \ref{thm:convexity, bounded step var} and the rest of Section \ref{sec:properties} ,we know that the step variance and the domain is bounded by some polynomial of ${\sigma_0,\beta},\frac{1}{a},1/\sigma_0$. By Theorem \ref{thm: lr -1/2}, if the final $(w^*,{\sigma^*}^2)\in D$, we have, after $n$ steps, $\Exp\left[{\bar\ell(\hat{v},\hat{\lambda})-\bar\ell(v^*,\lambda^*)}\right]\le \frac{\poly(\frac{\sigma_0 \cdot \beta}{a \cdot b}, \frac{1}{a\times \sigma_0}) \cdot \log(n)}{n^{1/2}}$. Notice that from the definition of the projection set, $\sigma_0/\sigma^*$ and $\sigma^*/\sigma_0$ both are bounded by $\poly(1/a)$. Therefore, we can rewrite this bound to $\Exp\left[{\bar\ell(\hat{v},\hat{\lambda})-\bar\ell(v^*,\lambda^*)}\right]\le \frac{\poly(\frac{\sigma^* \cdot \beta}{a \cdot b}, \frac{1}{a\times \sigma^*}) \cdot \log(n)}{n^{1/2}}$.

Our next step is to transform the optimality in function values to closeness to the true parameters. To do so we use the strong convexity of $\bar\ell$ at the optimum.

%\begin{lemma}
%\label{lem: Hessian at min} The Hessian at the true $(v^*, \lambda^*)$
%satisfies 
%$\mathbf{H}(v^*, \lambda^*)\succeq \frac{b\times\mathrm{poly}(a)}{\mathrm{poly}(1/a,\beta/\sigma_0^2)}\begin{pmatrix} {\sigma_0}^2 I & 0\\
%0 & {\sigma_0}^4
%\end{pmatrix}$.
%\end{lemma}
%\noindent \textbf{Main Proof of Theorem \hyperref[thm: main estimation]{3}: }
%We also utilize the following lemma to calculate an upper bound on $\bar\ell({v}^*,{\lambda}^*)$:
%\begin{lemma}\label{lem: bound on square}\emph{(Lemma 3 in \cite{daskalakis2019computationally})} Let
%$x, w\in \mathbb{R}^k$, $\sigma>0$, then 
%$\mathbb{E}_{y \sim \mathcal{N}(w^Tx,\sigma^2,S)}(y-w^Tx)^2 \le (4-2\log \alpha(w,\sigma,x))\sigma^2$.
%\end{lemma}
By  Lemma \ref{lem: bound on square}, we have: 
\begin{align}
    \bar\ell({v}^*,{\lambda}^*) = \Exp_{x\in \calP} \Exp_y \left[\frac{(y-{w^*}^Tx)^2}{2{\sigma^*}^2}+\log \alpha(w^*,\sigma^*,x)\right]
    \le 4-2\log \min_{x\in\calP}\alpha(w^*,\sigma^*,x) \le 4-2\log a\notag 
\end{align}

We receive the first inequality by ignoring the $\log \alpha(w^*,\sigma^*,x)$ term and the fact that the expectation is smaller then the maximum. By applying Markov's inequality we get
\begin{equation}\label{eq: A.1}
    \mathbb{P}\left(\bar\ell(\hat{v},\hat{\lambda})-\bar\ell(v^*,\lambda^*)  > 5\mathbb{E}\b{\bar\ell(\hat{v},\hat{\lambda})-\bar\ell(v^*,\lambda^*)}\right) < \frac{1}{7}.
\end{equation}

The inequality implies that, with a probability of at least $6/7$,  the actual difference is at most $7$ times its expectation. The constant $7$ can be absorbed by the $\poly$ term. Also, the probability that $(v^*,\lambda^*)\in D$ is at least $13/16$ by Lemma \ref{lem:prob in proj set}. Combining these two, we have at least $13/16-1/7>2/3$ probability to hold: 
\begin{equation}\label{eq: A.2}
    {\bar\ell(\hat{v},\hat{\lambda})-\bar\ell(v^*,\lambda^*)}\le \Exp\left[{\bar\ell(\hat{v},\hat{\lambda})-\bar\ell(v^*,\lambda^*)}\right] \le \frac{\poly(\frac{{\sigma^*} \cdot \beta}{ab}, \frac{1}{{a\times \sigma^*}}) \cdot \log(n)}{n^{1/2}}
\end{equation}
% Note that if $(v,\lambda)$ and $(v^*,\lambda^*)$ are close, the Hessian matrix can estimate the convexity. Also, we present in the proof of the confidence interval, that $(\hat v,\hat\lambda)$ converge to $(\lambda^*,\sigma^*)$ almost surely. So after a constant of time at a high probability, we have $||(v,\lambda)-(v^*,\lambda^*)||$ is small enough. Strictly speaking, there is a region $\varepsilon$ such that for all $v,\lambda,$ $||(v,\lambda)-(v^*,\lambda^*)||<\varepsilon$, 
Now by applying Lemma \ref{lem: Hessian at min}, we transform the difference in functions to the $L_2$ norm distance. Since $\hat{v},\hat{\lambda}$ converge to $v^*,\lambda^*$, when they have small distance, we can approximate the rate using the convexity of the minimum. Therefore, we have
\begin{equation}\label{eq: A.3}
    \norm{\hat{v}-v^*}^2+(\hat{\lambda}-\lambda^*)^2\le  \frac{2}{\delta}\left({\bar\ell(\hat{v},\hat{\lambda})-\bar\ell(v^*,\lambda^*)}\right) \le  \frac{\poly(\frac{\sigma^* \cdot \beta}{ab}, \frac{1}{a\times \sigma^*}) \cdot \log(n)}{\delta \cdot n^{1/2}}
\end{equation} 
%\[ \bar\ell({v},\hat{\lambda})-\bar\ell(v^*,\lambda^*)\ge \frac{\delta}{4}( ||v-v^*||^2+(\lambda-\lambda^*)^2). \]
where 
$$\mathbf{H}(v^*, \lambda^*)\succeq b\frac{a^8}{2^{13}\mathcal{C}^4}\frac{1}{3+2b\norm{w^*}^2/{\sigma^*}^2}\begin{pmatrix} {\sigma^*}^2 I & 0\\
0 & {\sigma^*}^4
\end{pmatrix}\succeq \delta I$$
Notice that $\norm{w}^2\le \beta$. Therefore $ \delta^{-1} = \poly\left(\frac{\sigma^* \cdot \beta}{a \cdot b}, \frac{1}{a\cdot\sigma^*}\right)$, and this term can be absorbed into the term $\poly(\frac{\sigma^* \cdot \beta}{a\times b}, \frac{1}{a\sigma^*})$.

Finally, by Cauchy-Schwartz Inequality, we have
\begin{equation}\label{eq: A.4}
    \norm{\hat{w} - w^*}+\abs{\hat{\sigma}^2 - {\sigma^*}^2}\le \sqrt{2\left(\norm{\hat{w} - w^*}^2+\abs{\hat{\sigma}^2 - {\sigma^*}^2}^2\right)}
\end{equation}
And for the $w$ part, we parameterize back and we get
\begin{align}\label{eq: A.5}
    &\norm{\hat{w} - w^*}^2\le \norm{\frac{\hat{v}}{\hat{\lambda}}-\frac{v^*}{ \lambda^*}}^2
    \le 2\left( \norm{\frac{\hat{v}}{\hat{\lambda}}-\frac{\hat{v}}{\lambda^*}}^2+ \norm{\frac{\hat{v}}{{\lambda}^*}-\frac{v^*}{\lambda^*}}^2\right)\notag\\
    \le&\frac{\norm{\hat{v}}^2}{{\lambda^*}^2\hat{\lambda}^2}(\hat\lambda-\lambda^*)^2+\frac{||\hat{v}-v^*||^2}{{\lambda^*}^2}
    \le \mathrm{poly} (1/a,{\sigma^*},\beta)\ \norm{(\hat{v},\hat{\lambda})-(v^*,\lambda^*)}^2 
\end{align}
The last inequality is because of the projection set, we have $\hat\lambda,\lambda^*=\poly(1/a){\sigma^*}^{-2}$. Also, by Assumption \ref{asm: norm} we have $\norm{\hat v}= \norm{\hat w}/\hat\sigma^2\le \sqrt{\beta}/(\poly(1/a){\sigma^*}^2)$. For the $\sigma$, we have
\begin{align}\label{eq: A.6}
    |\hat{\sigma}^2-{\sigma^*}^2|^2=\norm{\frac{1}{\hat{\lambda}}-\frac{1}{
    \lambda^*}}^2\le\frac{1}{{\lambda^*}^2\hat{\lambda}^2}(\bar\lambda-\lambda^*)^2
    \le  \poly(1/a,{\sigma^*})\ \norm{(\hat{v},\hat{\lambda}) - (v^*,\lambda^*)}^2
\end{align}
    
The last inequality is also derived from the projection set. Combining inequality \eqref{eq: A.1} with inequality \eqref{eq: A.6}, we have, with a probability of at least $2/3$,
\begin{align}
    &\norm{\hat{w} - w^*}+\abs{\hat{\sigma}^2 - {\sigma^*}^2}\le \sqrt{2\left(\norm{\hat{w} - w^*}^2+\abs{\hat{\sigma}^2 - {\sigma^*}^2}^2\right)}\notag\\
    \le &\sqrt{2\mathrm{poly} (1/a,{\sigma^*},\beta)\ \norm{(\hat{v},\hat{\lambda})-(v^*,\lambda^*)}^2 }
    \le \sqrt{2\mathrm{poly} (1/a,{\sigma^*},\beta)\frac{\poly(\frac{{\sigma^*} \cdot \beta}{ab},\frac{1}{{a\cdot\sigma^*}}) \cdot \log(n)}{\delta \cdot n^{1/2}}}\notag\\
    = &\frac{\poly(\frac{{\sigma^*} \cdot \beta}{ab},\frac{1}{{a\cdot\sigma^*}}) \cdot \log(n)}{ n^{1/4}}\notag
\end{align}
And therefore, inequality \eqref{eq: 3.1} of Theorem \ref{thm: main estimation} follows.

\textit{Remark:} We can derive in the theorem 2 in \cite{shamir2013stochastic} that the bound is $O(\rho^2\frac{1+\log n}{\sqrt{n}})$. We have already written the bound for $\mathbf{H}$ and step variance in the form of $\sigma^*$. Therefore, the right hand side of inequality \ref{eq: A.2} is
$$O\Bigg(\frac{(1-\log a)^{10}}{a^8}(1+\beta^2+{\sigma^*}^4)\frac{\log n}{\sqrt{n}}\Bigg)$$
And $\delta^{-1}=\frac{3+2b\beta/{\sigma^*}^2}{ba^8\min({\sigma^*}^2,{\sigma^*}^4)}=O\Big(\frac{({\sigma^*}^2+\beta)(1+{\sigma^*}^2)}{ba^8{\sigma^*}^6}\Big)$. Here, we used a trick that $\min(x^2,x^4)=\Omega(\frac{x^4}{1+x^2})$

By equations \ref{eq: A.5} and \ref{eq: A.6}, the factor to be multiplied from $v,\lambda$ to $w,$ is $\max(\frac{1}{{\lambda^*}^2},\frac{\norm{v}^2}{\hat{\lambda}^2{\lambda^*}^2})=\max({\sigma^*}^4,{\sigma^*}^4\norm{\hat w}^2)=O((\sigma^*)^4(1+\beta))$ the factor to be multiplied from $\lambda$ to $\sigma$ is $\frac{1}{{\hat\lambda}^2{\lambda^*}^2}={\sigma^*}^2{{\hat\sigma}^2}=O({\sigma^*}^4s^2)=O({\sigma^*}^4\frac{1-\log a}{a^2})$. So, the overall multiplying factor is $O({\sigma^*}^4\frac{1-\log a}{a^2}(1+\beta))$

Therefore, by multiplying these bounds and square root, we have a bound of
$$O(\sqrt{\frac{(1-\log a)^{10}}{a^8}(1+\beta^2+{\sigma^*}^4)\frac{\log n}{\sqrt{n}}\frac{({\sigma^*}^2+\beta)(1+{\sigma^*}^2)}{ba^8{\sigma^*}^6}{\sigma^*}^4\frac{1-\log a}{a^2}(1+\beta)})$$
Which can be simplify to 
% (for the parameters $a,\beta, b$ we just count their highest and lowest coefficients)
$$O\Bigg(\frac{(1-\log a)^{5.5}}{a^9}\frac{\sqrt{\log n}}{\sqrt[4]{n}}\frac{(1+\beta^2)({\sigma^*}^3+{\sigma^*}^{-1})}{\sqrt{b}}\Bigg)$$
%%%%%%Split 2 inequalities%%%%%%%

\noindent \paragr{Case $\eta_t = 1/(\zeta t)$.} We just plug in the result in Theorem \ref{thm:convexity, bounded step var} into Theorem \ref{thm: lr -1} and in Section \ref{sec: proof of 4.6} we have proved Corollary \ref{cor: diff of func lr -1}. That is, if $w^*,\sigma^*\in D$, we have
\begin{align}
  \Exp&\left[\norm{(\hat{v},\hat{\lambda}) - (v^*,\lambda^*)}^2\right]\le \mathrm{poly}\left(\beta,\sigma_0+\frac{1}{\sigma_0},\frac{1}{a}\right)\times \exp\left(\mathrm{poly}(1/a)\left(1+\frac{\beta}{\sigma_0^2}\right)\right) \frac{1}{b^2n}.\notag
\end{align}
Applying  Markov's inequality again as inequality \eqref{eq: A.1}, we have
\begin{equation}
    \mathbb{P}\left(\norm{(\hat{v},\hat{\lambda})-(v^*,\lambda^*)}^2<7\mathbb{E}\b{\norm{\hat{v},\hat{\lambda}-(v^*,\lambda^*)}^2}\right)<1/7
\end{equation}
Similarly to the proof of inequality \eqref{eq: 3.1} above, of Theorem \ref{thm: main estimation}, we have that with probability at least $2/3$ it holds that 
% \begin{align}
%   \norm{(\hat{v},\hat{\lambda})-(v^*,\lambda^*)}_2^2\le& \mathrm{poly}(\beta,\sigma_0+\frac{1}{\sigma_0},\frac{1}{a})\notag\\
%   \times &\exp\left(\mathrm{poly}(1/a)(1+\frac{\beta}{\sigma_0^2})\right) \frac{1}{b^2n}.\notag
% \end{align}
% The coefficient $120$ is absorbed in the $\mathrm{poly}$ term. 
% %Also using the strong convexity bound from Theorem \hyperref[10]{4.3}, we can prove
%$f(\hat{v},\hat{\lambda})-f(v^*,\lambda^*)\ge \frac{\delta}{2}( ||v-v^*||^2+(\lambda-\lambda^*)^2)$, and hence
%\begin{align}
%    ||v-v^*||_2^2+(\lambda-\lambda^*)^2\le& \mathrm{poly}(\beta,\sigma_0+\frac{1}{\sigma_0},\frac{1}{a})\notag \\\times& \exp\left(\mathrm{poly}(1/a)(1+\frac{\beta}{\sigma_0^2})\right) \frac{1}{b^2n}\notag
%\end{align}
%The $2/\delta$ is absorbed in the $\exp$ term. 
% Thus, we can rewrite as 
$$\norm{(\hat{v},\hat{\lambda})-(v^*,\lambda^*)}^2 \le % &\mathrm{poly}(\beta,\sigma_0+\frac{1}{\sigma_0},\frac{1}{a})\notag\\
    % \times& \exp\left(\mathrm{poly}(1/a)(1+\frac{\beta}{\sigma_0^2})\right) \frac{1}{b^2n}\notag\\
    \frac{\poly(\sigma_0) \cdot \exp\p{\poly(\frac{\beta}{a \cdot (\sigma_0)})}}{b^2 \cdot n}$$
Again, using the same inequality \eqref{eq: A.4} to \eqref{eq: A.6}, we can transform the bound from square bound to a linear one:
\begin{align}
    &\norm{\hat{w} - w^*}+\abs{\hat{\sigma}^2 - {\sigma^*}^2}
    \le \sqrt{2\mathrm{poly} (1/a,{\sigma^*},\beta)\ \norm{(\hat{v},\hat{\lambda})-(v^*,\lambda^*)}^2 }\notag\\
    \le& \sqrt{2\mathrm{poly} (1/a,{\sigma^*},\beta)\frac{\poly(\sigma_0) \cdot \exp\p{\poly(\frac{\beta}{a \cdot (\sigma_0)})}}{b^2 \cdot n}}\notag\\
    = &\frac{\poly(\sigma_0) \cdot \exp\p{\poly(\frac{\beta}{a \cdot \sigma_0})}}{b \sqrt{n}}\notag
\end{align}
We get the last inequality because the $\poly(1/a,\sigma^*,\beta)$ can be absorbed into $\poly(\sigma_0)$ and exponential term of $\poly(\beta/\alpha\sigma_0)$. Therefore, the inequality \eqref{eq: 3.2} of Theorem \ref{thm: main estimation} follows.

\textit{Remark}: 
Again, we can calculate the bound by plugging Theorem \ref{thm:convexity, bounded step var} into \ref{thm: lr -1}. This gives us a difference of $\frac{4\rho^2}{\zeta^2 t}$, where $\rho$ is the bound on the step variance in Section \ref{sec: proof of step var} and $\zeta$ is the lower bound in the \ref{sec: proof of strong convexity}. Specifically, we have
$$\rho^2=O\Bigg(\frac{(1-\log a)^{10}}{a^8}(1+\beta^2+{\sigma^*}^4)\Bigg)$$
and from the remark in the \ref{sec: proof of strong convexity}, we have
$$\zeta = \Omega\left( \exp\left(-16s^2 \left(-\log a+2+\frac{2\beta}{{\sigma^*}^2}\right)\right)\frac{b}{b(1+\beta/\sigma^2)+1}\right)\min(\sigma^2,\sigma^4) $$ for all the possible $\sigma$ in the projection set.
Since we know that $\sigma^2/{\sigma^*}^2$ we may say $\sigma^2/{\sigma^*}^2\ge O\Big(\frac{a^2}{1-\log a}\Big)$. Also $s=\frac{96\times8(5-2\log a)}{a^2}$. For $\frac{(b(1+\beta/{\sigma}^2)+1)}{b\min({\sigma}^2,{\sigma}^4)}$, the numerator is $O(1+\beta/\sigma^2)=O(1+\frac{\beta/(1-\log a){\sigma^*}^2}{a^2})=O((1+{\beta/{\sigma^*}^2})\frac{1-\log a}{a^2})$, and the denominator is $b\min(\sigma^2,\sigma^4)=\Omega\Big(b\frac{\sigma^4}{1+\sigma^2}\Big)=\Omega\Big(b\frac{{\sigma^*}^4}{1+{\sigma^*}^2}\frac{(1-\log a)^2}{a^4}\Big)$. Therefore, we can write $\frac{1}{\zeta}$ as
$$O\Bigg(\exp\Bigg(\frac{9\times 2^{14}(1-\log a)}{a^2} \Big(-\log a+2+\frac{2\beta}{{\sigma^*}^2}\Big)\Bigg)\frac{1+\beta/{\sigma^*}^2}{b{\sigma^*}^4}(1+{\sigma^*}^2)\frac{(1-\log a)^3}{a^6}\Bigg)$$
Since we know that $\rho^2=O(\frac{(1-\log a)^{10}}{a^8}(1+\beta^2+{\sigma^*}^4))$ and the coefficient $O({\sigma^*}^4\frac{1-\log a}{a^2}(1+\beta))=:K$ for transforming $v,\lambda$ to $w,\sigma$, we can take the square root of this term, and calculate a bound for the final estimation. This is given by taking $\sqrt{\frac{4\rho^2}{\zeta^2 n}K}$ and then counting the both highest and lowest degrees of $\beta,{\sigma^*},a$
$$O\left(\exp\left(\frac{3\times 2^{12}(5-2\log a)}{a^2} \Big(-\log a+2+\frac{2\beta}{{\sigma^*}^2}\Big)\right)\frac{(1-\log a)^{8.5}}{a^{15}}(1+\beta^{2.5})({\sigma^*}^{-1}+{\sigma^*}^3)\frac{1}{b\sqrt{n}}\right)$$

\subsection{Proof of Theorem \ref{thm: asym normal} and Corollary \ref{cor: asym normal}}\label{sec: proof of conf reg}

We cite a lemma in the multivariate version of Proposition 2 in \cite{leluc2020towards}.
\begin{definition}
  \label{21}A stochastic algorithm is a sequence $(x_k)_{k\ge 0}$ of random variables defined in a probability space $(\Omega, \mathcal{F}, P)$ and valued in $\mathbb{R}^d$. Define $(\mathcal{F}_{k})_{k\ge 0}$ as the natural $\sigma$-field associated to the stochastic algorithm $(x_k)_{k\ge 0}$, i.e., $\mathcal{F}_k = \sigma(x_0, x_1, \cdots , x_k), k \ge 0$. A policy is a sequence of random probability measures $(P_k)_{k\ge 0}$, each defined on a
measurable space $(S, \mathcal{S})$ that are adapted to $\mathcal{F}_k$.
\end{definition}
\begin{definition}
  \label{22} Given a policy $(P_k)_{k\ge 0}$ and a learning rates sequence $(\alpha_k)_{k\ge 1}$ of positive numbers, the SGD algorithm is defined by the update rule
$x_k = x_{k-1} - \alpha_k C_k g(x_{k-1}, \xi_k)$ where $\xi_k \sim P_{k-1}$ with $g: \mathbb{R}^d \times S \to \mathbb{R}^d$
\end{definition}
\begin{theorem}
    \label{23}{ Proposition 2 in \cite{leluc2020towards}. If we suppose that the assumptions below are fulfilled: 
\begin{enumerate}
    \item The gradient estimation of $F$ is unbiased, that is, for $k\ge 1$, the expected sampled gradient is equal to the total gradient. ($\mathbb{E}(g(x_{k-1}, \xi_k)|\mathcal{F}_{k-1})=\nabla F(x_{k-1})$)
    \item The sequence $(\alpha_k)_{k\ge1}$ is positive, decreasing to 0, and satisfies the Robbins-Monro condition: $\sum_{k\ge1} \alpha_k = +\infty$ and $\sum_ {k\ge1} \alpha^2_k<\infty$.
    \item The objective function is $L-$smooth.
    \item The objective function has only one minimum point $x^*$ and $\lim_{\norm{x}\to\infty} F(x)=\infty$ 
    \item With probability 1, there exist $0 \le \mathcal{L}, \sigma^2 <\infty$ such that
    \begin{align}
        \forall x \in \mathbb{R}^d, \forall k \in \mathbb{N}, \mathbb{E}(||g(x_{k-1}, \xi_k)||^2|\mathcal{F}_{k-1})
        \le 2\mathcal{L}(F(x)-F(x^*))+\sigma^2\notag
    \end{align}
    \item The sequence of step-size is equal to $\alpha_k = \alpha k^{-\beta}$ with $\beta \in (1/2, 1]$.
    \item $H = \nabla^2F(x^*)$ is positive definite and $x \mapsto \nabla^2F(x)$ is continuous at $x^*$.
    \item Denote $w_k = \nabla F(x_{k-1}) - g(x_{k-1}, \xi_k)
    $ and $\gamma_k = \mathbb{E}(w_{k+1}w_{k+1}^T|\mathcal{F}_k)$. There is a positive definite matrix $\Gamma_k\xrightarrow{k\to \infty}\Gamma$.
    \item And there exist $\delta,\varepsilon>0$ such that almost surely $\sup_{k\ge 1}\mathbb{E}(||w_k||^{2+\delta}|\mathcal{F}_{k-1})\mathbf{1}_{||x-x^*||\ge \varepsilon}<\infty$.
\end{enumerate}
Assume that $(H - \kappa I)$ is positive definite where $\kappa = 1_{\beta=1}1/2\alpha$. Let $(x_k)_{k\ge0}$ be obtained by the SGD rule, then $\frac{1}{\sqrt{\alpha_k}}(x_k - x^*)$ weakly converge to a multivariate normal $\mathcal{N}(0,\Sigma)$ where $\Sigma$ satisfy the following equation:
$$(H - \kappa I)\Sigma + \Sigma(H^T - \kappa I) = \Gamma$$}
\end{theorem}

In our settings, we have $x_k=(v ~ \lambda)^{(k)}$ and the $\xi_k$ is new data $(x_k,y_k)$. The gradient $g(x_k,\xi_k)$ is $\nabla -\ell(v^{(k)}, \lambda^{(k)}; x^{(k)},y^{(k)})$. Also, we have set $C_k=I,\alpha=\frac{1}{\zeta}, \beta=1$ where $\zeta I$ is the lower bound of the Hessian matrix $\mathbf{H}$ at $(v^*,\lambda^*)$. Here, we can see that $\mathbf{H}-\frac{1}{2\alpha} I$ is positive definite, since $\mathbf{H}\succeq \frac{1}{\alpha} I\succ \frac{1}{2\alpha}I$. In the section below, we prove that the nine assumptions hold in our settings. Also, in our settings, $\mathbf{H}$ is symmetric, so $\mathbf{H}^T=\mathbf{H}$. We will omit the transpose of $H$ in the future content.

\subsubsection{Assumptions 1-7}

\begin{itemize}
\item Assumption 1 holds for our algorithm: each time the sampled $u_i$ is an unbiased estimator of $\nabla \bar{\ell} (v^{(t - 1)}, \lambda^{(t - 1)})$.

\item Assumption 2 holds because of our chosen parameter. The step size is $\eta_t= 1/\zeta\cdot t$ which also satisfies assumption 6.

\item Assumption 3 is proved in the subsection \ref{sec: smoothness}.

\item Assumptions 5 and 7 are implied by Theorem \ref{thm:convexity, bounded step var} for the bounded step variance and strong convexity.

\item From the strong convexity of the Hessian Matrix, we derive that the solution of $\nabla \bar\ell$ is unique since $\bar\ell$ has an unique minimum. Also, since the Hessian matrix is strongly convex everywhere, it can be deduced that when $\norm{(v,\lambda)}\to\infty$, the $\bar\ell$ goes to infinity. This proves assumption 4.
\end{itemize}

\subsubsection{Assumptions 8 \& 9}

First, we propose another proposition, to state that $(v,\lambda)$ converge almost surely to $(v^*, \lambda^*)$:
\begin{theorem} \label{thm: lp proposition1}
    \label{24} Proposition 1 in \cite{leluc2020towards}: Assumptions 1 to 5 are fulfilled. Then the sequence of iterates
$(x_k)_{k\ge0}$ obtained by the SGD rule in Definition \ref{22} converges almost surely towards the minimizer $x_k \to x^*$
\end{theorem}

Since $(v,\lambda)$ converge almost surely to $(v^*, \lambda^*)$, the $\bar\ell$ is twice differentiable, and $x^{(i)},y^{(i)}$ are i.i.d, then we have that the distribution of $w_k$ converges to the case when $v,\lambda\to v^*, \lambda^*$. So, $\Exp[w_kw_k^T]$ converges to some matrix $\Gamma$. The matrix $\Gamma$ is positive definite, since it is a positive combination of some matrices of form $vv^T$. Since $n\to \infty$ the parameter $(v, \lambda)$ converge to $(v^*,\lambda^*)$ almost surely, and the gradient of $\bar\ell$ at the true parameter is zero, we have \begin{align}
    \Gamma=&\Exp_{x\sim\cal P}\Exp_{y}\left[(\frac{\partial l(x,y;v,\lambda)}{\partial(v,\lambda)}-\nabla\bar\ell )(\frac{\partial l(x,y;v,\lambda)}{\partial(v,\lambda)}-\nabla\bar\ell )^T\right]\notag\\
    =&\Exp_{x\sim\cal P}\Exp_{y}\left[\left(\frac{\partial l(x,y;v,\lambda)}{\partial(v,\lambda)}\right)\left(\frac{\partial l(x,y;v,\lambda)}{\partial(v,\lambda)}\right)^T\right]\notag
\end{align} when $(v,\lambda)=(v^*,\lambda^*)$. Since the assumption \ref{asm: thickness} gives $\Exp_{x\sim\calP} {x}{x^T}\succeq bI$, the rank of $x^{(i)}$ is full. Therefore, the gradients will not all be on the same linear subspace (since the gradient is some scalar times $x^{(i)}$ and $y^{(i)}$, which itself is generated with noise.) Hence, $\Gamma$ is positive definite.

For assumption 9, notice that if we bound $v,\lambda$ by the projection set, the $w,\sigma$ are both bounded. Thus, $\forall (v,\lambda)\in D_r$, the gradient is bounded if we set $\varepsilon$ such that $||(v, \lambda)-(v^{*}, \lambda^*)||$ to be inside the projection set. Also $\mathcal{P}$ is bounded. So, if $y\sim F_x=\mathcal{N}({w^*}^T x,{\sigma^*}^2,S)$, we have

\[ \frac{\partial {\ell}}{\partial v}= (\mathbb{E}_{z\sim Q_x}[z]-y)x, ~~ \frac{\partial {\ell}}{\partial \lambda}= \frac{1}{2} (y^2-\mathbb{E}_{z \sim Q_x}[z^2]) \]

Where $Q_x=\mathcal{N}({w}^T x, {\sigma}^2,S)$. Next, we show that $\mathbb{E}_{y}\left[\norm{\frac{\partial {\ell}}{\partial v}}^4+\left(\frac{\partial {\ell}}{\partial \lambda}\right)^4\right]$ is bounded. We know that the expression can be written as
\begin{align}
     \mathbb{E}_{y}\left[\norm{\frac{\partial {\ell}}{\partial v}}^4+\left(\frac{\partial {\ell}}{\partial \lambda}\right)^4\right]
    =  &\frac{ \mathbb{E}_{y\sim \mathcal{N}({w^*}^T x,{\sigma^*}^2)}\left[\mathbf{1}_{y\in S}\left(\norm{\frac{\partial {\ell}}{\partial v}}^4+\left(\frac{\partial {\ell}}{\partial \lambda}\right)^4\right)\right]}{\alpha({w^*},{\sigma^*}, x)^4}\notag\\
    \le & \frac{1}{a^4}{\mathbb{E}_{y\sim \mathcal{N}({w^*}^T x,{\sigma^*}^2)}\left[\norm{\frac{\partial {\ell}}{\partial v}}^4+\left(\frac{\partial {\ell}}{\partial \lambda}\right)^4\right]}\notag
\end{align}
Because $\mathbb{E}[z]$ and $\mathbb{E}[z^2]$ are bounded, and $x$ is also bounded. Also, ${w^*}^Tx$ and $\sigma^*$ are bounded. the expression inside the expectation can be written as a polynomial of $y$ of degree $\le 8$. Notice that for the polynomial $P(t)$, if $t$ is a given normal distribution $\mathcal{N}(\mu,\sigma)$ where $\mu$ and $\sigma$ are bounded, $\mathbb{E}_{y}(P(t))$ is bounded. Thus, we have $\mathbb{E}_{y}\left[\norm{\frac{\partial {\ell}}{\partial v}}^4+\left(\frac{\partial {\ell}}{\partial \lambda}\right)^4\right]$ is bounded if we choose $x$ in the projection set.

Therefore, $||w_k||^4=||\frac{\partial \ell(v,\lambda,x,y)}{\partial (v,\lambda)}-\nabla \bar\ell||^4\le 8(||\frac{\partial \ell(v,\lambda,x,y)}{\partial (v,\lambda)}||^4+||\nabla \bar\ell||^4)$ is also bounded, since the bound of the latter term can be derived by the bounded step variance in \ref{thm:convexity, bounded step var}.

Therefore, when we set $\varepsilon$ such that $||(v ~ \lambda)-(v^{*} ~ \lambda^*)||$ to be inside the projection set, and $\delta=2$, we can maintain a bounded expectation.

\subsubsection{Final Estimation}
Since all assumptions hold, we know that the difference of the estimator and true parameter $\left(\sqrt{k^{-1}/\zeta}\right)^{-1}((v~\lambda)-(v^*~\lambda^*))$ is asymptotically normal, and that the variance is given by equation $$(\mathbf{H} - \kappa I)\Sigma + \Sigma(\mathbf{H}^T - \kappa I) = \Gamma$$
Where $\mathbf{H}$ is the Hessian matrix at $(v^*,\lambda^*)$.

Notice that the $(v,\lambda)$ converge to $(v^*,\lambda^*)$ almost surely, so we can apply the plug-in confidence interval. 
We can substitute the true estimates by the empirical estimates as justified by the Slutsky's theorem.
Notice that the parameter $\hat{\mathbf{H}}={\mathbf{H}}(\hat v,\hat\lambda)$
converges to the true parameter $\mathbf{H}$ (because $v,\lambda$ is converging by Theorem \ref{thm: lp proposition1})
And $\hat\Gamma$ (defined below) converges to $\Gamma$ (by the last subsection.) 
Therefore, we can apply Slutsky's Theorem and we can substitute the estimated parameters 
for true parameters to calculate the confidence region.
We can substitute $\mathbf{H}$ to $\hat{\mathbf{H}}$  where $\hat{\mathbf{H}}=\mathbf{H}(\hat 
v,\hat\lambda)$ is the empirical Hessian matrix evaluated at $(\hat v,\hat \lambda)$. Also, the empirical $\Gamma$ can be evaluated by $\Gamma(\hat v,\hat\lambda)$, where
\[\Gamma(v,\lambda)=\frac{1}{n}\sum_{i=1}^n \frac{\partial l(v,\lambda;x^{(i)},y^{(i)})}{{\partial (v,\lambda)}}\frac{\partial l(v,\lambda;x^{(i)},y^{(i)})}{{\partial (v,\lambda)}}^T\]
That is, we can write
$$\sqrt{n}(\hat\theta-\theta')\to \mathcal{N}(0,\zeta^{-1}\Sigma)$$
Where $(\hat{\mathbf{H}} - \kappa I)\Sigma + \Sigma(\hat{\mathbf{H}} - \kappa I) = \Gamma(\hat v,\hat\lambda)$, and $\theta$ is the joined vector of $v,\lambda$. Notice that here we are converging to a random matrix, so we interpret this estimation as 
$$\zeta^{1/2}\Sigma^{-1/2}\sqrt{n}(\hat\theta-\theta')\to \mathcal{N}(0,I)$$
By the delta method, since we have $(w,\sigma)=(v/\lambda,1/\lambda)$, we know that
\begin{align}
    \nabla(\theta)=&J^T=\left.\frac{\partial( w,{\sigma}^2)}{\partial( v,\lambda)}\right|_{v=\hat v,\lambda=\hat\lambda}=\begin{pmatrix}
    1/\hat\lambda I & 0\\
    -\hat v^T/\hat\lambda^2 & -1/\hat\lambda^2
    \end{pmatrix}
    =\begin{pmatrix}
    {\hat\sigma}^2I & 0\\
    -{\hat w}^T{\hat\sigma}^2 & -{\hat\sigma}^4
    \end{pmatrix}\notag
\end{align}

So we have (now $w,\sigma^2$ are written in terms of $\theta$)
$$\sqrt{n}(\hat\theta-\theta)\sim \mathcal{N}(0,J(\zeta^{-1}\Sigma) J^T)$$
Or, in other terms, let $S=J(\zeta^{-1}\Sigma) J^T$, we have
$$\sqrt{n}S^{-1/2}(\hat\theta-\theta)\sim \mathcal{N}(0,I)$$
Thus, Theorem \ref{thm: asym normal} is proved. In addition, this yields
$$(\hat\theta-\theta)^T(\frac{1}{n}J(\zeta^{-1}\Sigma)J^T)^{-1}(\hat\theta-\theta)\sim \chi_{k+1}$$

Let $q_{\alpha}$ is the $1-\alpha$ quantile of the distribution $\chi_{k+1}$, so we have the confidence region
$$(\hat\theta-\theta)^T(J(\zeta^{-1}\Sigma)^{-1}J^T)^{-1}(\hat\theta-\theta) \le q_{\alpha}/n$$

Where we can calculate
$$J^{-1}=\begin{pmatrix}{1}/{\hat\sigma^2}I & -{\hat w}/{\hat\sigma^4} \\
0 & -{1}/{\hat\sigma^4}
\end{pmatrix}$$

To make it clearer, we define $R=J^{-1}$. So
Or, write back to $\hat w,\hat\sigma$, as
$$\left(\begin{pmatrix}w\\\sigma^2\end{pmatrix}-\begin{pmatrix}\hat w\\\hat\sigma^2\end{pmatrix}\right)^TR^T\Sigma^{-1}R\left(\begin{pmatrix}w\\\sigma^2\end{pmatrix}-\begin{pmatrix}\hat w\\\hat\sigma^2\end{pmatrix}\right)\le \frac{q_{\alpha}}{\zeta n}$$
\section{Experiment Setup}
In our main paper, we provide theoretical guarantees for truncated linear regression with unknown noise variance and test our procedure on synthetic data. Here, we give an overview of how we set up these experiments. In our experimental section, we mention that we perform each experiment for an specified number of trials. For each algorithm a trial means something different. For the \cite{truncreg} experiments, we write the dataset of interest to a csv file, which is then read in an R script and the procedure is run. For \cite{daskalakis2019computationally} and our algorithm, a trial is considered complete in a of couple ways. First off, every 100 steps, we check the $L^{2}$ norm of the validation set's gradient. If its magnitude is less than $1e-1$, we terminate the procedure, and return the current estimates. However, if after taking 2500 gradient steps, the validation set's gradient $L^{2}$ norm is greater than $1e-1$, we re-run the stochastic process. We do this a maximum of three times, and return the trial with the smallest gradient. 

For these experiments, we use a PyTorch SGD optimizer, starting our procedure with a learning rate of $1e-1$, and decaying it at a rate of $.9$ every 100 gradient steps. 

All of the experiments performed in this paper were performed on a 15-inch MacBookPro with a 2.2 GHz 6-Core Intel Core i7 with 16 GB of memory. It is of note to mention that 2500 gradients steps with batch size 10 ran in a maximum of 3 seconds for the experiments.

\newpage
\onecolumn

\section{Code}

We provide code for our experiment at the following GitHub repository: \url{https://github.com/pstefanou12/Truncated-Regression-With-Unknown-Noise-Variance-NeurIPS-2021}.

Below, we provide the gradient that we used for conducting all of our experiments.

\begin{lstlisting}[language=Python, caption={Truncated version of mean squared-error loss}]
class TruncatedUnknownVarianceMSE(ch.autograd.Function):
    """
    Computes the gradient of negative population log likelihood for 
    truncated linear regression with unknown noise variance.
    """
    @staticmethod
    def forward(ctx, pred, targ, lambda_, phi):
        ctx.save_for_backward(pred, targ, lambda_)
        ctx.phi = phi
        return 0.5 * (pred.float() - targ.float()).pow(2).mean(0)

    @staticmethod
    def backward(ctx, grad_output):
        pred, targ, lambda_ = ctx.saved_tensors
        # calculate std deviation of noise distribution estimate
        sigma = ch.sqrt(lambda_.inverse())
        stacked = pred[None, ...].repeat(args.num_samples, 1, 1)
        # add noise to regression predictions
        noised = stacked + sigma * ch.randn(stacked.size())
        # filter out copies that fall outside of truncation set
        filtered = ctx.phi(noised)
        z = noised * filtered
        lambda_grad = .5 * (targ.pow(2) - (z.pow(2).sum(dim=0) / 
        (filtered.sum(dim=0) + args.eps)))
        """
        multiply the v gradient by lambda, because autograd computes 
        v_grad*x*variance, thus need v_grad*(1/variance) to cancel 
        variance factor
        """
        out = z.sum(dim=0) / (filtered.sum(dim=0) + args.eps)
        return lambda_ * (out - targ) / pred.size(0), 
        targ / pred.size(0), lambda_grad / pred.size(0), None
\end{lstlisting}

\section{Future Work} 
There are many ways to build upon our work. One thing to point out is that, we use an SGD framework, while \cite{truncreg} uses an analytic
gradient and Hessian provided by the \cite{maxLik} package and the Newton-Raphson method. Since we do not explicitly calculate integrals in our method, our framework could be used to train non-linear models, including neural networks. Further, now that there are two methods to learn from truncated linear models, it would be interesting to explore an actual example where a dataset has been \textit{truncated} or \textit{censored} due to uncontrollable circumstances. With multiple methods for learning from truncated samples, it would interesting to see what results each of the method's give. An interesting field to explore would be environmental sciences, as there are a lot of examples where there is truncation due to measurement instrumentation failure or natural causes. One last to ponder would be what theoretical guarantees can be derived for an algorithm in the \textit{censored} setting. We emphasize that our algorithm for truncated data will work in the censored setting, but with additional knowledge of all the covariate features, can we design an algorithm with better error and/or run-time bounds?


\begin{thebibliography}{29}
\providecommand{\natexlab}[1]{#1}
\providecommand{\url}[1]{\texttt{#1}}
\expandafter\ifx\csname urlstyle\endcsname\relax
  \providecommand{\doi}[1]{doi: #1}\else
  \providecommand{\doi}{doi: \begingroup \urlstyle{rm}\Url}\fi

\bibitem[Aldrin(2004)]{pm10}
Aldrin, M.
\newblock Pm10 dataset, 2004.
\newblock URL \url{http://lib.stat.cmu.edu/datasets/PM10.dat}.
\newblock PM particles at Oslo, Norway.

\bibitem[Amemiya(1973)]{amemiya1973regression}
Amemiya, T.
\newblock Regression analysis when the dependent variable is truncated normal.
\newblock \emph{Econometrica: Journal of the Econometric Society}, pp.\
  997--1016, 1973.

\bibitem[Balakrishnan \& Cramer(2014)Balakrishnan and
  Cramer]{BalakrishnanCramer}
Balakrishnan, N. and Cramer, E.
\newblock \emph{The art of progressive censoring}.
\newblock Springer, 2014.

\bibitem[Breen et~al.(1996)]{breen1996regression}
Breen, R. et~al.
\newblock \emph{Regression models: Censored, sample selected, or truncated
  data}, volume 111.
\newblock Sage, 1996.

\bibitem[Carbery \& Wright(2001{\natexlab{a}})Carbery and Wright]{CarberyW01}
Carbery, A. and Wright, J.
\newblock Distributional and l\^{q} norm inequalities for polynomials over
  convex bodies in r\^{n}.
\newblock \emph{Mathematical research letters}, 8\penalty0 (3):\penalty0
  233--248, 2001{\natexlab{a}}.

\bibitem[Carbery \& Wright(2001{\natexlab{b}})Carbery and
  Wright]{carbery2001distributional}
Carbery, A. and Wright, J.
\newblock Distributional and ${L}^{q}$ norm inequalities for polynomials over
  convex bodies in $\mathbb{R}^n$.
\newblock \emph{Mathematical research letters}, 8\penalty0 (3):\penalty0
  233--248, 2001{\natexlab{b}}.

\bibitem[Croissant \& Zeileis(2018)Croissant and Zeileis]{truncreg}
Croissant, Y. and Zeileis, A.
\newblock truncreg: Estimation of models for truncated gaussian variables by
  maximum likelihood, 2018.
\newblock URL
  \url{https://cran.r-project.org/web/packages/truncreg/truncreg.pdf}.
\newblock R package version 1.8.0.

\bibitem[Daskalakis et~al.(2018)Daskalakis, Gouleakis, Tzamos, and
  Zampetakis]{daskalakis2018efficient}
Daskalakis, C., Gouleakis, T., Tzamos, C., and Zampetakis, M.
\newblock Efficient statistics, in high dimensions, from truncated samples.
\newblock In \emph{2018 IEEE 59th Annual Symposium on Foundations of Computer
  Science (FOCS)}, pp.\  639--649. IEEE, 2018.

\bibitem[Daskalakis et~al.(2019)Daskalakis, Gouleakis, Tzamos, and
  Zampetakis]{daskalakis2019computationally}
Daskalakis, C., Gouleakis, T., Tzamos, C., and Zampetakis, M.
\newblock Computationally and statistically efficient truncated regression.
\newblock In \emph{Conference on Learning Theory (COLT)}, pp.\  955--960, 2019.

\bibitem[Fisher(1931)]{fisher31}
Fisher, R.
\newblock Properties and applications of hh functions.
\newblock \emph{Mathematical tables}, 1:\penalty0 815--852, 1931.

\bibitem[Fotakis et~al.(2020)Fotakis, Kalavasis, and
  Tzamos]{fotakis2020efficient}
Fotakis, D., Kalavasis, A., and Tzamos, C.
\newblock Efficient parameter estimation of truncated boolean product
  distributions.
\newblock In \emph{Conference on Learning Theory}, pp.\  1586--1600. PMLR,
  2020.

\bibitem[Galton(1897)]{Galton1897}
Galton, F.
\newblock An examination into the registered speeds of american trotting
  horses, with remarks on their value as hereditary data.
\newblock \emph{Proceedings of the Royal Society of London}, 62\penalty0
  (379-387):\penalty0 310--315, 1897.

\bibitem[Hajivassiliou \& McFadden(1998)Hajivassiliou and
  McFadden]{hajivassiliou1998method}
Hajivassiliou, V.~A. and McFadden, D.~L.
\newblock The method of simulated scores for the estimation of ldv models.
\newblock \emph{Econometrica}, pp.\  863--896, 1998.

\bibitem[Hausman \& Wise(1977)Hausman and Wise]{hausman1977social}
Hausman, J.~A. and Wise, D.~A.
\newblock Social experimentation, truncated distributions, and efficient
  estimation.
\newblock \emph{Econometrica: Journal of the Econometric Society}, pp.\
  919--938, 1977.

\bibitem[Heckman(1979)]{heckman79}
Heckman, J.~J.
\newblock Sample selection bias as a specification error.
\newblock \emph{Econometrica: Journal of the econometric society}, pp.\
  153--161, 1979.

\bibitem[Henningsen \& Toomet(2011)Henningsen and Toomet]{maxLik}
Henningsen, A. and Toomet, O.
\newblock maxlik: A package for maximum likelihood estimation in {R}.
\newblock \emph{Computational Statistics}, 26\penalty0 (3):\penalty0 443--458,
  2011.
\newblock \doi{10.1007/s00180-010-0217-1}.
\newblock URL \url{http://dx.doi.org/10.1007/s00180-010-0217-1}.

\bibitem[Ilyas et~al.(2020)Ilyas, Zampetakis, and
  Daskalakis]{ilyas2020theoretical}
Ilyas, A., Zampetakis, E., and Daskalakis, C.
\newblock A theoretical and practical framework for regression and
  classification from truncated samples.
\newblock In \emph{International Conference on Artificial Intelligence and
  Statistics}, pp.\  4463--4473. PMLR, 2020.

\bibitem[Kontonis et~al.(2019)Kontonis, Tzamos, and
  Zampetakis]{kontonis2019efficient}
Kontonis, V., Tzamos, C., and Zampetakis, M.
\newblock Efficient truncated statistics with unknown truncation.
\newblock In \emph{2019 IEEE 60th Annual Symposium on Foundations of Computer
  Science (FOCS)}, pp.\  1578--1595. IEEE, 2019.

\bibitem[Leluc \& Portier(2020)Leluc and Portier]{leluc2020towards}
Leluc, R. and Portier, F.
\newblock Towards asymptotic optimality with conditioned stochastic gradient
  descent.
\newblock \emph{arXiv preprint arXiv:2006.02745}, 2020.

\bibitem[Mohan et~al.(2013)Mohan, Pearl, and Tian]{Pearl_data_missingness}
Mohan, K., Pearl, J., and Tian, J.
\newblock Graphical models for inference with missing data.
\newblock In \emph{Proceedings of the 26th Annual Conference on Neural
  Information Processing Systems (NeurIPS)}, 2013.

\bibitem[Moitra et~al.(2021)Moitra, Mossel, and Sandon]{moitra2021learning}
Moitra, A., Mossel, E., and Sandon, C.
\newblock Learning to sample from censored markov random fields.
\newblock \emph{arXiv preprint arXiv:2101.06178}, 2021.

\bibitem[Pearson(1902)]{Pearson1902}
Pearson, K.
\newblock On the systematic fitting of frequency curves.
\newblock \emph{Biometrika}, 2:\penalty0 2--7, 1902.

\bibitem[Pearson \& Lee(1908)Pearson and Lee]{PearsonLee1908}
Pearson, K. and Lee, A.
\newblock On the generalised probable error in multiple normal correlation.
\newblock \emph{Biometrika}, 6\penalty0 (1):\penalty0 59--68, 1908.

\bibitem[Plevrakis(2021)]{plevrakis2021learning}
Plevrakis, O.
\newblock Learning from censored and dependent data: The case of linear
  dynamics.
\newblock \emph{Conference on Learning Theory -- COLT}, 2021.

\bibitem[Rakhlin et~al.(2011)Rakhlin, Shamir, and Sridharan]{rakhlin2011making}
Rakhlin, A., Shamir, O., and Sridharan, K.
\newblock Making gradient descent optimal for strongly convex stochastic
  optimization.
\newblock \emph{arXiv preprint arXiv:1109.5647}, 2011.

\bibitem[Shamir \& Zhang(2013)Shamir and Zhang]{shamir2013stochastic}
Shamir, O. and Zhang, T.
\newblock Stochastic gradient descent for non-smooth optimization: Convergence
  results and optimal averaging schemes.
\newblock In \emph{International conference on machine learning}, pp.\  71--79.
  PMLR, 2013.

\bibitem[Tobin(1958)]{tobin1958estimation}
Tobin, J.
\newblock Estimation of relationships for limited dependent variables.
\newblock \emph{Econometrica: journal of the Econometric Society}, pp.\
  24--36, 1958.

\bibitem[Tropp(2015)]{tropp2015introduction}
Tropp, J.~A.
\newblock An introduction to matrix concentration inequalities.
\newblock \emph{arXiv preprint arXiv:1501.01571}, 2015.

\bibitem[Tukey(1949)]{tukey49}
Tukey, J.~W.
\newblock Sufficiency, truncation and selection.
\newblock \emph{The Annals of Mathematical Statistics}, pp.\  309--311, 1949.

\end{thebibliography}
\end{document}